# AN EMPIRICAL ANALYSIS OF INTERNET PROTOCOL VERSION 6 (IPV6)

by

**IOAN RAICU**

**THESIS**

Submitted to the Graduate School

of Wayne State University,

Detroit, Michigan

in partial fulfillment of the requirements

for the degree of

**MASTER OF SCIENCE**

2002

MAJOR: COMPUTER SCIENCE

Approved by:

____________________________________

Advisor Date

## ACKNOWLEDGMENTS

Dr. Sherali Zeadally, my research advisor, is truly a one of a kind professor and mentor. I do not even want to think where I might have been if I would not have met him; I owe him everything for overseeing my research and advising me throughout my last two years of graduate school. Special thanks go to the three most important professors in my graduate career, Dr. Sherali Zeadally, Dr. Loren Schwiebert, and Dr. Monica Brockmeyer for everything they have taught me about computer networks and the academic world in general. It was their patience, expertise, and teaching ability that shaped me into what I am today and made my thesis possible.

I especially want to thank my fiancée, Daniela Stan, for creating all the ideal conditions for my academic career to succeed. It is obvious that without my parents, Margareta and Ion Raicu, and my brother, Mihai, none of my current achievements would have been possible. Family is truly the most important thing in my life. Although many would like to believe that careers and family are independent of each other, they are very intertwined to the point where neither one can prevail unless they both happily coexist.

Last but not least, I thank my many colleagues, Dr. Klaus-Peter Zauner, Naveed Ahmad, Davis Ford, Scott Fowler, Daniel Frank, Liqiang Zhang, Joe Ahmed, DV Sreenath, Jia Lu, Pan Juhua, Zhaoming Zhu, and Prarthana Kukkalli, for their friendship, support, and help in my various projects and discussions throughout my graduate career.

## PREFACE

The work presented in this thesis is intended to introduce the reader to both Internet Protocol version 4 (IPv4), 6 (IPv6), and its transition mechanisms. I have designed this thesis to be a comprehensive guide to the evaluation of IPv6 from beginning to end and, therefore, it is meant for audiences of varying expertise from beginners to experts who wish to learn about the next generation internet protocol (IPv6), the various transition mechanisms available, and have a clear unbiased performance overhead of the new internet protocol. Network architects and administrators might find our evaluation and information very insightful if they have to design and implement an IPv6 infrastructure.

## TABLE OF CONTENTS

## LIST OF TABLES

## LIST OF FIGURES

# CHAPTER 1

# INTRODUCTION

It is a well known fact today's networks, mainly the Internet, has surpassed IPv4's (Internet Protocol Version 4) [1, 9] capabilities. In the simplest definition of IP, the Internet Protocol is the heart of most of the modern networks. Without IP, the Internet as we know it would not have existed and therefore the fundamentals that made the original IP possible need to be preserved, enhanced, and redeployed in order for the Internet to survive. The shortcomings of IPv4 were seen well in advance, and therefore work started almost a decade ago. Its successor will be IPv6 (Internet Protocol Version 6) [2, 8, 9], and according to most experts, over the next five to ten years, IPv6 will be slowly integrated into the existing IPv4 infrastructure. [9]

IPv6 hopes to solve numerous problems that IPv4 has been plagued by over the past two decades, however it will accomplish it at a performance overhead. This performance loss is only using traditional data transfers in which many of IPv6's features supporting QoS (Quality of Service) traffic are not used. Although the QoS support features of IPv6 will be briefly discussed, it is a topic in itself which ultimately is outside the scope of our work for this thesis.

We tried to keep the experimentation as simple as possible to have a good base of comparisons before we attempt to repeat similar test using more features. Our experiments were conducted over an unloaded network using two routers and various workstations. Most of the experiments were conducted with the workstations running both Windows 2000 and Solaris 8.0, however some were only performed on Windows 2000.

Chapter 2 covers some background information about IPv4 and IPv6 in general, and some of the fundamental differences between the two network protocols; it also delves into the various transition mechanisms that are available when upgrading from IPv4 to IPv6. Chapter 3 explains the various test-bed configurations and each one's respective important characteristics. Chapter 4 offers the performance metrics as well as the experimental results and explanations of IPv4 versus IPv6. Chapter 5 offers a similar evaluation as chapter 4, except that it focuses on the transition mechanisms. Chapter 6 will cover future work and the conclusions drawn from our evaluation. Chapter 7 is Appendix A which is mainly a glossary of important terms with definitions. Finally, chapter 8 is Appnedix B which includes sample source code which clearly outlines the source code that is needed to implement both IPv4 and IPv6 ready applications.

Our goal was to perform an unbiased empirical performance evaluation between the two protocol stacks (IPv4 and IPv6), and at the same time, compare two different implementations (Windows 2000 and Solaris 8.0) on identical hardware and under identical settings. Through our experiments, we hope to emphasize the benefits and drawbacks to either of these network protocol stacks.

# CHAPTER 2

# BACKGROUND INFORMATION

This chapter's main goals are to familiarize the reader with the subject at hand. First some related work is discussed, in terms of what work has been done in the research community that is most similar to our work. With no surprise, we found nothing that was even close to the wide range of performance metrics and comparing two different implementations, two different network protocols, and two different transport protocols. By covering some of other people's work, our own motivation will prevail in terms of why we pursued the avenue we did and what is the value of our findings. An in-depth description of IPv4, IPv6, and the transition mechanisms are presented. In order to better understand the Internet Protocol, layering principles are first described in section 2.2.

## 2.1 Related Work

Our work was driven by the fact that there was no good comparison between IPv4 and IPv6 that was conducted in a scientific method and tried to depict the real world scenario in both IPv4 and IPv6 protocol stacks would have to traverse routers to reach their ultimate destination. Even if some of the experiments in the research community used routers, they were always software routers built from conventional PCs and had installed FreeBSD in order to handle the necessary routing.

Most of the industry wide routers implement most of their functionality in hardware and therefore are much more efficient than a software router approach which was taken by most researchers. Obviously there is a very good explanation to why nobody has tested IPv6's performance using real routers: hardware based routers supporting dual stack IPv4/IPv6 are rare and expensive. As an example, the two routers which we used for our experiments cost a total of US $60,000, which is a price tag out of reach for most research laboratories.

Furthermore, we also tested two different implementations, namely Windows 2000 and Solaris 8.0, side by side, throughout all of our experiments; we covered both TCP and UDP transport protocols. Our metrics included throughput, latency, CPU utilization, socket creation time, TCP connection time, the number of TCP connections per second, and the performance of a video application designed by our lab. This is in essence our contribution that nobody else has been able to accomplish: an unbiased empirical evaluation of two different implementations of IPv6 covering all the basic performance metrics and transport protocols under a realistic test-bed configuration.

The next few paragraphs will cover some of the work that had similar goals as our own; however they stopped short of accomplishing the task at hand when compared to our results. In [34], the first attempt at developing an IPv6 protocol stack for Windows NT is shown. The work presented is very old (early 1998) and offers some performance evaluation of a very specific instance of the wide range of tests we performed. I am sure they choose the best case scenario in order to show that their IPv6 implementation was almost as good as its IPv4 counterpart. They never mentioned what packet size they used for their transmission and they only utilized the TCP transport protocol. They also had no router and hence only connected the two PCs with a direct cable link. Most likely, there were no routers supporting IPv6 back in early 1998. They also did not do many tests such as latency, CPU utilization, socket creation time, etc.

In [36], the author evaluated the MSR IPv6 BETA protocol stack for Windows NT 4.0. The author evaluated the performance of MSR IPv6 protocol stack by measuring and analyzing its network latency, throughput, and processing overheads. Their test-bed consisted of two Pentium machines with 100Mbps fast Ethernet connected via an unloaded switch. The work presented seemed interesting and contained only a small part of our work. First of all, it only evaluated IPv6 and did not compare it with IPv4. Secondly, they only evaluated the Windows NT implementation and did not compare it with any other implementations. Notice that there were no routers involved in their experimentation and only connected their hosts with a switch. Obviously the findings they made are nearly obsolete since IPv6 and computing hardware evolved so much since 1999. For example the MSR IPv6 protocol stack has been replaced by the Windows 2000 IPv6 Preview Protocol Stack. Regardless, their work showed very interesting initial results on IPv6.

In [35], the authors evaluate the performance of data transmission over IPv4 and IPv6 suing various security protocols. The authors choose a particular application, namely digital video (DV) transmission in order to execute their experiments. They utilized end hosts with FreeBSD 2.2.8 and KAME IPv6 protocol stack and a router implemented in a PC platform also running FreeBSD 2.2.8 and KAME IPv6 protocol stack. The criticism of this work lies in the fact that the routers utilized obviously did not support most of the router functions in the hardware and therefore the depicted performance is lower than the performance in a real world scenario in which actual hardware routers would be utilized. One of the other criticisms is that they only covered small sample of the test we performed. They utilized two different buffer sizes (57344 bytes and 32769 bytes), which makes no sense; it is a known fact that when performing experiments of this nature, the buffer size is kept constant throughout all the experiments. They claim that the MTU size they used was either 1024 or 4096 bytes, however IP routers do not support MTU sizes above 1514 bytes. They might have had the functionality to change the MTU size beyond the maximum due to the software router implementation they were using. Obviously such a large MTU size might yield falsely higher than usual results. The only place where they mentioned the packet size, they specified 32 KB packets, but they called it the socket size. As an overall evaluation, the depicted results are interesting, but not complete in the sense of depicting real world performance.

In both [37, 33], the authors presented an evaluation of IPv6 compared to IPv4 using the dual stack implementation of KAME over FreeBSD OS using the ping utility and a FTP application; their metrics were latency and throughput. The major criticism of the work presented in [33, 37] is that the experiments were not done in a scientific manner. They used a ported FTP application to find out the throughput rates of the IPv6 protocol; they used the ping utility to find the latency. In [33], they had no router, but rather connected the two end hosts via a hub. In [37], they had a router which was a conventional PC using the FreeBSD router software. They obviously could not control any critical parameters, such as buffer size, packet size, and of course they could not perform any UDP tests due to the nature of FTP.

After reading all the related work performed in the research community, it should be clear that there was a need for the evaluation we performed in our research endeavors.

## 2.2 Layering Principles

Layering is one of the major reasons network architectures have been so successful. One great success story is the Internet, which shows how robust and scalable it has been despite the initial design goals which did not foresee the exponential growth that it indured.

Layering helps break complex problems into smaller more manageable pieces. It helps reduce design complexity and it simplifies the design and testing protocols. Sender and receiver software can be tested, designed and implemented independently. Layering prevents changes in software from propagation to other layers. It allows designers to construct protocol suites and allows ease of change regarding an implementation of a service. Some of its drawbacks include some performance loss, time delay, and perhaps having more than 1 copy of data at any given moment. Obviously, these drawbacks are quickly overshadowed by all the advantages of a layered approach to designing protocols.

The basic definition of layering is that the layer N software on the receiving machine should receive the exact message sent by the layer N software at the sender machine. It should satisfy whatever transformation was applied to the packet should be completely reversible at the receiving side.

### 2.2.1 OSI Reference Model

The OSI model is not a network architecture because it does not specify exact services and protocols. It is designed for open system interconnection. Each layer should represent a well defined function and a new layer is needed when a new level of abstraction is required. The layers should be chosen in order to minimizing flow of information across layers. And last of all, each layer should be chosen towards standardizing protocols.

I will be concentrating my efforts on three layers, namely the network layer, the transport layer, and the application layer. Figure 1 depicts the OSI reference model and its 7 layers. Since we will mainly concentrate on IPv4 and IPv6, it is relevant to discuss the TCP/IP reference model, in which we will describe some of the necessary layers in more detail.

The OSI model is composed of 7 layers:

| | |
|---|---|
| Layer 1: | Application layer |
| Layer 2: | Presentation layer |
| Layer 3: | Session layer |
| Layer 4: | Transport layer |
| Layer 5: | Network layer |
| Layer 6: | Data link layer |
| Layer 7: | Physical layer |

**Figure 1: OSI Reference Model**

### 2.2.2 TCP/IP Reference Model

The TCP/IP model is composed of 4 layers:

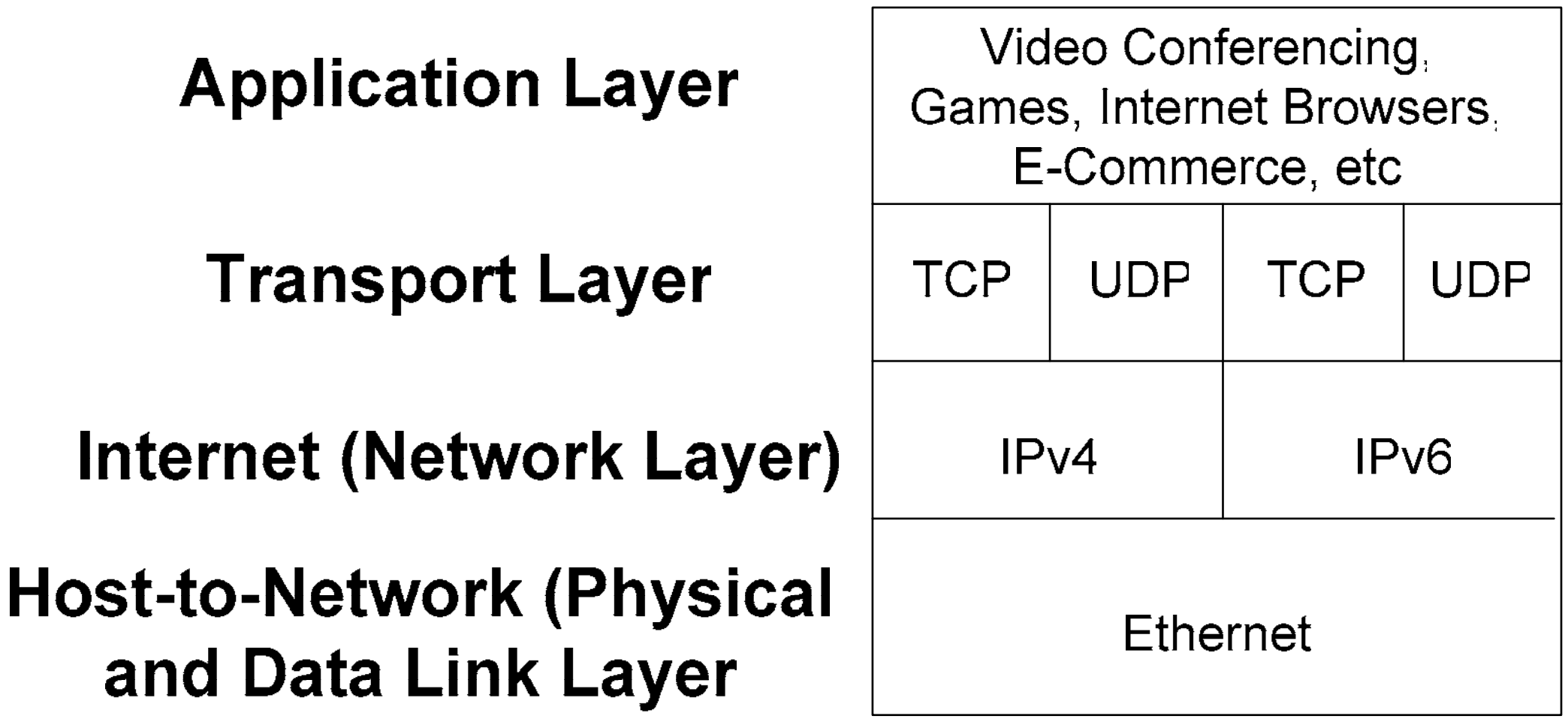

**Figure 2: TCP/IP Reference Model; on the left the various levels are identified while on the right examples of functionality/protocol at each respective layer**

The Internet layer, known as the network layer in the OSI model, allows heterogeneous networks to be connected. It provides congestion control, it establishes, maintains, and tears down connections, and most important of all, it determines the route of packets transmitted. Both the IP protocol versions, IPv4 and IPv6, are found in the network layer. Due to the diverse functionality of this layer, it should be obvious that it is very important that the services maintained by the network layer are the key to the entire protocol stack's triumph.

The transport layer provides reliable, transparent data transfers between senders and receivers. It provides error recovery mechanism and flow control in order to throttle the sending rates. It also fragments data into smaller pieces, and passes them down to the network layer. Both TCP and UDP are found in the transport layer.

The application layer has many protocols used in conjunction with the application. TELNET, FTP, and DNS are only a few that are among the protocols that applications can use.

### 2.3 IPv4 and IPv6 Architecture

Internet Protocol was first developed in the early 1980s. Its intent was to interconnect few nodes and was never expected to grow to the size of the Internet has become today. IPv4 was initially designed for best-effort service and only scaled to today's Internet size because of its state-less design. One of the few things that the creators of the Internet Protocol never envisioned was the exhaustion of a 32 bit address space. In the early 1990s, it became pretty evident that if the Internet will continue to grow at the exponential rate of doubling every eighteen months, the IPv4 address space would be depleted by the turn of the millennium. Some temporary solutions were offered, such as NAT (Network Address Translator) [4] or CIDR (Classless InterDomain Routing) [9], however work began on a new Internet Protocol, which was first called IPnG from Internet Protocol Next Generation, but later became known as IPv6, Internet Protocol version 6 (IPv6). IPv6 is the main focus of our work and hence this thesis.

The most evident reason for a new version of an IP was to increase the address space; IPv6 was designed with a 128 bit address schema, enough to label every molecule on the surface of the earth with a unique address ($7x10^{23}$ unique IP addresses per square meter) [9]. Even in the most pessimistic scenario of inefficient allocation of addresses, there would still be well over 1000 unique IP addresses per square meter of the earth [5]. There were other reasons that were a bit more subtle, such as better support for inelastic traffic and real time applications, and without doubt will most likely drive the deployment of IPv6 just as hard as the address space depletion problem. Twenty years ago, the only kind of traffic that existed on the internet was elastic traffic, such as emails or file transfers. Elastic traffic enjoys having high bandwidth and low latency, however if the network can only deliver a small percentage of its capacity, than the transmission will still deliver the data just as good, but just at a later time. On the other hand, inelastic traffic has much more stringent restrictions in which bad network performance can render the data useless. In the past five years, multimedia applications have emerged and have mostly dominated the Internet's growth and demand for more bandwidth and processing power. IPv6 was designed for both elastic and inelastic traffic in its vision scope. That does not mean that IPv6 is not a best effort service anymore, but merely that it has the potential to interoperate much easier with Quality of Service (QoS) architectures such as RSVP [15], Integrated Services (Intserv) [16,17], and Differentiated Services (Diffserv) [18] in order to make end-to-end QoS over IP-based networks a reality. These features of IPv6 are outside the context of this paper, so please refer to Chapter 5 in regards to future work.

Some of the differences between IPv4 and IPv6 features are outlined in the next few statements. Keep in mind that most of the improvements on IPv6 were done with three things in mind: scalability, security, and support for multimedia transmissions. First of all, the address space is increased from 32 bits to 128 bits. Obviously, this increase in address space means more capacity for nodes, but it also enlarges the header overhead and the routing tables' size. Unlike IPv4, IPSec support has become a requirement in the IPv6 header. This was a much needed improvement to at least offer basic security features. Payload identification for QoS handling by routers is now supported by the flow label field. This was introduced primarily because of the earlier statements about multimedia applications that require more stringent guarantees of data delivery. Fragmentation support has been moved from both routers and sending hosts to just sending hosts. This is an important fact due to the amount of work that the routers have been alleviated by, and therefore it improves scalability. The IPv6 header does not include a checksum and has no options included in the header, but rather introduces extension headers. This allows faster processing at the routers by performing the checksum less often and analyzing only the header information needed. Finally, IPv6 requires no manual configuration or DHCP, which will become more and more important as the number of nodes increases. Overall, IPv6 was carefully thought out and was designed with future applications in mind. [9]

Theoretically, taking a close look at the brake-down of the various headers in both IPv4 and IPv6, it is evident that the overhead incurred is minimal between IPv4 and IPv6. As a quick overview of Table 1 found below, the primary difference between IPv4 and IPv6 is that IPv4 has a 20 byte header while IPv6 has a 40 byte header. Although the address space in IPv6 is four times the size of its counterpart, IPv6 has decreased the number of required fields and made them optional as extension headers. Let's take the IPv4 UDP packet as an example to better understand Table 1. The total Ethernet MTU (Maximum Transfer Unit) is 1514 bytes, from which 14 bytes are the Ethernet header, 20 bytes are the IP header, and 8 bytes are the UDP header. The payload for a UDP packet in IPv4 is 1472 bytes, and is computed by Equation 1:

$$\mathbf{MTU = Payload + TLH + NLH + DLLH}$$

**Equation 1: MTU calculation; the formula used in deriving Table 1; payload is the application layer data size; TLH is the transport layer (TCP/UDP) header size; NLH is the network layer (IP) header size; DLLH is the data link (Ethernet) layer**

**header size; MTU is the total Ethernet MTU size that is transmitted on the physical medium.**

| | IPv4 TCP | IPv6 TCP | IPv4 UDP | IPv6 UDP |
|---|---|---|---|---|
| **TCP/UDP Payload** | 1460 | 1440 | 1472 | 1452 |
| **TCP/UDP Header** | 20 | 20 | 8 | 8 |
| **IP Payload** | 1480 | 1460 | 1480 | 1460 |
| **IP Header** | 20 | 40 | 20 | 40 |
| **Ethernet Header** | 14 | 14 | 14 | 14 |
| **Total Ethernet MTU** | 1514 | 1514 | 1514 | 1514 |
| **Overhead %** | 3.7% | 5.14% | 2.85% | 4.27% |

**Table 1: Packet breakdown and overhead incurred by header information; please refer to Equation 1 for obtaining the information above**

The difference between IPv4 and IPv6 would most obviously be the IP header, which instead of being 20 bytes, would now be 40 bytes. The overhead that is incurred by having header information can be figured out by taking the total Ethernet MTU and dividing by the TCP or UDP payload. For example, the difference between IPv4 UDP and IPv6 UDP is a mere 1.42 %, while for TCP it is almost the same at 1.44 %.

In theory, the performance overhead between these two protocols is so minimal that the benefits of IPv6 should quickly overshadow the negatives. In Chapter 4 and 5, I will discuss the performance evaluation in reality between IPv4 and IPv6, which proved to be quite a bit larger than the theoretical difference.

In order to better visualize the layering principles, we captured a screen shot of Microsoft Network Monitor as it displays a packet and all its header information and placed in Figure 3 and Figure 4. Figure 3 displays a ping echo (ICMP) message and its header information. Notice that the IP version is 4 and the IP header length is 20 bytes. Notice also the source and destination addresses, as they are all part of the packet header information.

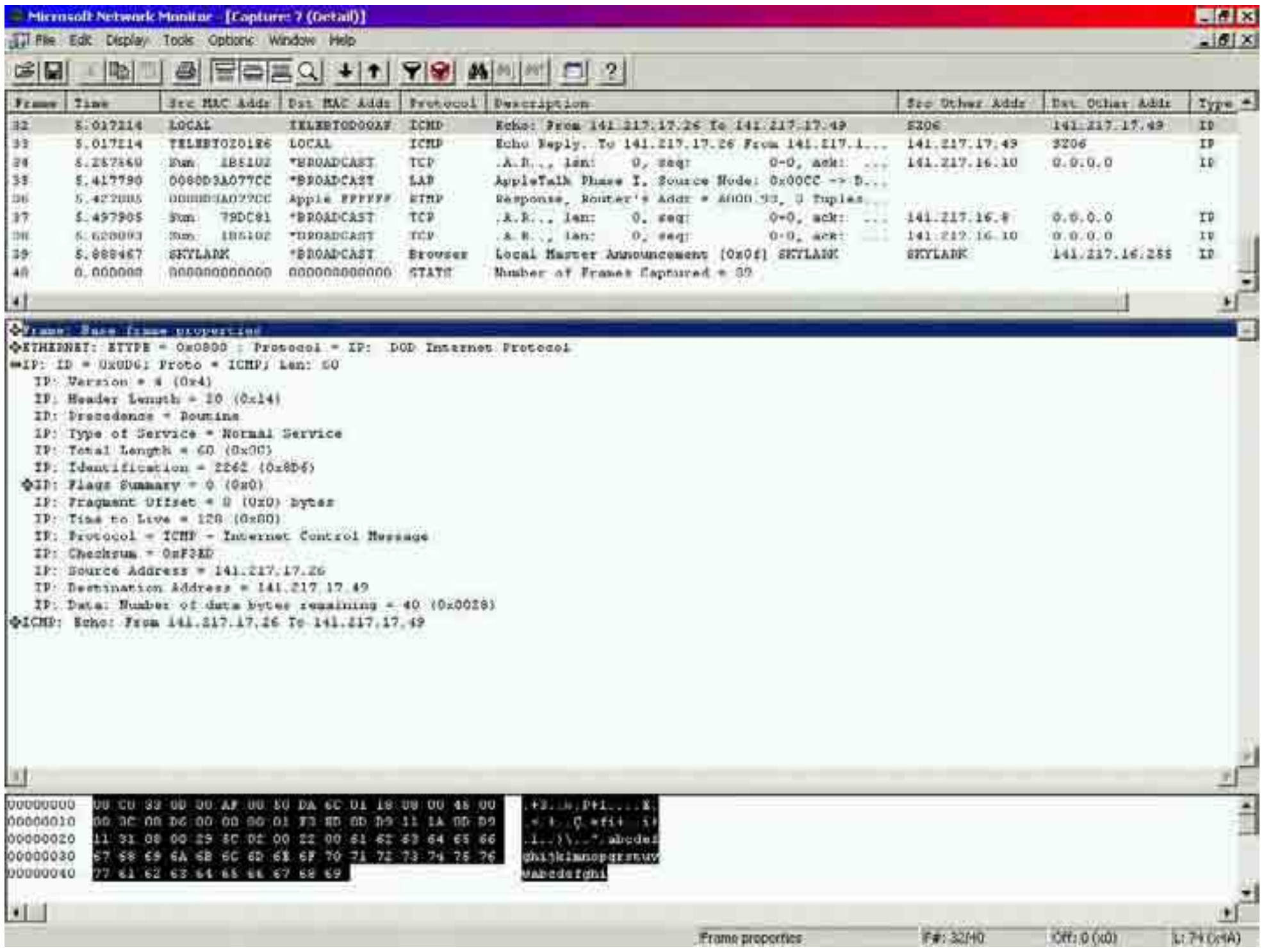

**Figure 3: IPv4 Packet as depicted by the Microsoft Network Monitor**

Figure 4 will show a similar screen shot, but this time presenting an IPv6 packet. Figure 4 displays a ping echo (ICMP) message and its header information for an IPv6 packet. Notice that the IP version is 6 and the IP header length is 40 bytes. Notice also the IPv6 128 bit source and destination addresses, as they are all part of the packet header information. Some new fields can also be seen, such as priority, flow label, and next header. We will discuss these in more detail later.

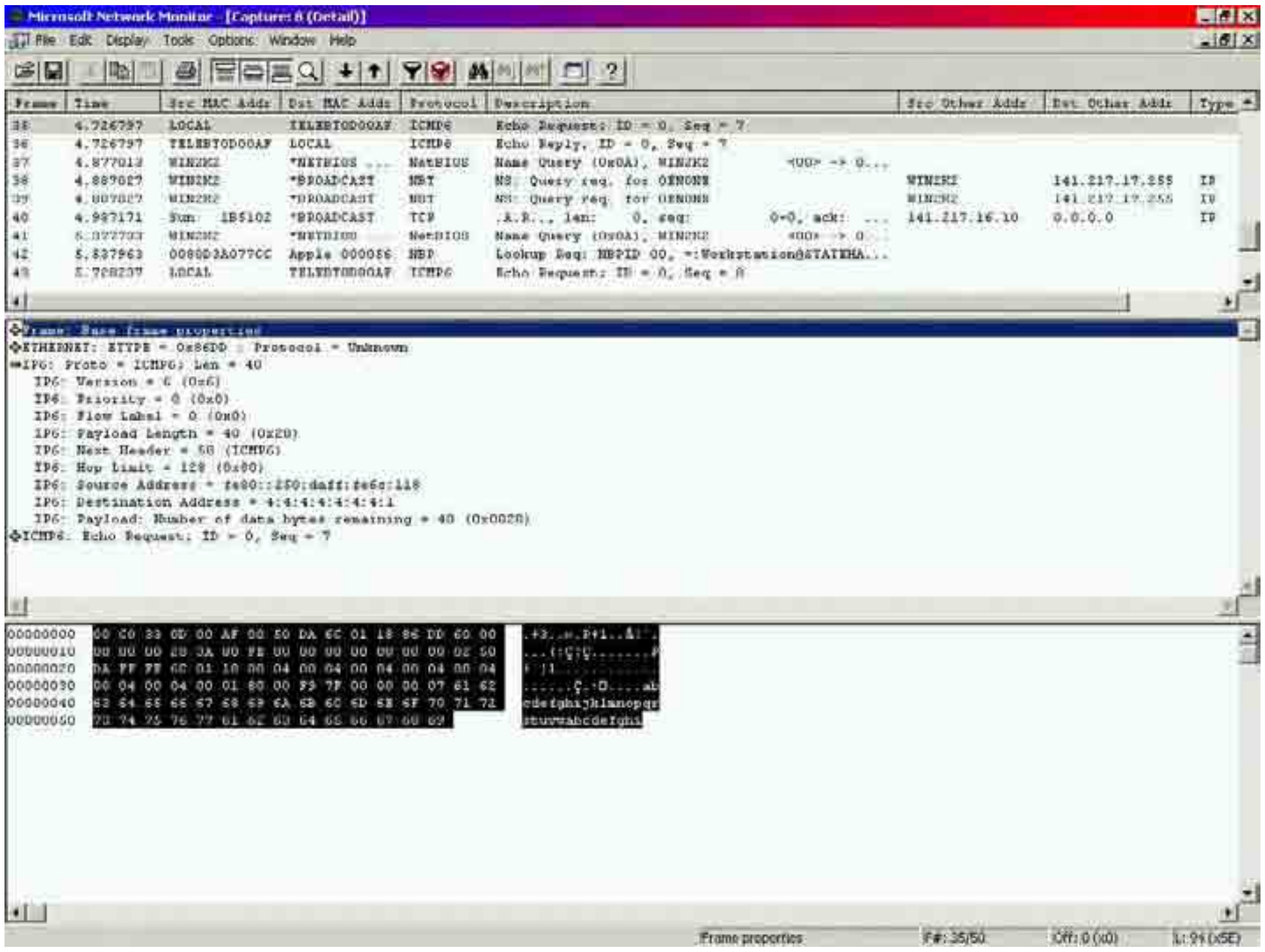

**Figure 4: IPv6 Packet as depicted by the Microsoft Network Monitor**

### 2.3.1 IPv4 Specifications

Internet Protocol version 4 is the current version of IP, which was finally revised in 1981. It has a 32 bit address looking like 255.255.255.255, and it supports up to 4,294,967,296 ($4.3x10^9$) addresses. The IPv6 header is a streamlined version of the IPv4 header. It eliminates fields that are unneeded or rarely used and adds fields that provide better support for real-time traffic. An overview of the IPv4 header is helpful in understanding the IPv6 header.

The “Version” field indicates the version of IP and is set to 4 in the case of IPv4; the size of this field is 4 bits.

The “Internet Header Length” field indicates the number of 4-byte blocks in the IP header. The size of this field is 4 bits. The minimum IP header size is 20 bytes, and therefore the smallest value of the Internet Header Length field is 5. IP options can extend the minimum IP header size in increments of 4 bytes. If an IP option does not use all 4 bytes of the IP option field, the remaining bytes are padded with 0’s, making the entire IP header an integral number of 32-bits (4 bytes). With a maximum value of 0xF, the maximum size of the IP header including options is 60 bytes (15*4).

The “Type of Service” field indicates the desired service expected by this packet for delivery through routers across the IP internetwork. The size of this field is 8 bits, which contain bits for precedence, delay, throughput, and reliability characteristics. Unfortunately, this field was not widely utilized, and only recently with the coming of RSVP did it see much activity. For example, RSVP uses the type of service field in order to setup flow labels.

The “Total Length” field indicates the total length of the IP packet (IP header + IP payload) and does not include link layer framing. The size of this field is 16 bits, which can indicate an IP packet that is up to 65,535 bytes long.

The “Identification” field identifies the specific IP packet. The size of this field is 16 bits. The Identification field is selected by the originating source of the IP packet. If the IP packet is fragmented, all of the fragments retain the Identification field value so that the destination node can group the fragments for reassembly.

The “Flags” field identifies flags for the fragmentation process. The size of this field is 3 bits, however, only 2 bits are defined for current use. There are currently two flags: one indicates whether the IP packet might be fragmented, while the other indicates whether more fragments follow the current fragment.

The “Fragment Offset” field indicates the position of the fragment relative to the original IP payload; the size of this field is 13 bits.

The “Time-to-Live” (TTL) field indicates the maximum number of links on which an IP packet can travel before being discarded. The size of this field is 8 bits. The TTL field was originally used as a time count with which an IP router determined the length of time required (in seconds) to forward the IP packet, decrementing the TTL accordingly. Modern routers almost always forward an IP packet in less than a second and are required by RFC 791 [1] to decrement the TTL by at least one. Therefore, the TTL becomes a maximum link count with the value set by the sending node. When the TTL equals 0, the packet is discarded and an ICMP Time Expired message is sent to the source IP address.

The “Protocol” field identifies the upper layer protocol; the size of this field is 8 bits. For example, TCP uses a protocol value of 6, UDP uses a protocol value of 17, and ICMP uses a protocol value of 1. The Protocol field is used to demultiplex an IP packet to the upper layer protocol.

The “Header Checksum” field provides a checksum on the IP header only. The size of this field is 16 bits. The IP payload is not included in the checksum calculation as the IP payload and usually contains its own checksum. Each IP node that receives IP packets verifies the IP Header Checksum and silently discards the IP packet if checksum verification fails. When a router forwards an IP packet, it must decrement the TTL. Therefore, the Header Checksum is recomputed at each hop between source and destination.

The “Source Address” field stores the IP address of the originating host; the size of this field is 32 bits.

The “Destination Address” field stores the IP address of the destination host; the size of this field is 32 bits.

The “Options” field stores one or more IP options. The size of this field is a multiple of 32 bits. If the IP options do not use all 32 bits, padding options must be added so that the IP header is an integral number of 4-byte blocks that is indicated by the Internet Header Length field.

### 2.3.2 IPv6 Specifications

Internet Protocol version 6 is designed as an evolutionary upgrade to the Internet Protocol (IPv4) and will, in fact, coexist with the older IPv4 for some time. IPv6 is designed to allow the Internet to grow steadily, both in terms of the number of hosts connected and the total amount of data traffic transmitted; it will have a 128 bit address looking like 1234:5678:90AB:CDEF:FFFF:FFFF:FFFF:FFFF, and it will support up to 340,282,366,920,938,463,463,374,607,431,768,211,456 ($3.4x10^{38}$) unique addresses.

The IPv6 header is always present and is a fixed size of 40 bytes. The fields in the IPv6 header are described briefly below.

The “Version” field is used to indicate the version of IP and is set to 6 in the case of IPv6; the field size is 4 bits.

The “Traffic Class” field indicates the class or priority of the IPv6 packet. The size of this field is 8 bits. The Traffic Class field provides similar functionality to the IPv4 Type of Service field. In RFC 2460, the values of the Traffic Class field are not defined. However, an IPv6 implementation is required to provide a means for an application layer protocol to specify the value of the Traffic Class field for experimentation.

The “Flow Label” field indicates that this packet belongs to a specific sequence of packets between a source and destination, requiring special handling by intermediate IPv6 routers. The size of this field is 20 bits. The Flow Label is used for non-default quality of service connections, such as those needed by real-time data (voice and video). For default router handling, the Flow Label is set to 0. There can be multiple flows between a source and destination, as distinguished by separate non-zero Flow Labels.

The “Payload Length” field indicates the length of the IP payload. The size of this field is 16 bits. The Payload Length field includes the extension headers and the upper layer PDU. With 16 bits, an IPv6 payload of up to 65,535 bytes can be indicated. For payload lengths greater than 65,535 bytes, the Payload Length field is set to 0 and the Jumbo Payload option is used in the Hop-by-Hop Options extension header.

The “Next Header” field indicates either the first extension header (if present) or the protocol in the upper layer PDU (such as TCP, UDP, or ICMPv6, etc). The size of this field is 8 bits. When indicating an upper layer protocol above the Internet layer, the same values used in the IPv4 Protocol field are used here.

The “Extension Header” field is utilized for additional functionality that might be needed, such as jumbo packet sizes, security, etc. Zero or more extension headers can be present and are of varying lengths. A Next Header field in the IPv6 header indicates the next extension header.

Within each extension header is another Next Header field that indicates the next extension header. The last extension header indicates the upper layer protocol (such as TCP, UDP, or ICMPv6) contained within the upper layer protocol data unit. The IPv6 header and extension headers replace the existing IPv4 IP header with options. The new extension header format allows IPv6 to be augmented to support future needs and capabilities. Unlike options in the IPv4 header, IPv6 extension headers have no maximum size and can expand to accommodate all the extension data needed for IPv6 communication.

The “Hop Limit” field indicates the maximum number of links over which the IPv6 packet can travel before being discarded. The size of this field is 8 bits. The Hop Limit is similar to the IPv4 TTL field except that there is no historical relation to the amount of time (in seconds) that the packet is queued at the router. When the Hop Limit equals 0, the packet is discarded and an ICMP Time Expired message is sent to the source address.

The “Source Address” field stores the IPv6 address of the originating host; the size of this field is 128 bits.

The “Destination Address” field stores the IPv6 address of the destination host; the size of this field is 128 bits. In most cases the Destination Address is set to the final destination address. However, if a Routing extension header is present, the Destination Address might be set to the next router interface in the source route list.

### 2.3.3 IPv4 vs. IPv6

Table 2 shows the highlights in the differences between IPv4 and IPv6 protocols. There are many other differences; however, it depends on what level of detail we wish to examine the matter. There have been entire books written on the IPv6 protocol and all the differences down to the minutest detail from the old IPv4 protocol. One such book is "IPv6 Networks" [22] by Marcus A. Goncalves; it offers an excellent in-depth explanation of any material covered here in this chapter regarding IPv6 networks and much more.

An important aspect of the following information is that the facts presented in Table 2 and Table 3 are all theoretical. These are all the proposed changes that have been outline in the various Requests for Comments (RFC) lead by IETF. The actual implementation of all the features is still in the infancy stages of development and there still lacks maturity, as will be presented in our experimental results. Most likely, by the time that IPv6 will be deployed worldwide and will replace IPv4, all the features stated below should be implemented. Most experts predict that in the next five years, most of the Internet will have support for IPv6.

The left hand side of the table represents features of IPv4 while the right hand side represents features of IPv6; they are interrelated and depict how the particular feature of IPv4 was upgraded to support IPv6. Definitions of the terminology or acronyms can be found in Appendix A.

| IPv4 | IPv6 |
| --- | --- |
| Source and destination addresses are 32 bits (4 bytes) in length. | Source and destination addresses are 128 bits (16 bytes) in length. |
| IPSec support is optional. | IPSec support is required. |
| No identification of payload for QoS handling by routers is present within the IPv4 header. | Payload identification for QoS handling by routers is included in the IPv6 header using Flow Label field. |
| Fragmentation is supported at both routers and the sending host. | Fragmentation is only supported at the sending host. |
| Header includes a checksum. Must be computed at every intervening node on a per packet basis. | Header does not include a checksum. It relies on other layers to find erroneous packets. |
| Header includes options. Potential inefficient use of header bits. | All optional data is moved to IPv6 extension headers. |
| Address Resolution Protocol (ARP) broadcast ARP Request to resolve an IPv4 address to the link layer. | ARP Request frames are replaced with multicast Neighbor Solicitation messages. |
| Internet Group Management Protocol (IGMP) is used to manage local subnet group membership. | IGMP is replaced with Multicast Listener Discovery (MLD) messages. |
| ICMP Router Discovery is used to determine the IPv4 address of the best default gateway. | ICMPv4 Router Discovery is replaced with ICMPv6 Router Solicitation and Router Advertisement. |
| Broadcast addresses are used to send traffic to all nodes on a subnet. | There are no IPv6 broadcast addresses; a link-local scope all-nodes multicast address is used. |
| Must be configured either manually or through DHCP. | Does not require manual configuration or DHCP. |
| Uses host address (A) resource records in the DNS to map host names to IPv4 addresses. | Uses host address (AAAA) resource records in the DNS to map host names to IPv6 addresses. |
| Pointer resource records (PTR) in IN-ADDR.ARPA DNS domain map IPv4 addresses to host names. | Uses pointer (PTR) resource records in the IP6.INT DNS domain to map IPv6 addresses to host names. |

**Table 2: Differences between IPv4 and IPv6 protocol [22]**

Now that the main differences in the protocols are clear, Table 3 will describe the differences between the IPv4 and IPv6 header fields. The left column names the header field while the right side describes the change which IPv6 incurred from its IPv4 predecessor.

| IPv4 Header Field | IPv6 Header Field |
|---|---|
| **Version** | Same field but with different version numbers. |
| **Internet Header Length** | Removed in IPv6. IPv6 does not include a Header Length field because the IPv6 header is always a fixed size of 40 bytes. Each extension header is either a fixed size or indicates its own size. |
| **Type of Service** | Replaced by the IPv6 Traffic Class field. |
| **Total Length** | Replaced by the IPv6 Payload Length field, which only indicates the size of the payload. |
| **Identification** | Removed in IPv6. Fragmentation information is not included in the IPv6 header. It is contained in a Fragment extension header. |
| **Fragmentation Flags** | Removed in IPv6. Fragmentation information is not included in the IPv6 header. It is contained in a Fragment extension header. |
| **Fragment Offset** | Removed in IPv6. Fragmentation information is not included in the IPv6 header. It is contained in a Fragment extension header. |
| **Time to Live** | Replaced by the IPv6 Hop Limit field. |
| **Protocol** | Replaced by the IPv6 Next Header field. |
| **Header Checksum** | Removed in IPv6. In IPv6, bit-level error detection for the entire IPv6 packet is performed by the link layer. |
| **Source Address** | The field is the same except that IPv6 addresses are 128 bits in length. |
| **Destination Address** | The field is the same except that IPv6 addresses are 128 bits in length. |
| **Options** | Removed in IPv6. IPv4 options are replaced by IPv6 extension headers. |

**Table 3: Differences between IPv4 and IPv6 headers [22]**

## 2.4 IPv4 to IPv6 Transition Mechanisms

As IPv6 is finally beginning to mature and IPv4 is approaching its limits, it is evident that methods of upgrading the Internet from IPv4 to IPv6 need to be found. One idea would be to turn off the entire Internet at 12AM, upgrade the network infrastructure (routers, protocol stacks, etc), and turn the Internet back on at 6AM and hope everything works. This is unrealistic due to the astronomical price and the high probability that it will not work as well as the theoretical prediction. Hence, more gradual transition methods have evolved, ones which are likely to happen over the course of the next 10 years.

Some transition mechanisms are: Dual Stacks [3], DTI & Bump-in-dual-stack, NAT Protocol Translator [27], Stateless IP/ ICMP Translator (SIIT), Assignment of IPv4 Global Addresses to IPv6 Hosts (AIIH), Tunnel Broker [28], 6-to-4 Mechanism [29], and IPv6 in IPv4 tunneling [30,31].

Dual Stacks are easiest to implement, however complexity increases due to both infrastructures and the cost is higher due to a more complex technology stack. NAT Protocol Translator has scaling and DNS issues, and has single point of failure disadvantage. The Tunnel Broker dynamically gains access to tunnel servers, but has authentication and scaling issues. 6-to-4 mechanism creates dynamic stateless tunnels over IPv4 infrastructure to connect 6-to-4 domains. IPv6 in IPv4 tunneling allows existing infrastructure to be utilized via manually configured tunnels.

We chose to pursue the IPv6 in IPv4 tunneling as a transition mechanism because it would be the most cost effective and can implement islands of IPv6 networks that can be connected over the existing ocean of IPv4 networks, the existing infrastructure. With time, as the islands grow, the ocean will diminish to a point that all the islands will touch, at which point it is evident that native IPv6 networks will finally reign and benefit 100% from its new features. There are two transition mechanisms which we will discuss: host-to-host encapsulation and router-to-router encapsulation, which is also know as tunneling. The router-to-router tunneling is the more interesting of the two since entire LANs can be upgraded to IPv6 while maintaining connectivity to the rest of the Internet. Host-to-host encapsulation is also addressed mainly because of its simplicity of implementation, and offers another method of making the transition from IPv4 to IPv6 as smooth as possible.

Encapsulation of IPv6 packets within IPv4 packets, better known as tunneling, is one of the easiest transition mechanisms by which two IPv6 hosts / networks can be connected with each other while running on existing IPv4 networks through establishing some special routes called tunnels. In this technique, IPv6 packets are encapsulated in IPv4 packets and then are sent over IPv4 networks like ordinary IPv4 packets through tunnels. At the end of tunnel these packets are de-capsulated to the original IPv6 packets.

When encapsulating a datagram, the TTL in the inner IP header is decremented by only one if the tunnel is being done as part of forwarding the datagram; otherwise the inner header TTL is not changed during encapsulation. If the resulting TTL in the inner IP header is zero, the datagram is discarded and an ICMP Time Exceeded message is returned to the sender. Therefore, an encapsulator will not encapsulate a datagram with TTL=0. When encapsulating IPv6 packets in IPv4 packets, only IPv4 routing properties will be utilized and hence the IPv6 packet will loose any special IPv6 features until it is de-capsulated at the receiving host/router. Another drawback is that it requires a hole in a firewall to allow protocol 41 (IP in IP) passage.

If a tunnel falls entirely within a routing domain, it will be considered as plain serial link by interior routing protocol such as RIP or OSPF. But if it lies between two routing domains it needs exterior protocols such as BGP. In case of congestion in the tunnel, an ICMP Source Quench message will be issued in order to inform the previous node of the congestion.

In different two different types of tunneling, only de/encapsulation points are varied depending on the start and end of tunnels, however the basic idea remains the same. Once again, the two tunneling mechanisms are Host-Host Tunneling and Router-Router Tunneling.

### 2.4.1 Host-to-Host Encapsulation

In host-to-host tunneling method, encapsulation is done at the source host and the de-capsulation is done at the destination host. So the tunnel is created in between two hosts supporting both IPv4 and IPv6 stacks. Therefore, the encapsulated datagrams are sent through a native IPv4 network that has no knowledge of the IPv6 network protocol.

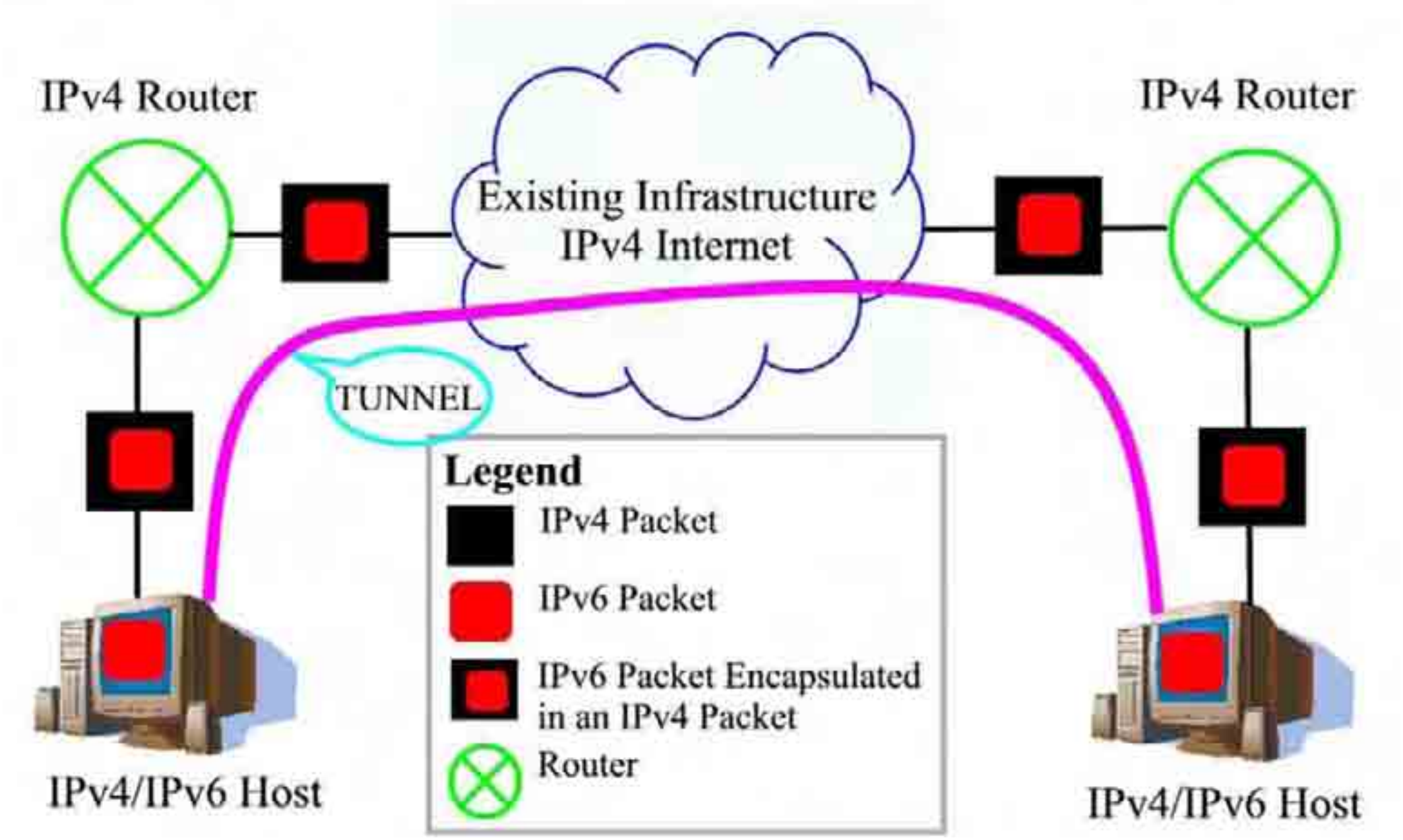

**Figure 5: Host-to-Host tunneling; packet traversal across a network**

In Figure 5, it is clear that both hosts having dual stack encapsulate the packets of IPv6 in IPv4 packets and transmit over the network as an IPv4 packet utilizing all the characteristics and routing mechanisms of IPv4. With this transition mechanism, it is possible to support IPv6 simply by upgrading the end hosts protocol stacks to IPv6 while leaving the IPv4 infrastructure unchanged.

The black bigger square depicts an IPv4 packet while the red smaller rounded square depicts an IPv6 packet. The red square overlaid on top of the black square means that the IPv6 packet is encapsulated inside the IPv4 packet. Host-to-host tunneling will be consistently referred to as IPv4(IPv6) in the later performance evaluation of Chapter 5.

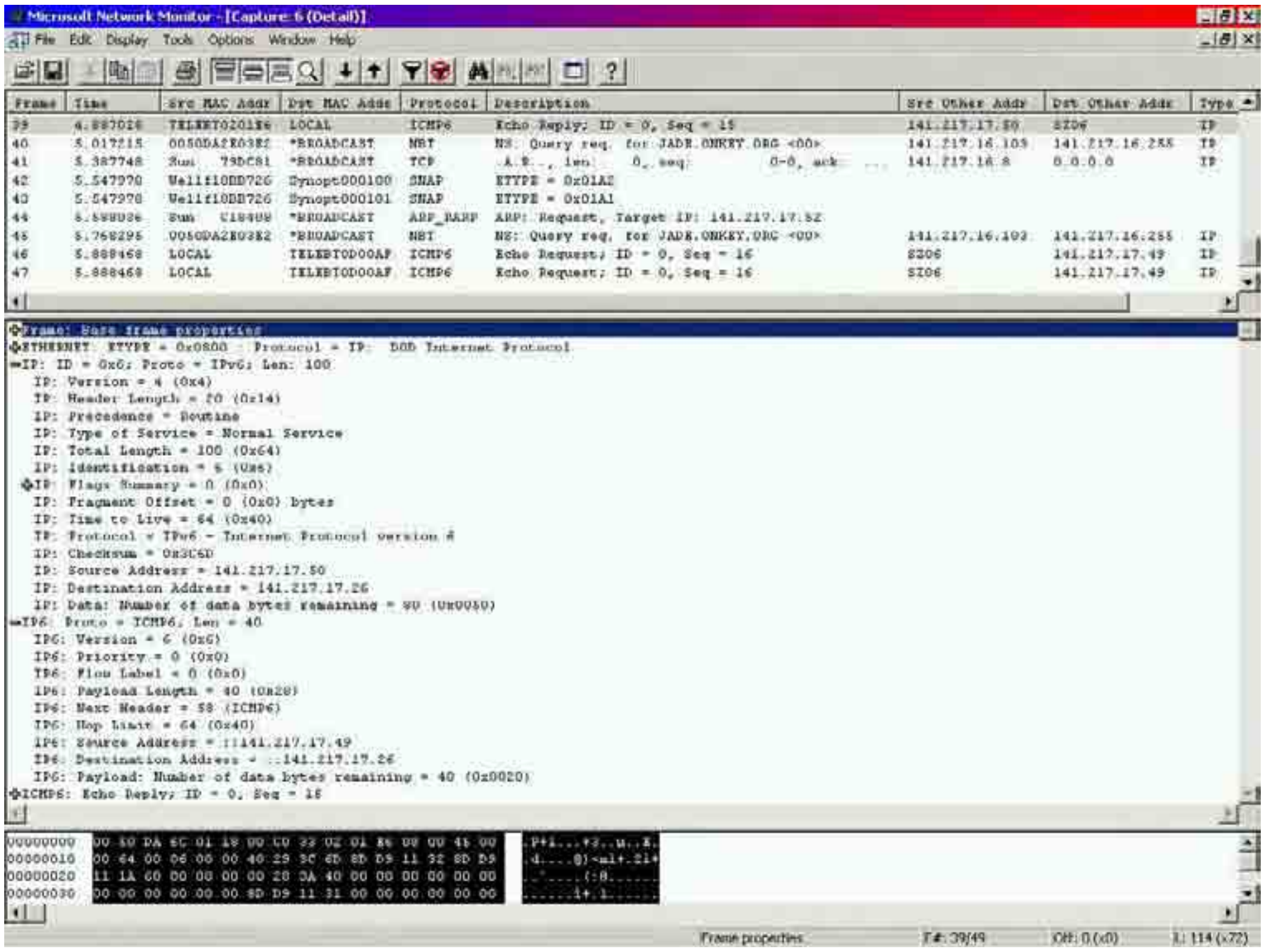

**Figure 6: IPv6 packet encapsulated in an IPv4 packet depicted by the Microsoft Network Monitor**

Just as we displayed a picture of the IPv4 and IPv6 packet alone in Figure 3 and Figure 4, we will show an IPv6 packet encapsulated in an IPv4 packet in Figure 6 to illustrate to the reader how the hosts sees the encapsulation. All the various header fields are clearly visible just as we had described them in the earlier section.

### 2.4.2 Router-to-Router Tunneling

In router to router tunneling mechanism, encapsulation is done at the edge router of the origination host and de-capsulation is done in the same way at the edge router of the destination host. The tunnel is created in between two edge routers supporting both IPv4 and IPv6 stacks. Therefore, the end hosts can support native IPv6 protocol stack while the edge routers create the tunnels and handle the encapsulation and de-capsulation in order to transmit the packets over the existing IPv4 infrastructure in between the two edge routers.

Figure 7 shows a tunnel established between two edge routers, which supports both (IPv4 / IPv6) stacks. The IPv6 packets are forwarded from host to edge routers while encapsulation takes place at the router level; similarly at the other end, the reverse process takes place. In this method, both edge routers need to support dual stacks and established a tunnel prior to transmission.

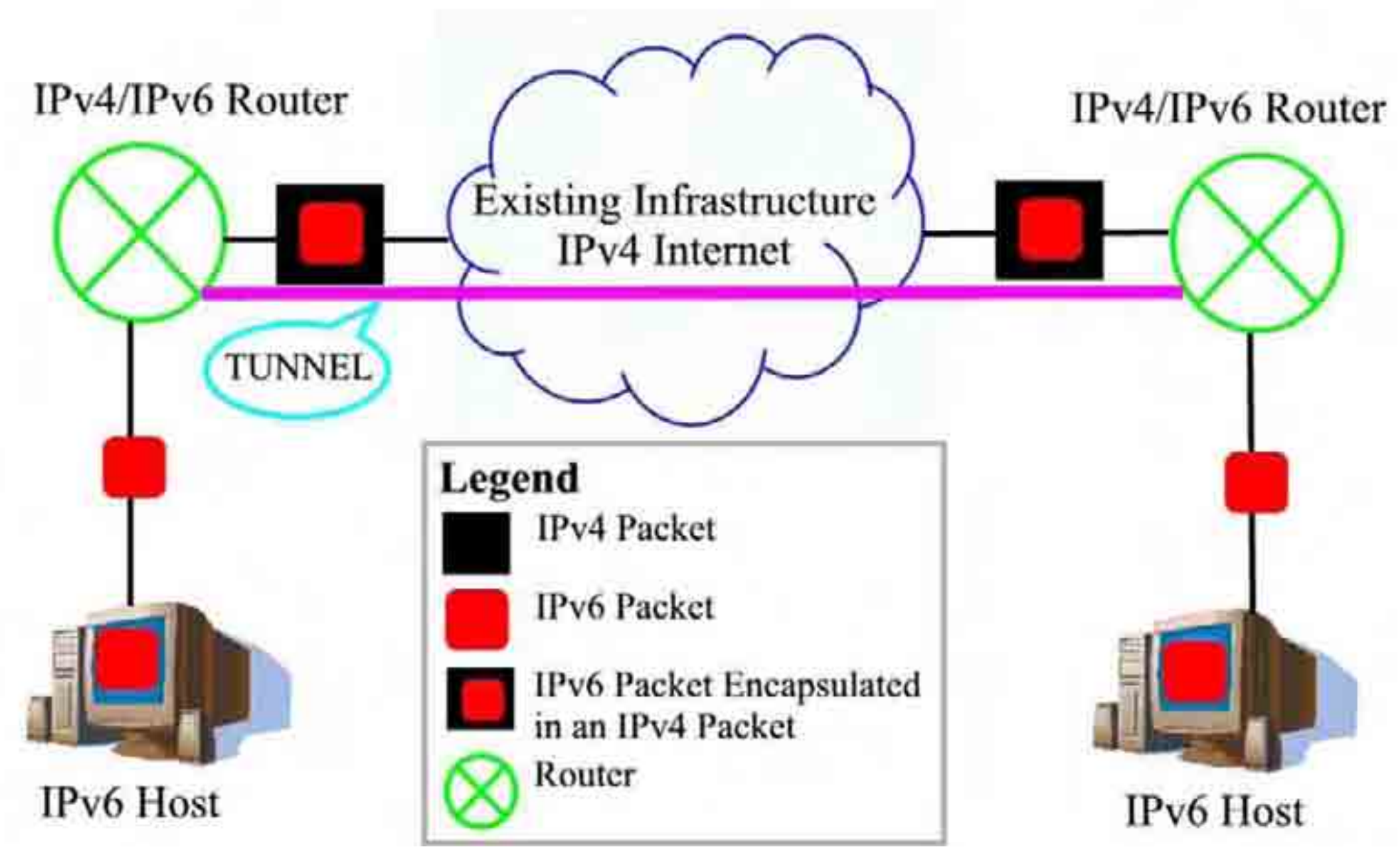

**Figure 7: Router-to-Router Tunneling; packet traversal across a network**

# CHAPTER 3

# TEST-BED CONFIGURATION

Our test-bed consisted of two dual stack (IPv4/IPv6) routers: an Ericsson AXI 462, and an IBM 2216 Nways Multiaccess Connector Model 400. Dual stack implementation specifications can be found in RFC 1933 [3]. We had two identical workstations that were connected directly to the routers and were configured to be on separate networks. Each router supported two separate networks each.

Both workstations were equipped with Intel Pentium III 500 MHz processors, 256 megabytes of SDRAM PC100, two 30GB IBM 7200 RPM IDE hard drive, and COM 10/100 PCI network adapters. The workstations were loaded with both Windows 2000 Professional and Solaris 8.0 as a dual boot configuration on two separate and identical hard drives. Windows 2000 had the IPv4 stack as a standard protocol; however in order to get IPv6 support, an add-on package was installed. There were two choices, both written by Microsoft and they were both in Beta testing. We chose the newer release of the two, “Microsoft IPv6 Technology Preview for Windows 2000” [6] which is supported by Winsock 2 as its programming API. It was evident that Microsoft’s IPv6 stack for Windows 2000 is not in production yet since it had various deficiencies. It did not seem to handle fragmentation well for the UDP transport protocol, and therefore we limited our test to message sizes less than the Ethernet MTU size of 1514 bytes. It also does not support IPSec yet, but that was outside of the scope of this paper and therefore is not important. On the other hand, Solaris 8.0 came with a dual production level IPv4/IPv6 stack. Because of Microsoft's IPv6 limitation with fragmentation, the tests on Solaris were limited to 1514 byte UDP messages as well.

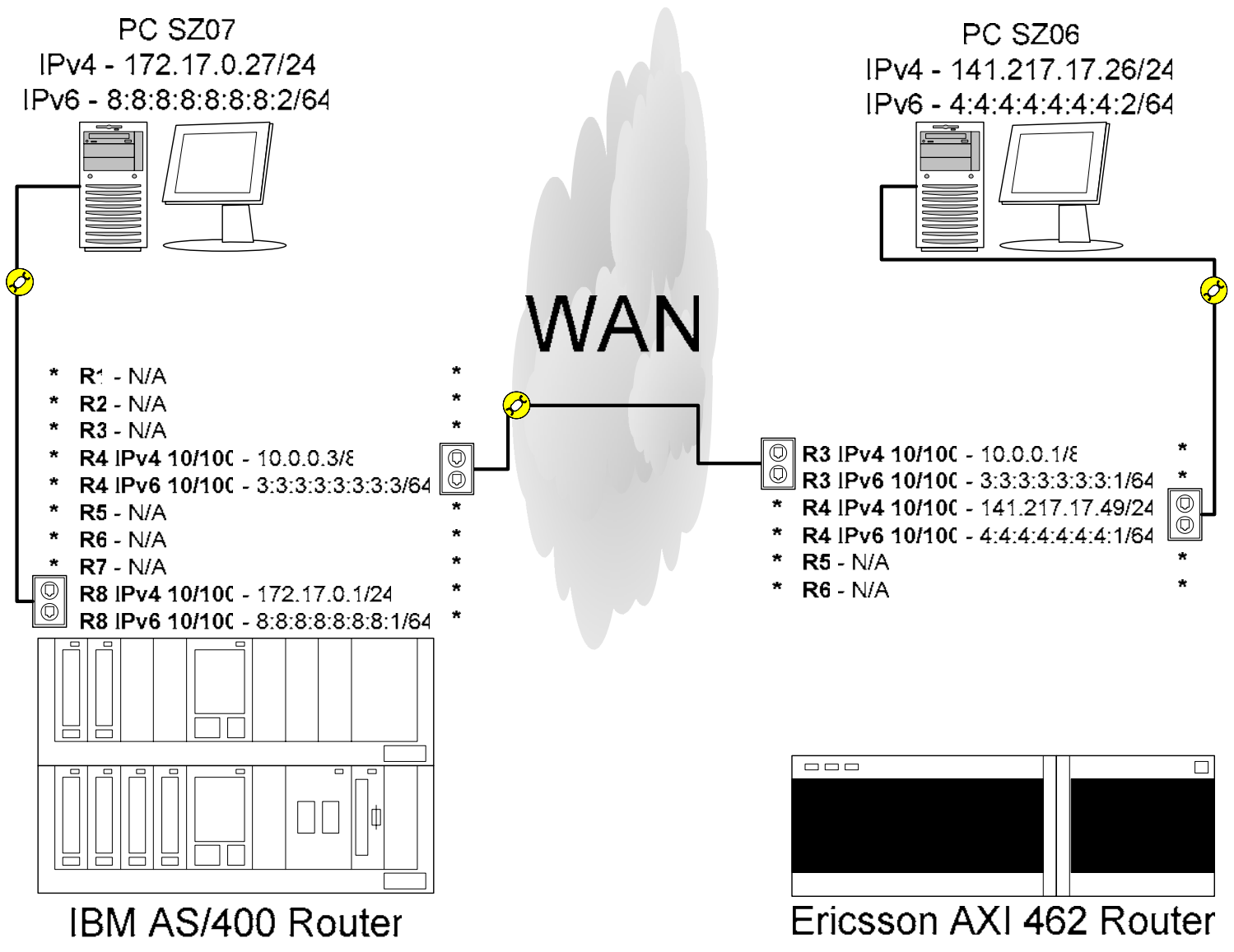

**Figure 8: Test-bed architecture named IBM-Ericsson; two routers are depicted, an IBM 2216 Nways Multiaccess Connector Model 400 and an Ericsson AXI 462**

Figure 8 depicts the entire test-bed as we had it configured in our laboratory. On the IBM router, R1 through R8 are the various network cards that are available while R3 through R6 are the various network cards available on the Ericsson router; each interface card has both an

IPv4 and an IPv6 address.

The important thing to notice is that each router has two network cards configured to be two separate networks. For the IBM router, we utilize network 172.17.0.xxx/24 and network 10.xxx.xxx.xxx/8. Notice that these are two different networks because one is a class D address with a subnet mask of 24 while the other is a class A address with a subnet mask of 8; a subnet mask of 8 means that the first 8 bits are considered to be the network address. Since the two network cards lie on separate networks, the router must utilize its functionality and forward packets from one card to another and vice versa. Similarly, we have the Ericsson router with networks 10.xxx.xxx.xxx/8 and 141.217.17.xxx/24 for the separate networks. Notice that the workstations, SZ06 and SZ07 lie in the same networks as their respective routers; SZ06 has an IP address of 141.217.17.26/24 while SZ07 has an IP address of 172.17.0.27/24.

Obviously, these workstations could communicate with their respective routers without a problem since they lie on the same respective networks. In order for the routers to pass packets between various networks, a protocol such as RIP [20] is needed to forward packets to their corresponding destination.

A quick example should explain the path of any message from either host. Let us assume that host SZ06 transmits a packet to host SZ07. The packet is sent from host SZ06 (141.217.17.26) to the Ericsson router on card R4 (141.217.17.49). Card R4 forwards the packet to card R3 (10.0.0.1), which then forwards the packet to the IBM router on card R4 (10.0.0.3). Card R4 forwards the packet to R8 (172.17.0.1) which is the intended destination network, and therefore the packet is finally forwarded to host SZ07 (172.17.0.27). If the packet would have been sent using the IPv6 stack, it would have followed the same path, except that it would have made its decisions based on the IPv6 128 bit addresses rather than the 32 bit IPv4 addresses.

Notice that both of the above examples utilized a single protocol, whether it was IPv4 or IPv6, in order to transmit a packet from one host to another. This is a necessary and fundamental configuration in our evaluation of IPv6. However, the Internet is far from being at the point where all routers and all hosts to guarantee the support for both IP protocols. In order to make the transition from IPv4 to IPv6 easier, several transition mechanisms have been proposed and implemented. In chapter 5, we will discuss some various transition mechanisms, their benefits and drawbacks, and most important of all, how much overhead will it incur on top of the already high overhead of IPv6.

In order to better understand our results from the above described test-bed, we developed three more configurations which would allow us to better analyze the results. We utilize a very similar setup, but we take out the IBM router. We are therefore left with the test-bed depicted in Figure 9 that has two end PCs (SZ06 and SZ07) that are directly connected to the Ericsson router. Notice that some of the IP addresses have changed from the first test-bed configuration named IBM-Ericsson. Messages now traverse the network similar to the first test-bed, except that there is now one less hop or router to cross.

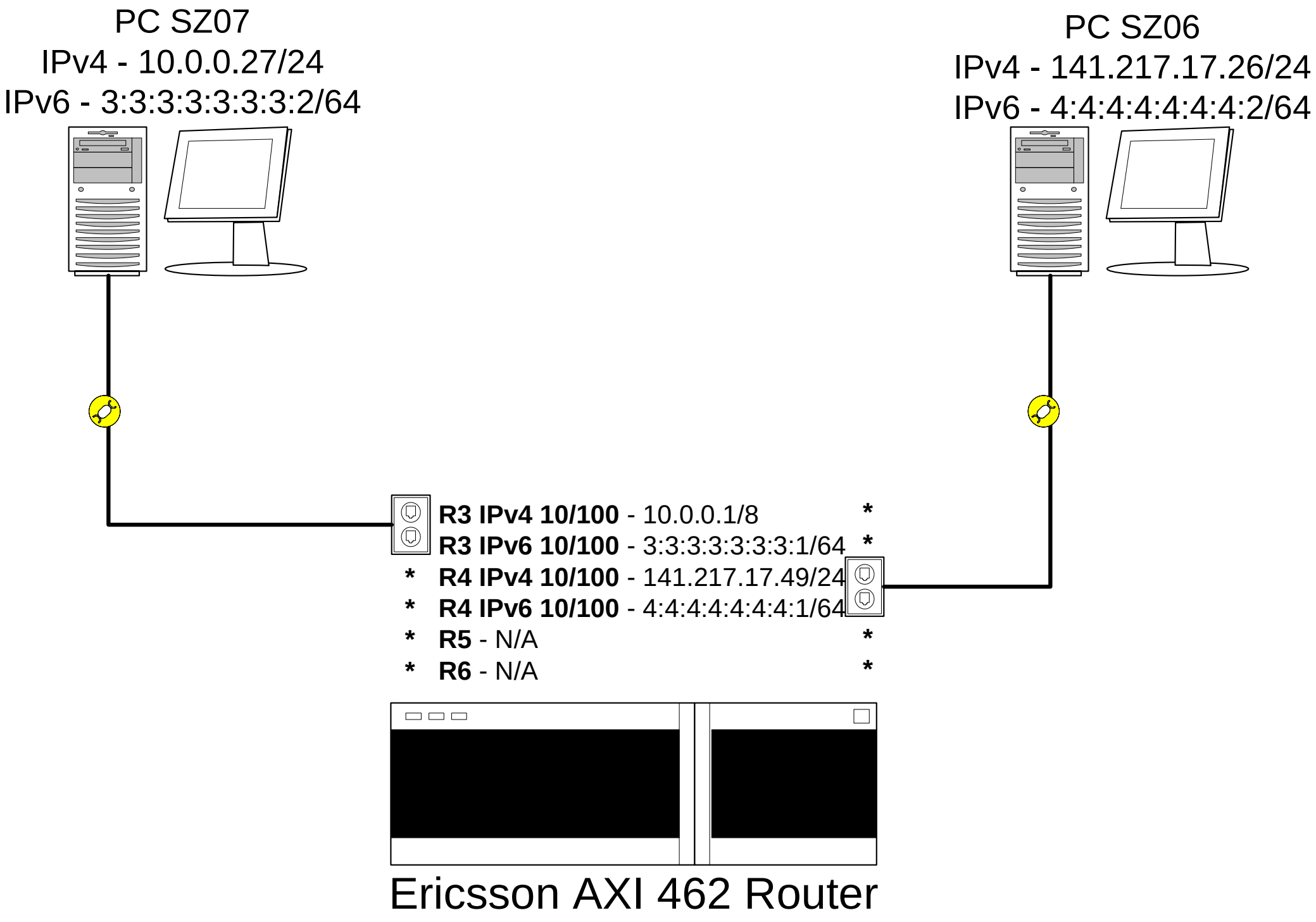

**Figure 9: Test-bed architecture named Ericsson; one router configuration is depicted using the Ericsson AXI 462**

The next test-bed depicted in Figure 10 is the opposite of the above, in which we leave out the Ericsson router and hence only the IBM router connects the workstations. Everything works just like in Figure 9, except that we have the IBM router in the place of the Ericsson router. Notice again that some of the IP addresses might be different to accommodate the new router.

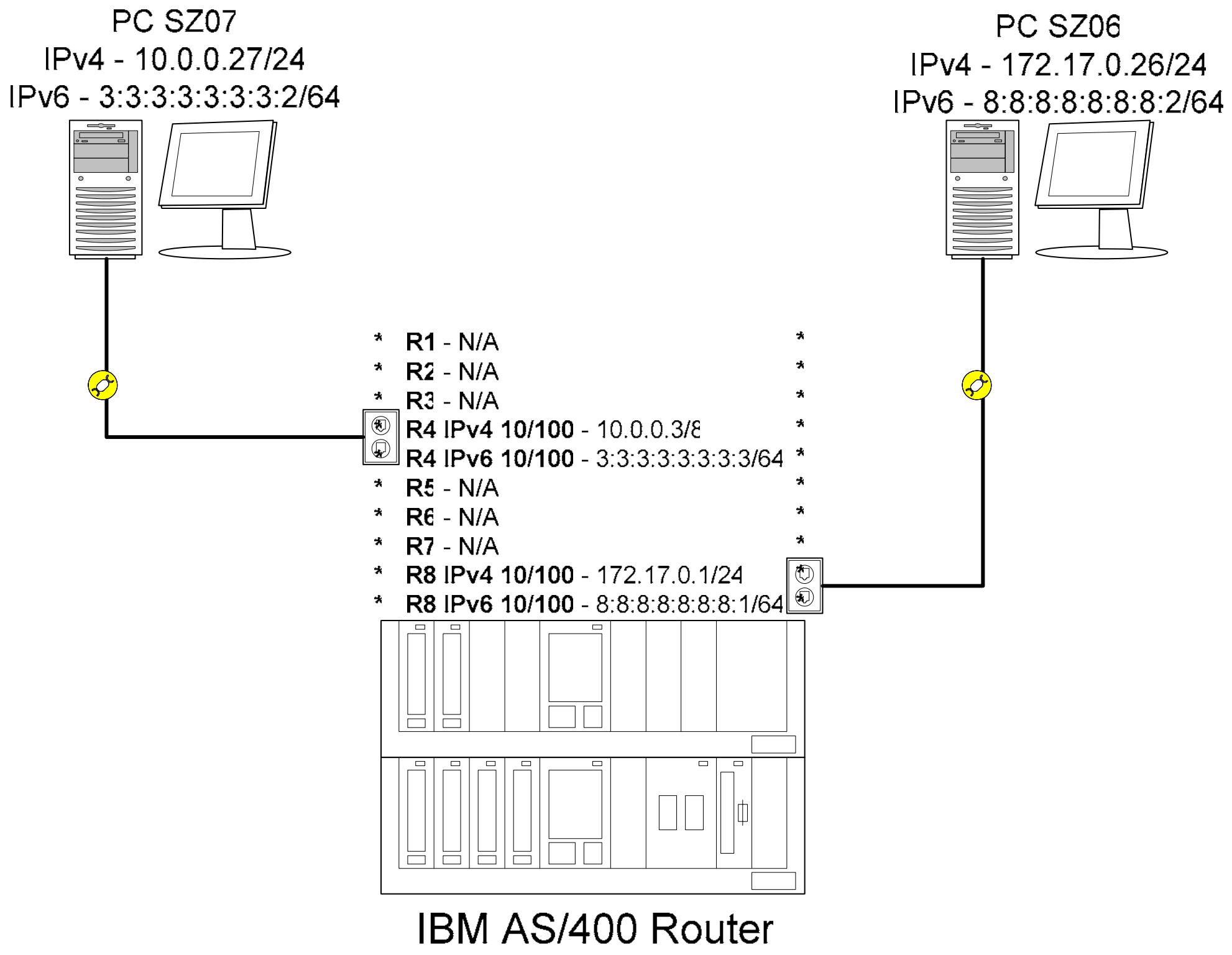

**Figure 10: Test-bed architecture named IBM; one router configuration is depicted using the IBM 2216 Nways Multiaccess Connector Model 400**

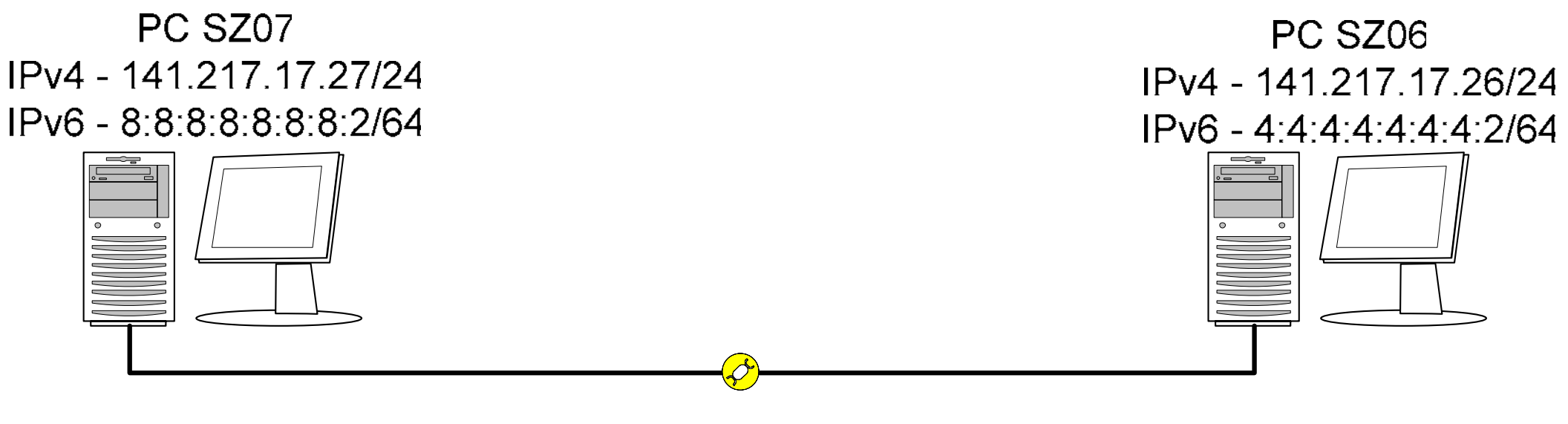

**Figure 11: Test-bed architecture named P2P for point-to-point; PCs are directly connected to each other via a twisted pair Ethernet cable**

Finally, we have our last configuration depicted in Figure 11 in which we took out both routers completely and were left with the two workstations directly connected to each other with no routers between them. Notice that both PCs are on the same network (141.217.17.xxx/24) and therefore can communicate between each other with no router in between them. The value of this configuration lies in the ability to take out as many variables (the routers) from the experiments and observe the behavior of the various tested protocols with just the PC hardware and Operating System as the only variables. It should be no surprise that the performance results based on the various test-beds will not be identical.

# CHAPTER 4

# IPv4 AND IPV6 PERFORMANCE EVALUATION

In this chapter, we first discuss the performance metrics in detail in order for the reader to understand the relevance of our findings. We then present the results for IPv4 and IPv6 network protocols using both TCP and UDP transport protocols under Windows 2000 and Solaris 8.0 operating systems.

As described in chapter 3, we have four different test-bed configurations. The same experiments were performed on each test-bed and the results are displayed incrementally in such a way that it maximizes the understanding of the results. We first cover the P2P (Point-to-point) Test-bed which had no routers between the end nodes. This test-bed is aimed at taking out as many variables as possible and leaving just the end nodes hardware and OS determine the performance. However, it is very unlikely that any two nodes have a direct physical cable connection between them throughout any network, and therefore the results do not do justice to the common view of present day networks.

We then performed the same experiments on the IBM-Ericsson Test-bed which had both routers in between the two end nodes. This is more representative of the present day networks, in which packets flowing from one node to another throughout any significant size network must traverse routers along the way. Due to some very surprising results, we decided to isolate the variance of the results by creating two more test-beds, namely the IBM Test-bed and the Ericsson Test-bed. Each of these test-beds was configured using only one respective router in between the two end nodes. The same experiments as before were performed on each of these new test-beds as well.

For more information regarding the specifics of each test-bed's configuration, please refer to chapter 3 in which all four test-beds (P2P, IBM-Ericsson, IBM, and Ericsson) are discussed in detail.

## 4.1 Performance Metrics

Our metrics of evaluation were: throughput, latency, CPU utilization, socket creation time, TCP connection time, the number of TCP connections per second, and the performance of a video application designed by our lab. All the performance measurement software was written in C++.

The majority of the tests were done for a period of about 60 seconds, which netted about 50,000 packets to about 1,000,000 packets, depending on the size of the packets sent and what tests were being completed. The tests dealing with testing the throughput of the UDP transport protocol were limited to 1472 byte datagrams because of a potential undocumented fragmentation bug in the IPv6 protocol stack. All other tests were done using various packet sizes ranging from 64 bytes to 64 Kbytes. Each test was repeated three times in order to rule out any inconsistencies. On occasion when the three different tests were not consistent enough to have a solid conclusion, the experiments were performed several more times until there was enough data to conclude our findings.

### 4.1.1 Throughput

Throughput offers a very clear representation of the real overhead incurred by the header information. Throughput tests push computer hardware to its limits from most points of view since many variables such as OS design, memory allocation/speed, and network link speed can radically alter the performance of the network.

A network link only has the total bandwidth capacity to transmit its packets which include all the headers for the different layers and the final payload of usable data. Obviously, no system can ever achieve throughputs of 100% of the bandwidth due to the overhead of header information.

For example, using IPv4 UDP, in a best case scenario, the system would only achieve 97.2% capacity of the bandwidth (see Table 1 for details). A great example of the throughput metric utilized in our every day lives is the typical download over a dial-up modem. Let us assume that the connected speed is 56 Kbits/s, which is about 7 Kbytes/s. How many people have seen anything close to 7 Kbytes/s download speeds on their home computers? My personal experience has been that I get about 3 to 4 Kbytes/s at best, which means that only about 50% of the link bandwidth is usable. Through our work, we are trying to establish exactly that kind of metrics in which we can deduce how much of the link bandwidth can be usable for data transmissions, which by definition is throughput.

Throughput was calculated by sending XX number of packets of YY bytes from a client to a server. At the beginning of the test, the time would be recorded; at the end of the test, again the time would be noted. The two timestamps, which have a microsecond resolution, would be subtracted from each other, and what remained would be the duration of the test in microseconds noted as $\Delta$. Since we knew the size of the messages we were sending (YY) and the number of messages we eventually sent (XX), we knew that we sent XX*YY number of bytes over $\Delta$ microseconds, and therefore were could translate the result into Mbit/s.

### 4.1.2 Latency

Latency, or better know as RTT (round trip time), is very important since many applications are sensitive to any kind of delays. Having better latency could mean that the protocol would perform better for real time applications such as video or audio.

The best example of latency is when a TELNET application is running and the user types something on the keyboard. If anyone has ever used TELNET over a dial-up modem that has relatively low connection speeds and high latency, you might have noticed that it sometimes takes on the order of half a second to several seconds sometimes to display to the screen the characters typed on the keyboard. This is so because the characters typed must first travel to the server before they come back to be displayed on the screen, and hence what is depicted is the RTT of the characters that were typed. Another example is the PING utility that most operating systems provide that allows a user to verify that another computer on the network is functional. The output to the PING utility most likely includes a RTT value for reaching the queried computer.

Latency was calculated by sending a message of XX bytes from a client to a server; upon the receipt of the message, the server sent back the same message back to the original client; when the client received its message back, the whole process would start all over again reiterating the same process for YY number of times. At the end of the test, we knew that we had YY iterations with a duration of $\Delta$ microseconds, and therefore could derive the RTT by dividing the $\Delta$ by YY.

### 4.1.3 Socket Creation Time

Another metric was measuring the socket creation time between IPv4 and IPv6. This is important because servers that handle many requests and create a new socket for every connection they make would benefit greatly by having a shorter socket creation time by alleviate precious resources. This was calculated merely by time stamping the instance right before the socket creation and right after it. Along the same line of tests, our fourth test was measuring the amount of time it would take to establish a TCP connection. This involves initiating the connect() function from the client and the server accepting the connection, and waiting for it to complete. The connect() function is a blocking function that will only return when either it gave up in trying to establish a connection, or that the connection was successfully established. By measuring the time it takes for the connect() function to execute, we can have insight at the length of time it takes to set up a TCP connection. For both of these previous two tests, were repeated the same procedure 10,000 times in order to rule out any inconsistencies.

### 4.1.4 Web Client/Server Simulation

Another metric, which is related to the previous test, was seeing how many TCP connections a server could handle per second. This is important because a web client will create a new TCP connection for each in-line image (icon, drawing, photo, etc.) to the web server. Needless to say, this creates many TCP connections that are probably short lived for each web page browsed [7]. We believe that if one protocol proves to be more efficient at handling TCP requests, it could mean that the overall efficiency of the World Wide Web would increase or decrease as the Internet would be converted to IPv6. We calculated the number of TCP connections by having a single threaded server that waits for incoming connections; a client will then setup the socket from scratch, connect with the TCP server, transmit 1 byte, and close the connection. This entire process repeats itself XX number of times. At the end, we know how many connections the client ended up having and how long it took, and therefore we come up with number of TCP connections per second.

#### 4.1.5 Video Client/Server Application

Our final test was porting an existing IPv4 high quality video application server/client to send its data over the IPv6 stack. The application was written for Windows 2000, and therefore we were not able to compare between Windows and Solaris, but at least we were able to compare between the IPv4 and IPv6 stack under Windows. The application consisted of a server which would take either a live signal or a file source, segment it, and send it over the network to a client, which would then display the received video stream. Our evaluation consisted of achieved throughput, frame rate, and CPU utilization at which it was able to display the video stream.

## 4.2 Performance results

For all figures depicting performance results such as throughput or latency, we will use the following consistent conventions as explained in the legends of each figure. The dotted lines represent IPv6 performance results while the solid lines represent IPv4 results. Solaris 8.0 is denoted by a triangle (▲) while Windows 2000 is denoted by a square (■). The x-axis is the packet size in the corresponding experiment, while the y-axis represents the measured metric. For each test we have two figures: one represents the large global view with packet sizes ranging from 64 bytes to 64 Kbytes, while the other represents only a small part of the bigger graph displaying the results for packet size between 64 bytes and 1408 bytes. For the tests representing CPU utilization, we utilize the same conventions except for the triangle (▲) denotes the UDP transport protocol while the square (■) denotes the TCP transport protocol.

### 4.2.1 P2P Test-bed Performance Results

Just as a reminder, the tests performed in this sub-section reflect the P2P Test-bed configuration which had no routers between the end nodes. The PCs had a direct communication link via twisted pair Ethernet cable from one end to the other. These tests are important in order to eliminate as many variables as possible and get a base performance evaluation of IPv4 and IPv6. For each experiment, we will be briefly reiterating the results depicted in the graph in case that it is not evident from the figures what the particular outcome may be.

#### *4.2.1.1 Throughput*

As Figure 12 indicates, it can be clearly seen that Solaris 8.0 does slightly better than Windows 2000 over the entire packet size spectrum.

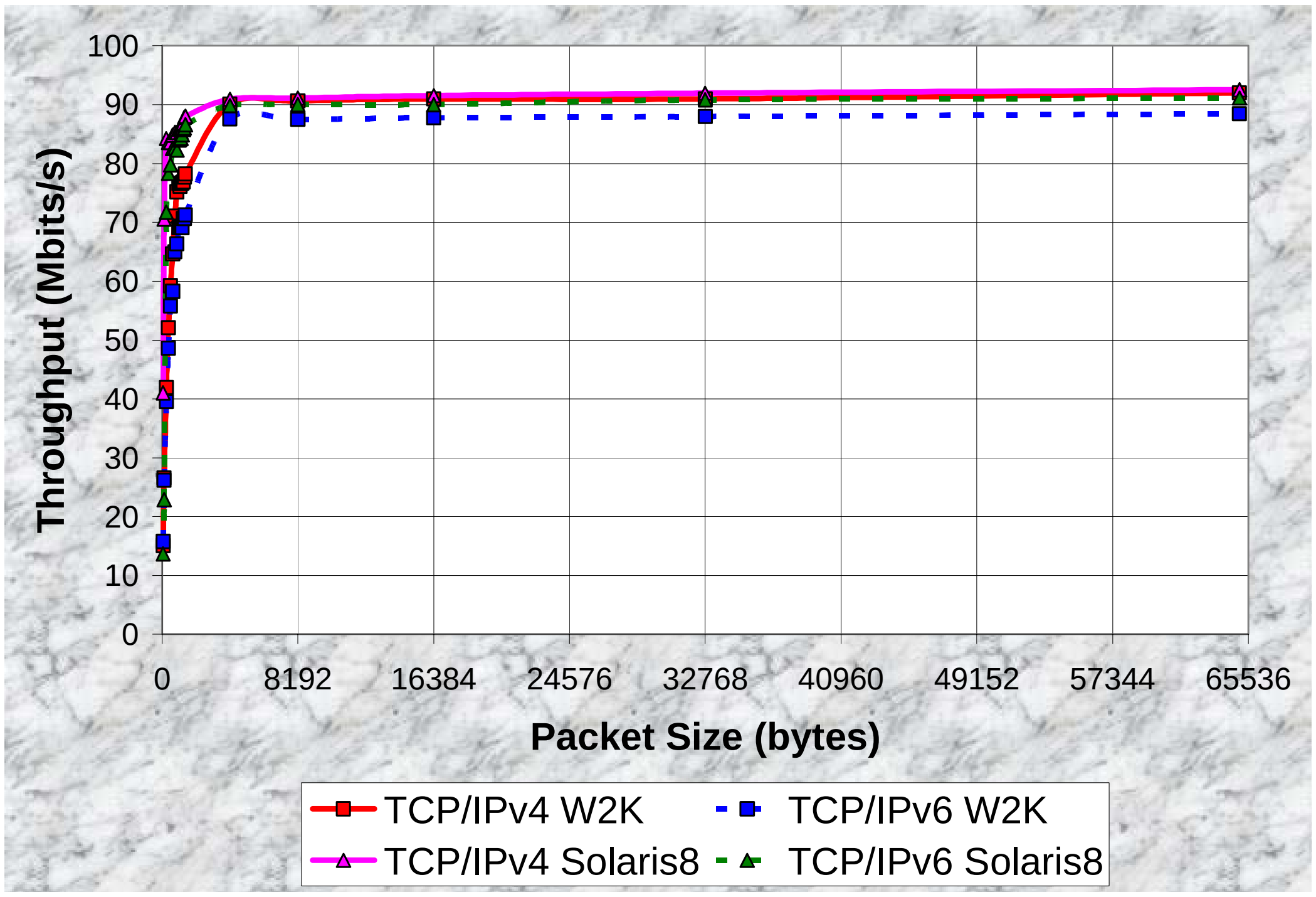

**Figure 12: P2P Test-bed: TCP throughput results for IPv4 & IPv6 over Windows 2000 & Solaris 8.0 with packet size ranging from 64 bytes to 64 Kbytes**

Under Windows 2000, although the IPv4 and IPv6 stacks present similar trends, IPv6 incurs an additional 6% to 13% overhead in the smaller packet sizes and 2% to 4% in the larger one. Under Solaris, IPv6 incurs a similar overhead, except that it is slightly less in the larger packet sizes. Figure 13 below depicts the same results from Figure 12 above, however only packet sizes ranging from 64 bytes to 1408 bytes are represented in order to detail that was just not possible in the global view of the packet size range.

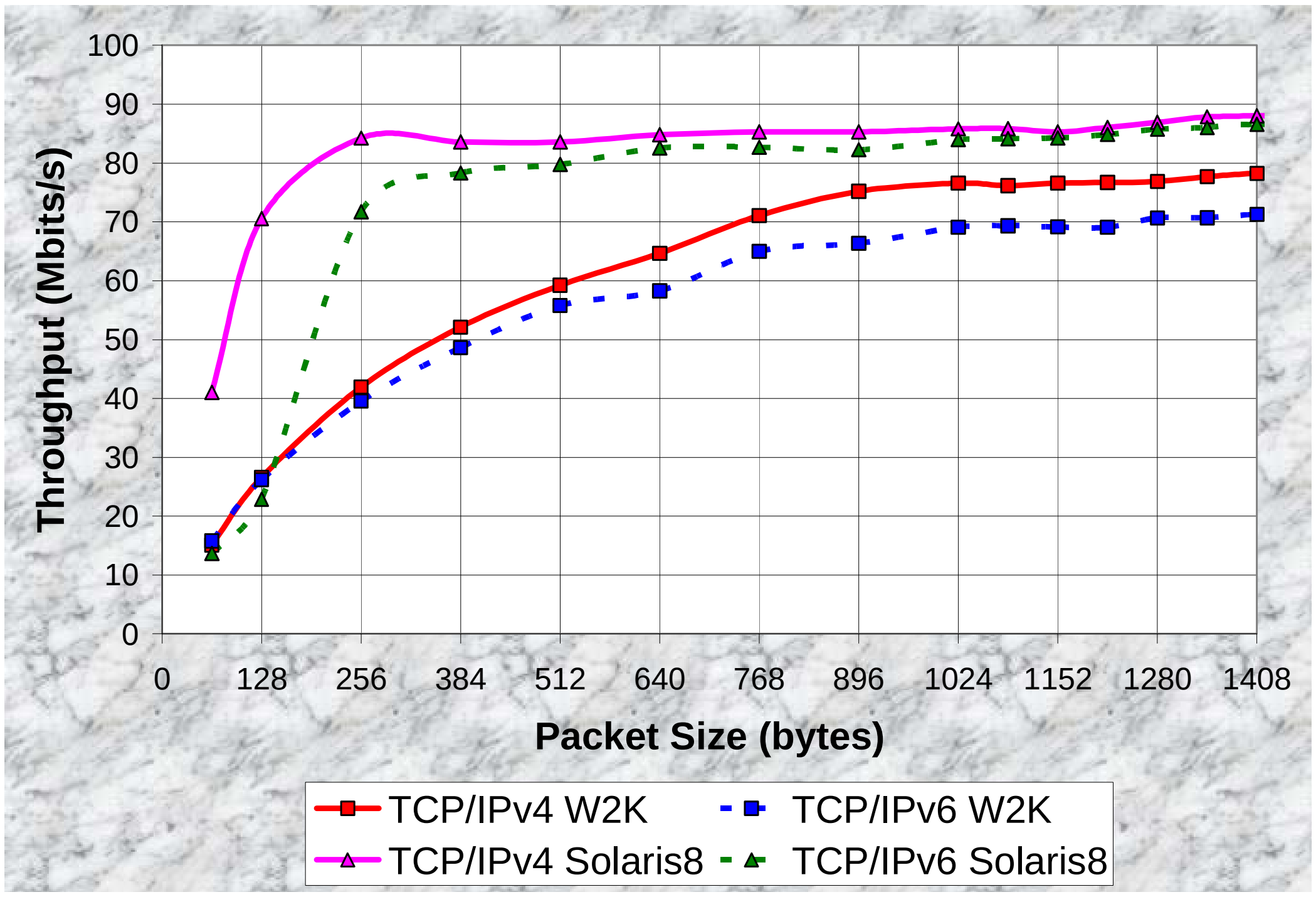

**Figure 13: P2P Test-bed: TCP throughput results for IPv4 & IPv6 over Windows 2000 & Solaris 8.0 with packet size ranging from 64 bytes to 1408 bytes**

As Figure 14 and Figure 15 indicate, it can be clearly seen that Solaris 8.0 again performs better than Windows 2000 over the entire packet size spectrum. Notice that the IPv6 protocol for both Windows and Solaris are barely visible. As we discussed earlier, we found a bug in the IPv6 protocol stack which prevents us from performing and throughput tests for UDP under IPv6 for packet sizes greater than the Ethernet MTU size of 1514 bytes.

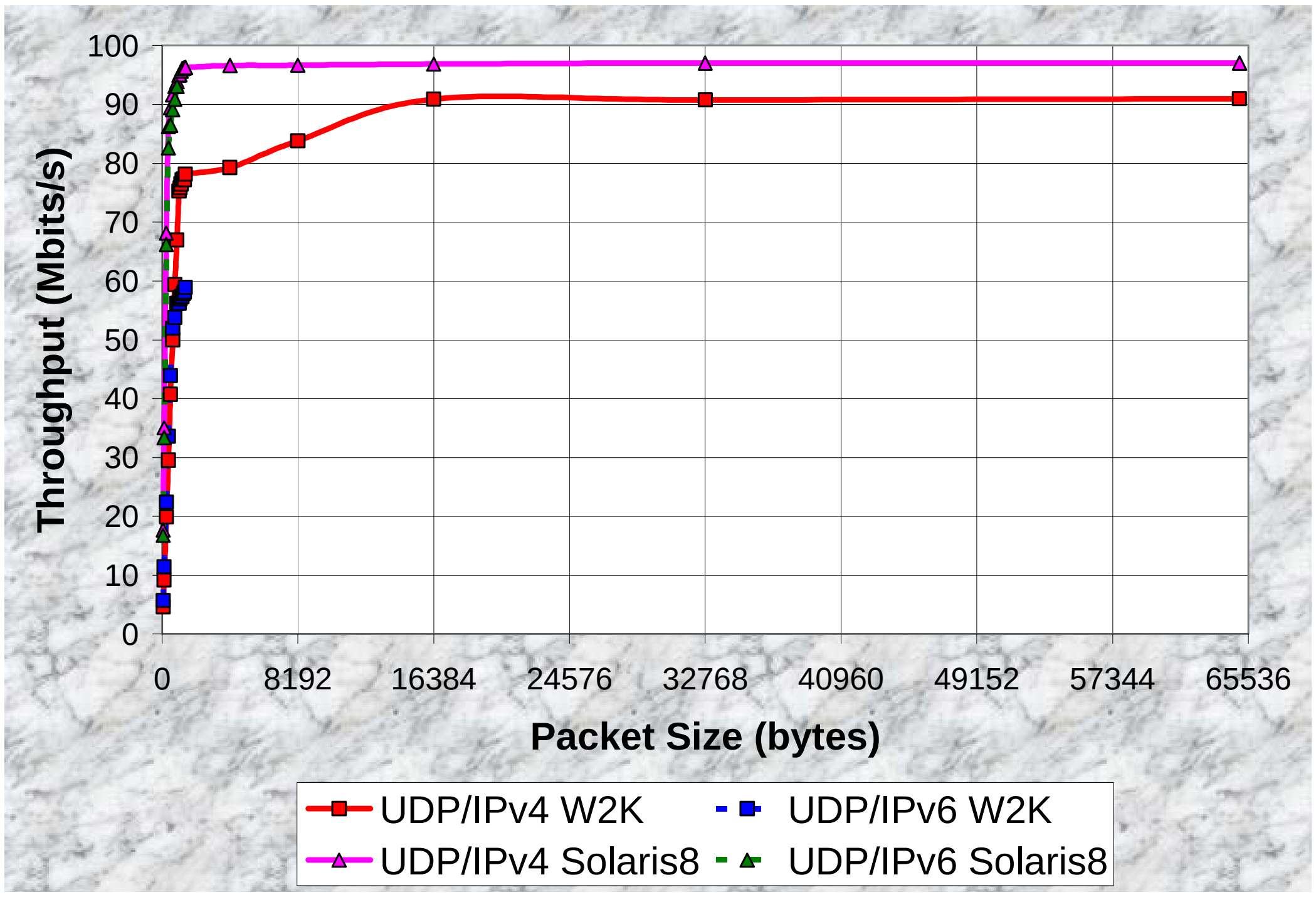

**Figure 14: P2P Test-bed: UDP throughput results for IPv4 & IPv6 over Windows 2000 & Solaris 8.0 with packet size ranging from 64 bytes to 64 Kbytes**

Figure 15 below clearly shows that under Solaris 8.0, IPv6 only incurs a 6% to 1% overhead over IPv4 ranging from the smaller packets to the larger ones. On the other hand, under Windows 2000, IPv6 incurs no overhead to up to 35% on top of IPv4 ranging from the smaller packets to the larger ones.

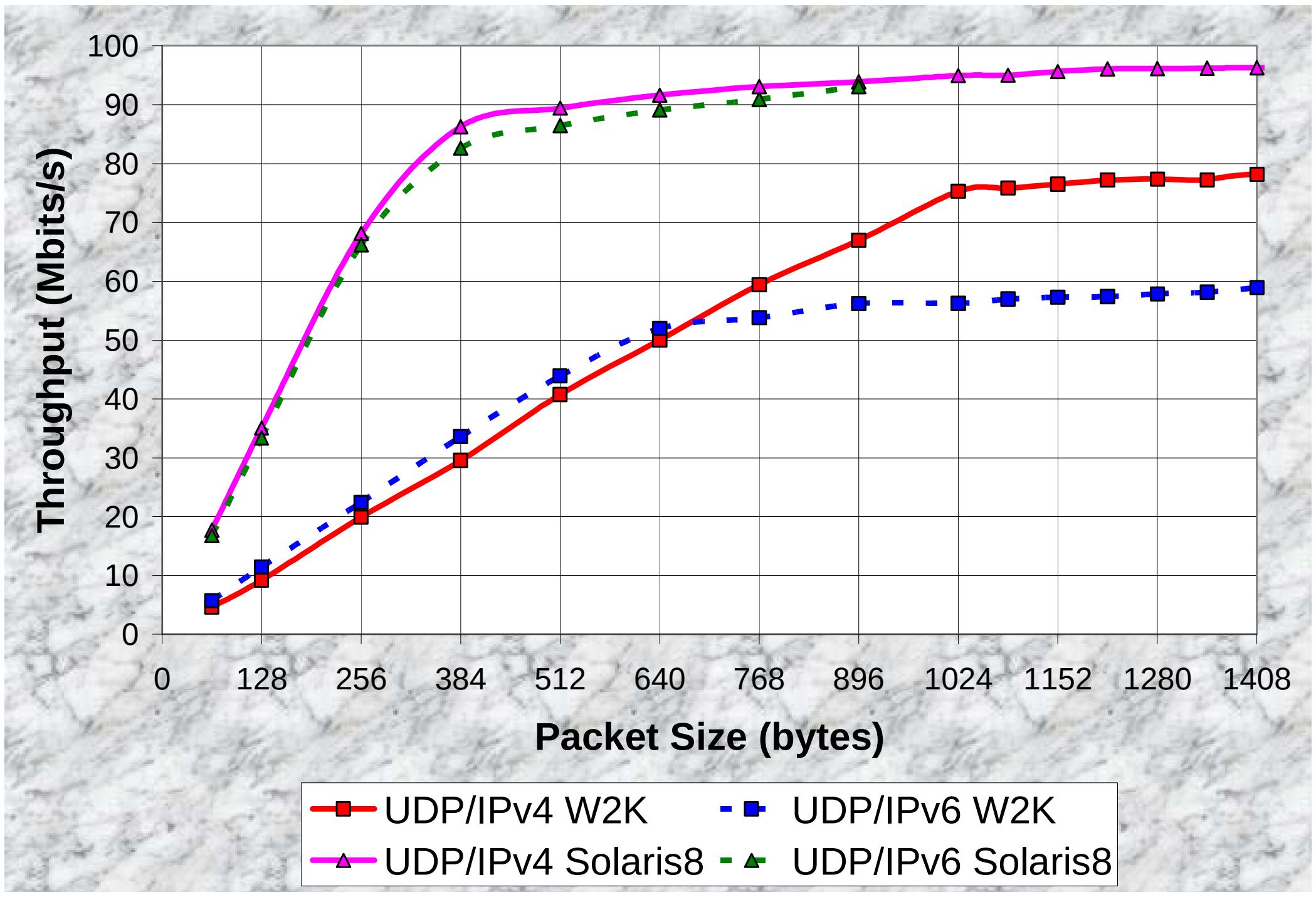

**Figure 15: P2P Test-bed: UDP throughput results for IPv4 & IPv6 over Windows 2000 & Solaris 8.0 with packet size ranging from 64 bytes to 1408 bytes**

As we were able to see throughout the results of the throughput experiments, the summary of the last few figures is that Solaris 8.0 performs slightly better than Windows 2000 and that IPv6 has little (0% to 5%) to significant (up to 35%) performance overhead.

Finally, we conclude this subsection with the results from the CPU utilization of the various protocols. The CPU utilization performance numbers were observed from the Windows Task Manager under the performance monitor. Figure 16 clearly shows that the TCP transport protocol incurs more CPU overhead than the UDP transport protocol. This was expected since UDP is known to be a lightweight protocol that only has minimal functionality while TCP is rather complex and utilizes many features that are much more CPU intensive.

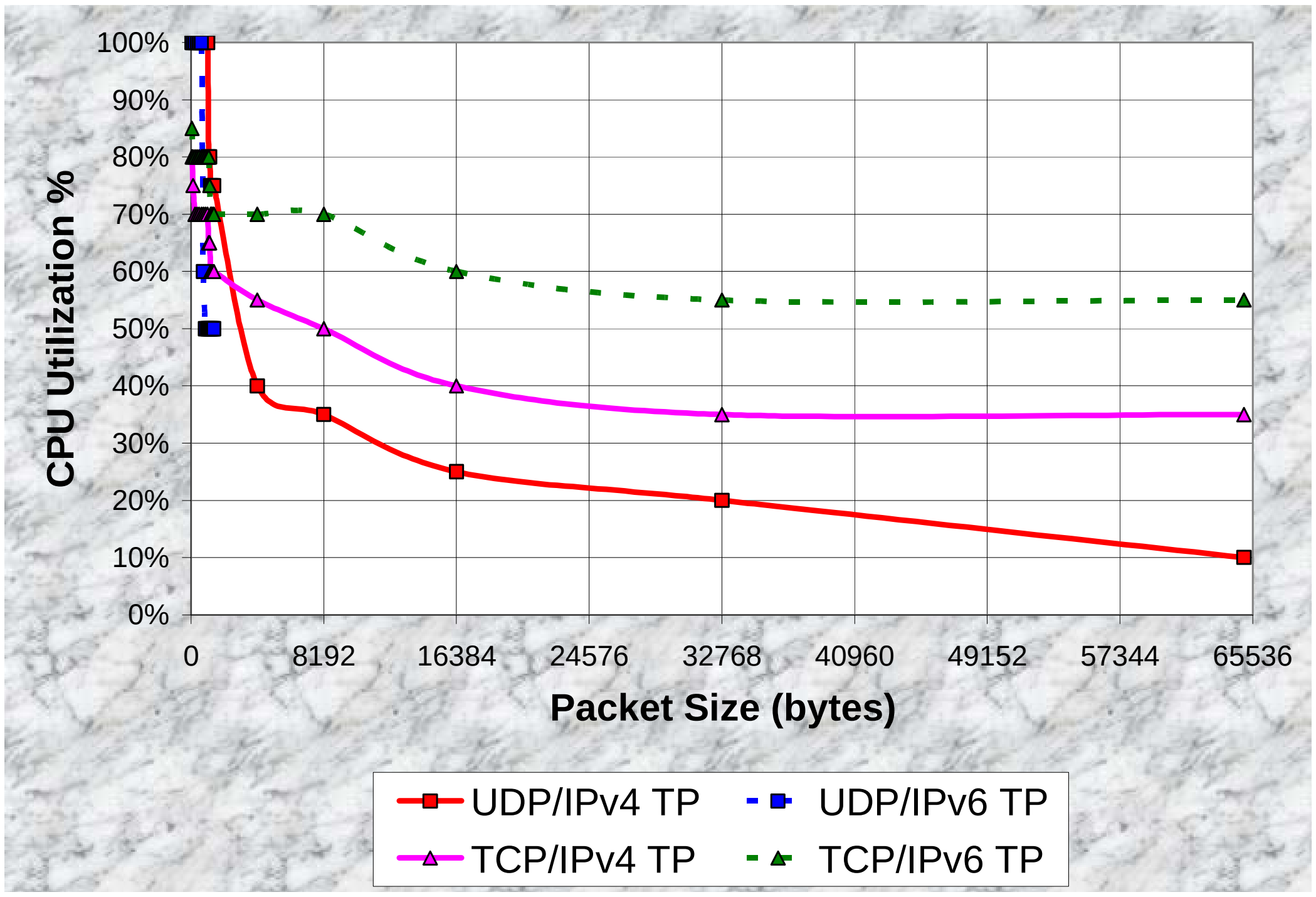

**Figure 16: P2P Test-bed: CPU Utilization results for the throughput experiments in IPv4 and IPv6 running TCP and UDP over Windows 2000 with packet size ranging from 64 bytes to 64 Kbytes**

Furthermore, it is clear that just as expected, IPv6 also incurs more overhead than IPv4. Remember that IPv6 has an IP header that is twice as large as its IPv4 counterpart, and therefore it makes sense that it would take more CPU cycles to process an IPv6 packet than an IPv4 packet as long as the performance characteristics were similar. In the later experiments, it should be obvious that as the host's performance drops, so does its CPU utilization.

### *4.2.1.2 Latency*

As Figure 17 and Figure 18 indicate, both Windows and Solaris offer comparable performance for the latency test, although Windows 2000 seems to perform slightly better than Solaris in the larger packet sizes.

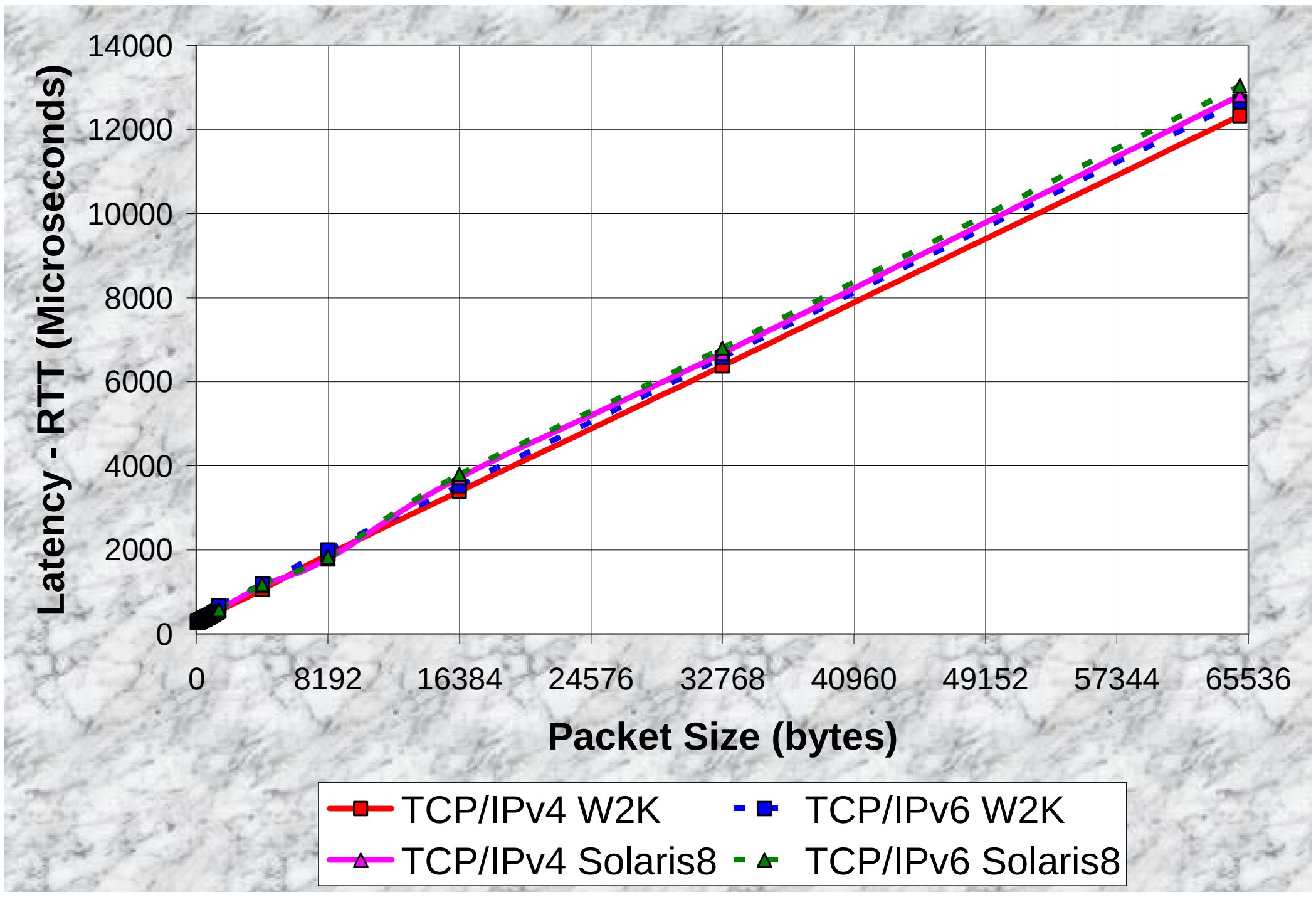

**Figure 17: P2P Test-bed: TCP latency results for IPv4 & IPv6 over Windows 2000 & Solaris 8.0 with packet size ranging from 64 bytes to 64 Kbytes**

The notable difference between IPv4 and IPv6 under Windows was 15% higher RTTs (the lower the better) for small packets and as little as 2% overhead for larger packets. Solaris closed the gap to only 5% overhead for small packets while having as little as 1% overhead for larger packets.

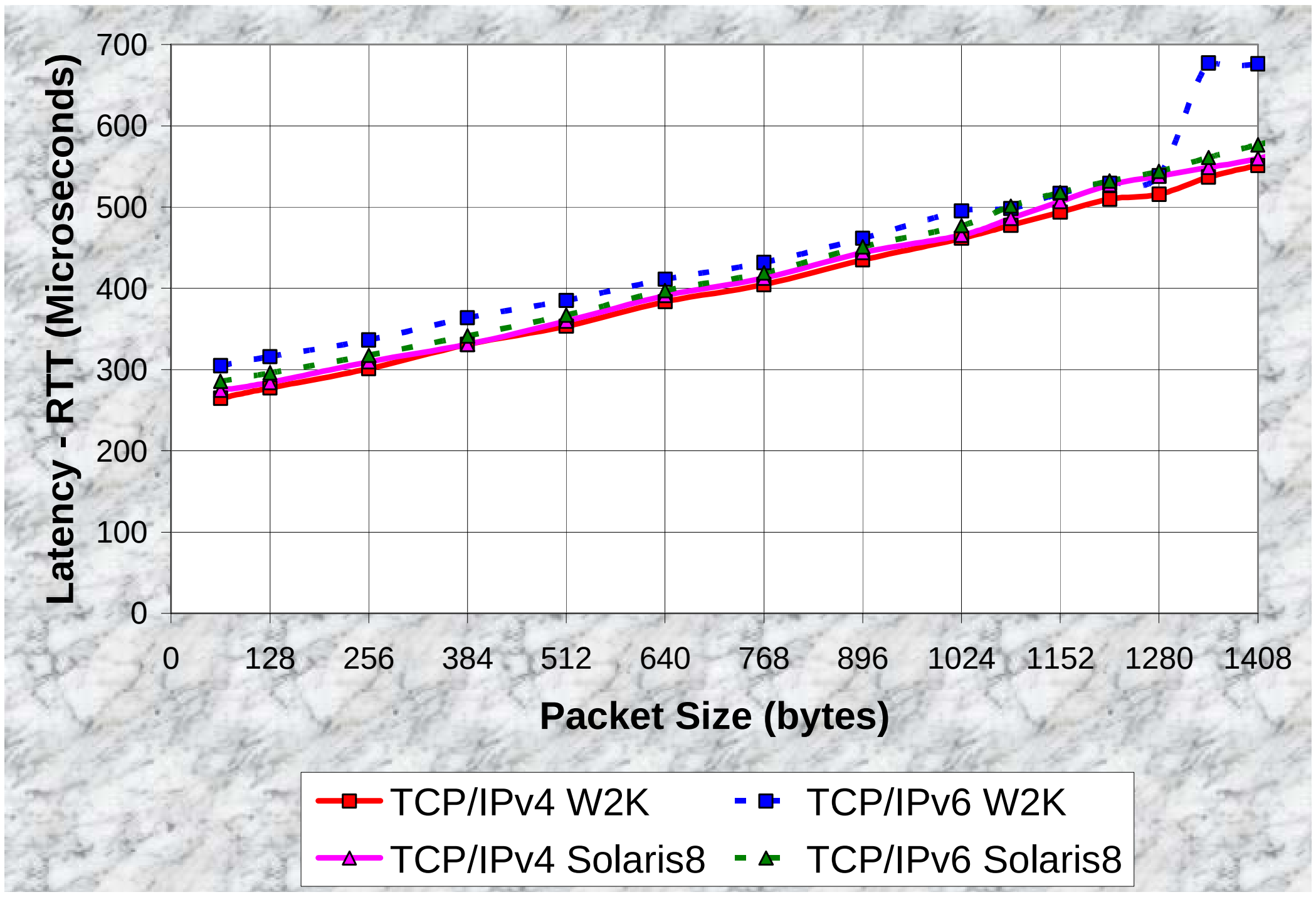

**Figure 18: P2P Test-bed: TCP latency results for IPv4 & IPv6 over Windows 2000 & Solaris 8.0 with packet size ranging from 64 bytes to 1408 bytes**

In Figure 18 above, the odd spike in RTT times for packet sizes of 1344 and 1408 byte packets in IPv6 under Windows 2000 is most likely due to a buffer allocation issue in which the contents of the packet plus the larger overhead of IPv6 cause the packet not to fit with the MTU of 1514 bytes. Therefore, the fragmentation mechanism probably caused the spike to occur. This kind of behavior is exactly the reason why we choose to display two different figures for each experiment; this way, we have enough detail at each respective level to see any odd behaviors.

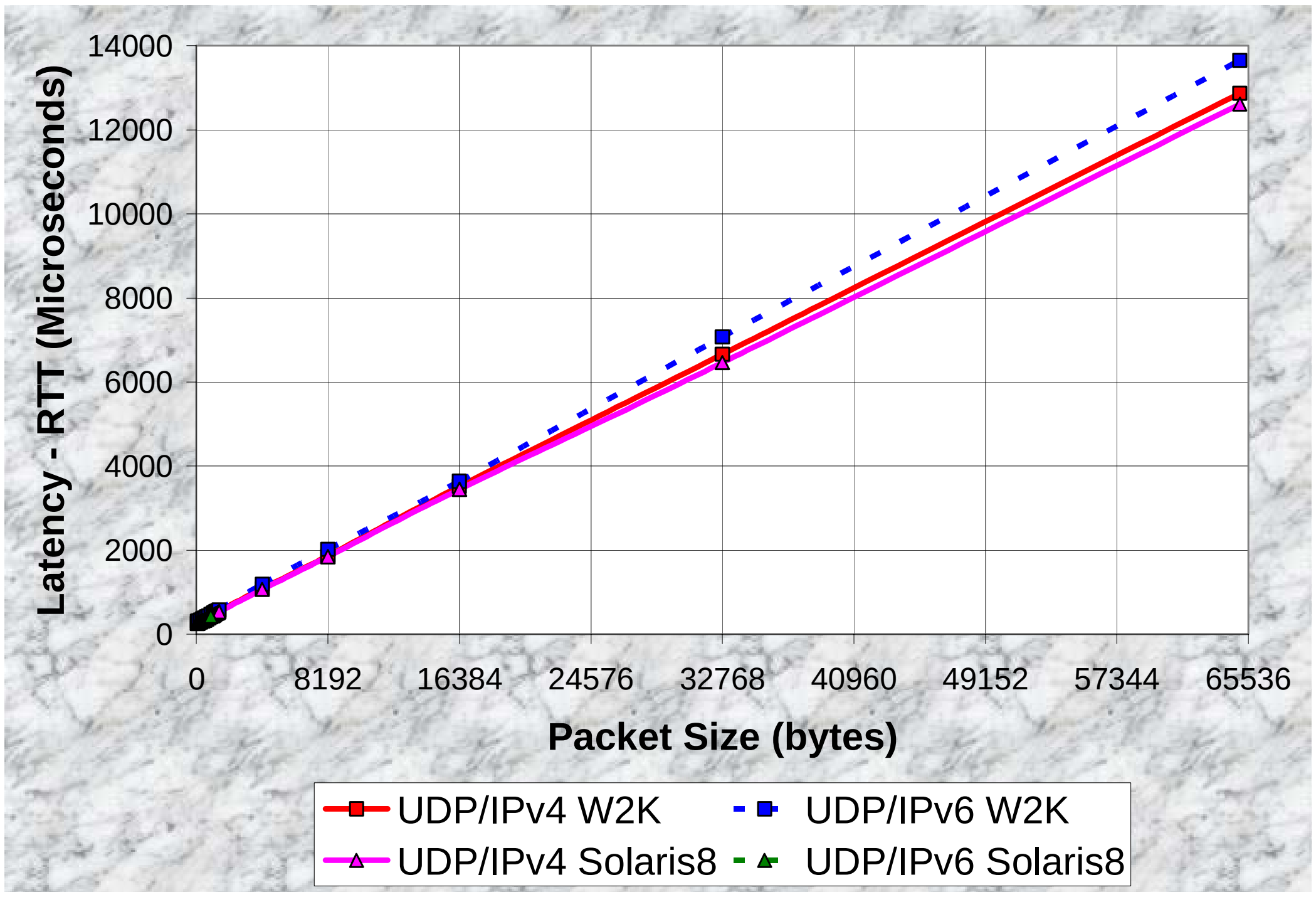

**Figure 19: P2P Test-bed: UDP latency results for IPv4 & IPv6 over Windows 2000 & Solaris 8.0 with packet size ranging from 64 bytes to 64 Kbytes**

For the UDP latency tests depicted in Figure 19 and Figure 20, we have similar behavior as the TCP latency in terms of the IPv6 overhead. Specifically, we have a 7% to 2% overhead for Solaris and an 18% to 4% overhead for Windows when comparing IPv6 with IPv4. However, unlike for the TCP latency tests, Solaris marginally performs better in IPv4 than Windows. For IPv6, Solaris performs quite better when compared to the Windows implementation of IPv6.

It is relatively important to notice the different scales for the RTT used between Figure 19 and Figure 20; hence the same tests and data might look differently although they are only different because of the varying scales. This is off course valid throughout the thesis and therefore the scale should always be checked before conclusions are made between various figures. We tried to be as consistent as possible, but sometimes much detail would be left unnoticed if we were to keep scales identical throughout all the figures in the entire thesis.

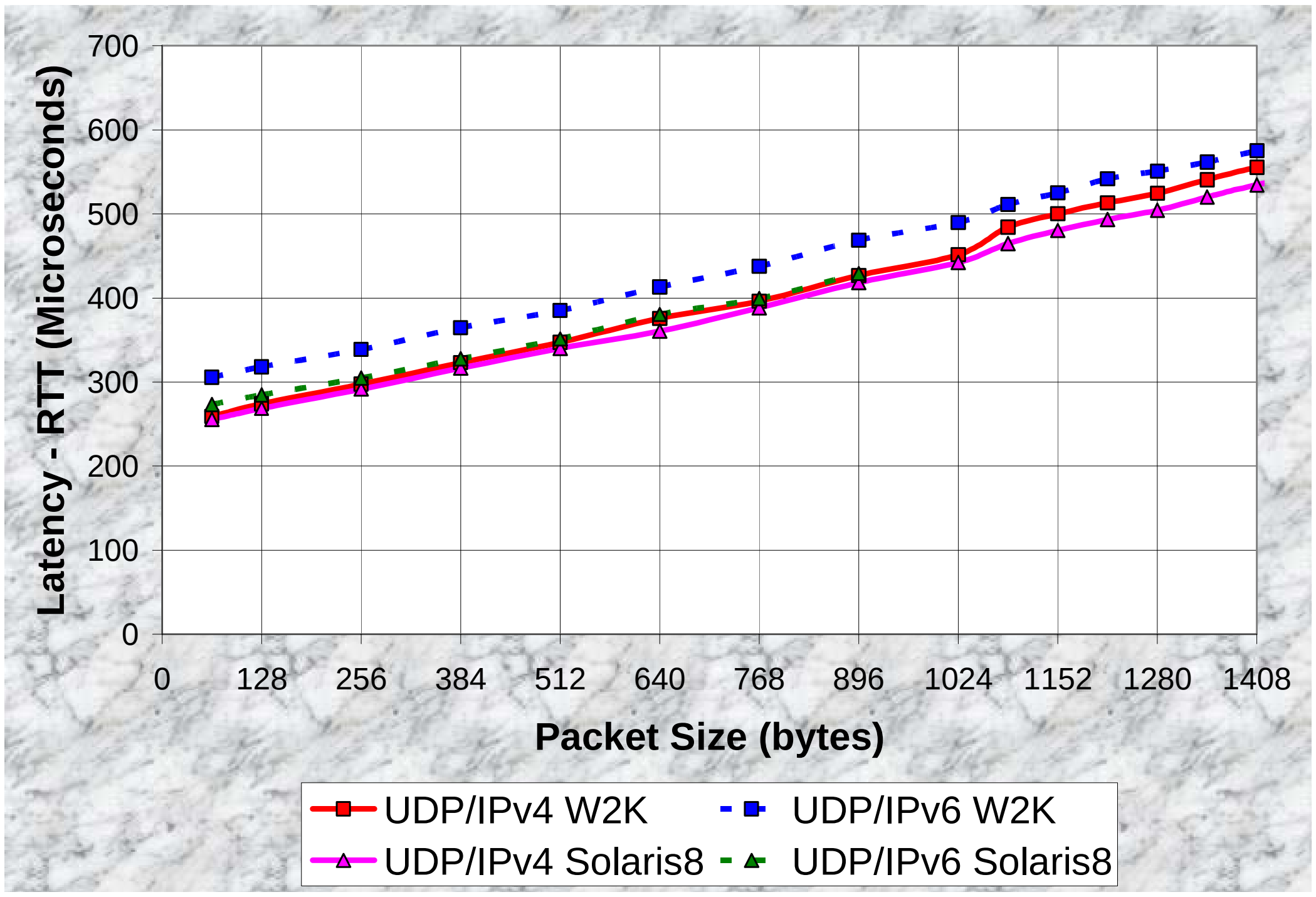

**Figure 20: P2P Test-bed: UDP latency results for IPv4 & IPv6 over Windows 2000 & Solaris 8.0 with packet size ranging from 64 bytes to 1408 bytes**

For the CPU utilization for the latency tests depicted in Figure 21, it is again obvious that TCP has a higher CPU utilization overhead over UDP, and that IPv6 has a higher overhead above each IPv4 protocol respectively.

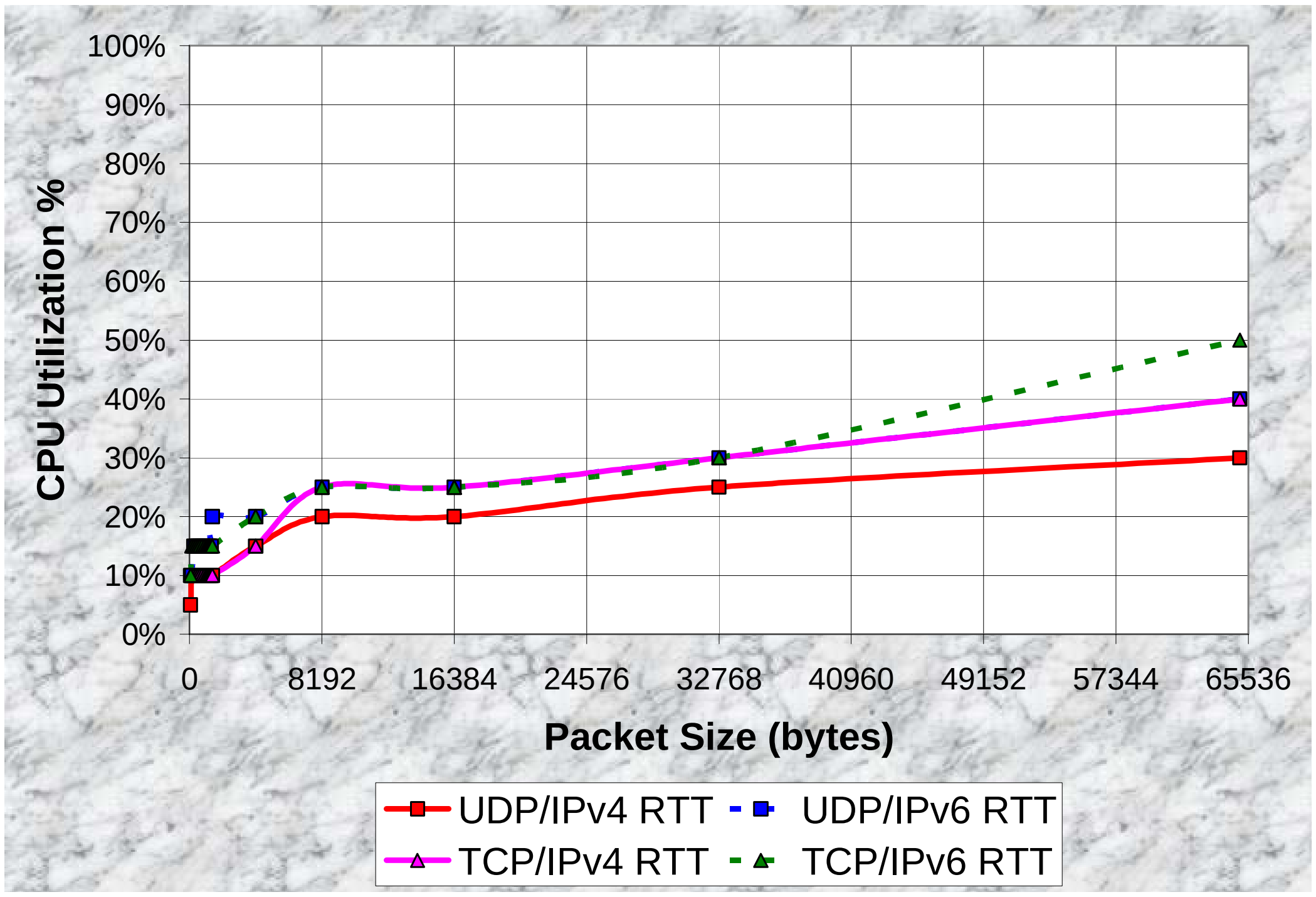

**Figure 21: P2P Test-bed: CPU Utilization results for the latency experiments in IPv4 and IPv6 running TCP and UDP over Windows 2000 with packet size ranging from 64 bytes to 64 Kbytes**

#### *4.2.1.3 Socket Creation Time and TCP Connection Time*

According to Table 4, it can be clearly seen that Solaris 8.0 outperforms Windows 2000 in both the TCP/UDP socket creation time and TCP connection time. It is worth noting that the socket creation time did not change significantly between IPv4 and IPv6 under Solaris, but it did under Windows. The connection time increased in both Solaris and Windows in IPv6, which is most likely the overhead in the address size when setting up the connection. The UDP transport protocol clearly shows similar results as in the TCP results. Obviously, the UDP connection time was irrelevant since UDP is a connection-less protocol. Since UDP does not have a connection mechanism such as TCP, we cannot measure the connection time.

| OS | IP Version | Transport Protocol | Sock. Cr. Time (μs) | Con. Time (μs) |
|---|---|---|---|---|
| **Solaris 8.0** | IPv4 | TCP | 1622.51 | 576.86 |
| **Solaris 8.0** | IPv6 | TCP | 1736.45 | 611.55 |
| **Solaris 8.0** | IPv4 | UDP | 1908.21 | N/A |
| **Solaris 8.0** | IPv6 | UDP | 2041.74 | N/A |
| **Windows 2000** | IPv4 | TCP | 6128.74 | 675.93 |
| **Windows 2000** | IPv6 | TCP | 8006.51 | 1012.13 |
| **Windows 2000** | IPv4 | UDP | 6002.74 | N/A |
| **Windows 2000** | IPv6 | UDP | 6812.13 | N/A |

**Table 4: P2P Test-bed: TCP and UDP socket creation time and TCP connection time in microseconds for both IPv4 and IPv6 running Windows 2000 and Solaris 8.0**

#### *4.2.1.4 Number of TCP Connections per Second*

Once again, Solaris outperforms Windows by a considerable margin. IPv6 in both operating systems seems to incur a considerable overhead as well and will most likely affect the performance of the Internet significantly. The significance of these findings is the most relevant to web servers and clients which have many TCP transactions as many clients access web pages from servers in a short period of time.

An important aspect of the results from Table 5 is the fact that the experiments were performed over the P2P Test-bed. Most likely this kind of performance statistics will not happen in a real world scenario and therefore make sure to see the difference between these results and the results that will be presented in the IBM-Ericsson Test-bed configuration.

| OS | IP Version | Number of Connections |
|---|---|---|
| **Solaris 8.0** | IPv4 | 430 |
| **Solaris 8.0** | IPv6 | 404 |
| **Windows 2000** | IPv4 | 147 |
| **Windows 2000** | IPv6 | 115 |

**Table 5: P2P Test-bed: the number of TCP connections per second for IPv4 and IPv6 running Windows 2000 and Solaris 8.0**

#### *4.2.1.5 Video Client/Server Application*

As a quick overview, the video application was designed for Windows only and therefore we do not have performance metrics for Solaris for this experiment. The performance numbers depicted below and in Table 6 were taken at the client side which had to retrieve the data stream and process it to display the video stream.

This test was pretty clear that IPv6 was incurring an extra overhead over the original IPv4 implementation, although it was a very tolerable 6%. With IPv4, the program netted about 68 Mbit/s and about 9.2 frames per second. Under IPv6, the transfer rates dropped to about 64 Mbit/s and about 8.7 frames per second. Notice that both IPv4 and IPv6 are nowhere near the bandwidth of 100 Mbit/s mainly because of the processing overhead for the client to render the high quality uncompressed video.

| Video Client/Server Application under Windows 2000 | | | |
|---|---|---|---|
| IP Version | Frame Rates (fps) | Transfer Rates (Mbit/s) | Client CPU Utilization |
| IPv4 | 9.2 | 68.05 | 25% |
| IPv6 | 8.7 | 64.12 | 60% |

**Table 6: P2P Test-bed: frame rates and transfer rates for the video client/server application for both IPv4 and IPv6 running Windows 2000**

### 4.2.2 IBM-Ericsson Test-bed Performance Results

#### *4.2.2.1 Throughput*

As Figure 22 and Figure 23 indicate, it can be seen that Solaris 8.0 does slightly better than Windows 2000 over the entire packet size spectrum similar to the performance trend in the P2P Test-bed presented in section 4.2.1. The odd poor performance of the IPv6 network protocol under the TCP throughput experiments presented in these two figures is not consistent with our previous findings. Figure 12 and Figure 13 from section 4.2.1 which represented the same experiment but over the P2P Test-bed showed that IPv6 only incurred only a 2% to 13% overhead depending on the packet size and operating system utilized.

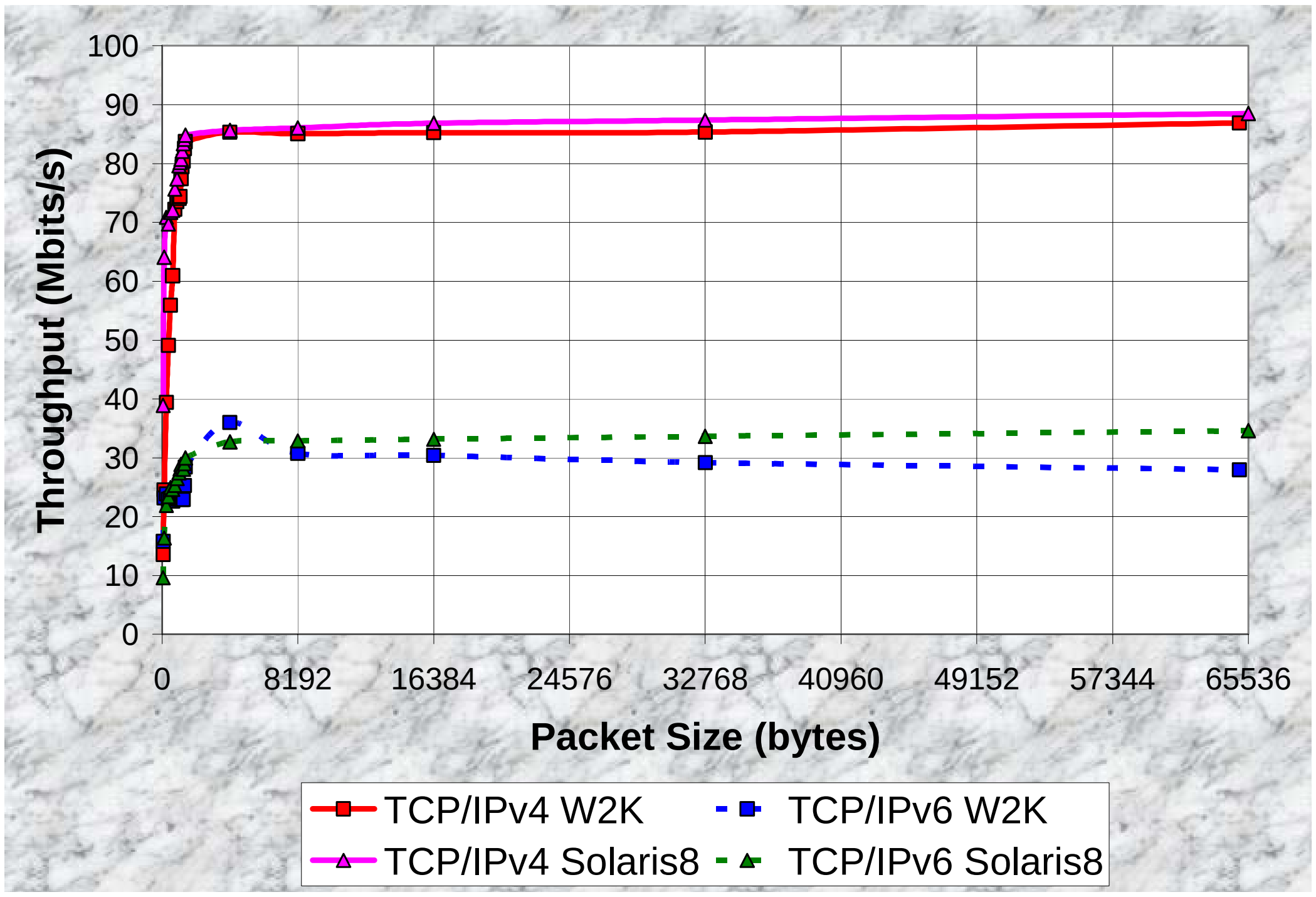

**Figure 22: IBM-Ericsson Test-bed: TCP throughput results for IPv4 & IPv6 over Windows 2000 & Solaris 8.0 with packet size ranging from 64 bytes to 64 Kbytes**

As a quick overview, the dotted lines represent the IPv6 protocol while the solid lines represent the IPv4 protocol. It should be evident that if IPv4 achieves throughput rates surpassing 88 Mbit/s while IPv6 barely gets over 34 Mbit/s under Solaris and 28 Mbit/s under Windows, the performance overhead incurred will render IPv6 as unappealing. It is obvious that the new larger IPv6 header which we have evidence showing IPv6 incurring an overhead of up to 35% over IPv4 in the worst case in the P2P Test-bed cannot be responsible for an overhead surpassing 250% for the TCP throughput experiment performed under Windows. For Solaris, the overhead is as high as 300% for very small packet sizes and the best it can do is about 150% overhead for larger packet sizes.

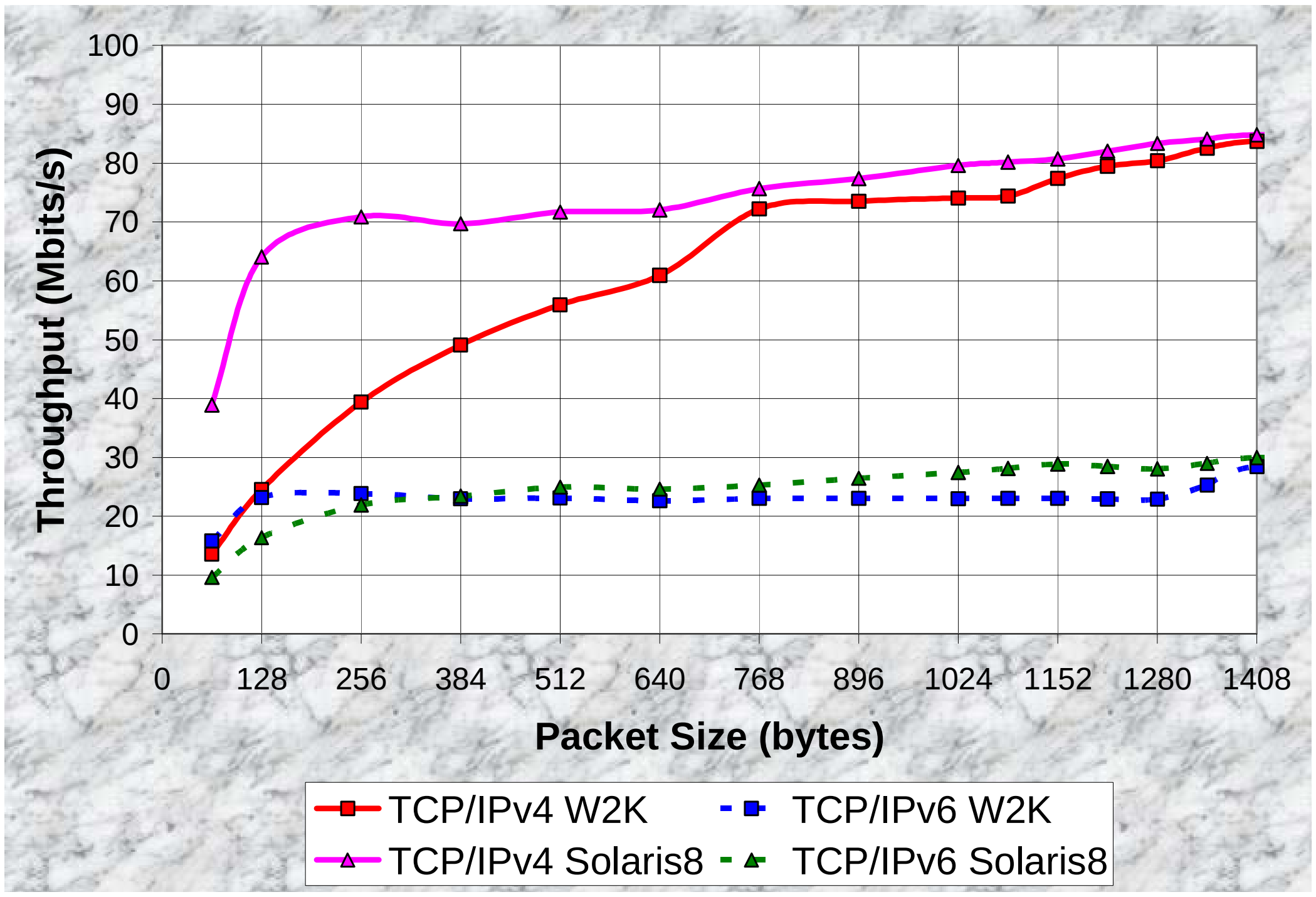

**Figure 23: IBM-Ericsson Test-bed: TCP throughput results for IPv4 & IPv6 over Windows 2000 & Solaris 8.0 with packet size ranging from 64 bytes to 1408 bytes**

As the figure above indicates, it is evident that starting from small packet sizes, the IPv6 protocol performs very poorly under both Windows and Solaris. Since we know that without any routers, IPv6 incurs a minimal performance overhead, we must deduce that the routers are the main cause of the poor performance of IPv6 in this experiment. In order to examine this farther, we tried to measure each router's individual performance by repeating the same experiments with only one router instead of both. Those findings will be presented in section 4.2.3 and 4.2.4.

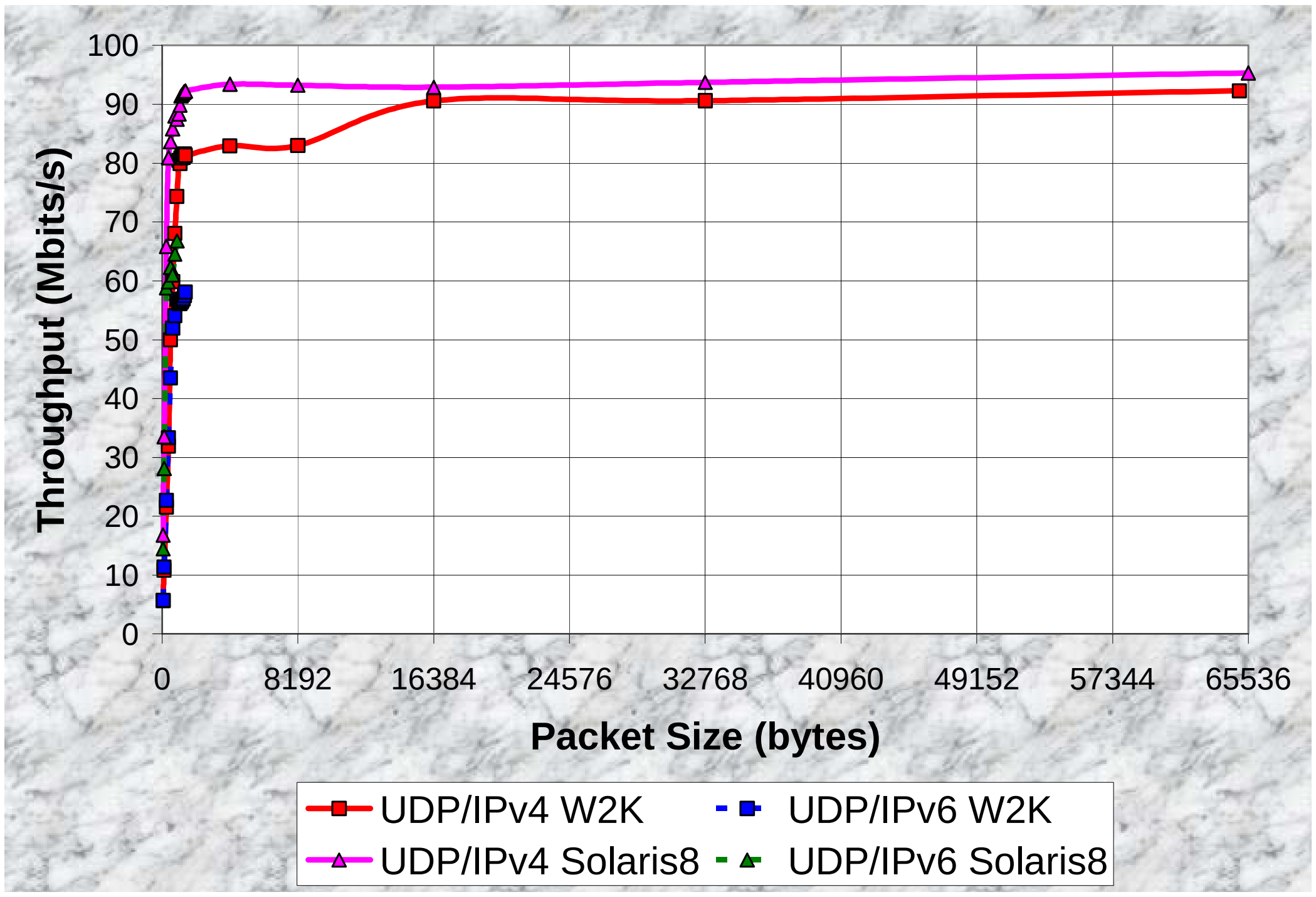

**Figure 24: IBM-Ericsson Test-bed: UDP throughput results for IPv4 & IPv6 over Windows 2000 & Solaris 8.0 with packet size ranging from 64 bytes to 64 Kbytes**

In regards to Figure 24, things look as we would expect them. Do not forget that the UDP throughput experiments were not performed for packet sizes greater than the Ethernet MTU size of 1514 bytes and therefore the two solid lines that are depicted in the larger packet ranges both represent the IPv4 network protocols of Windows and Solaris. Once again, Solaris outperforms Windows by several Mbit/s.

On the other hand, Figure 25 is beginning to show a similarly odd behavior as did TCP in Figure 22 and Figure 23. However, it is not quite as bad since the performance overhead is revolving around 40% in both Windows and Solaris. Regardless, 40% overhead for the majority of the packet size is much worse than the typical 10% of TCP and 30% of UDP in the P2P Test-bed experiments.

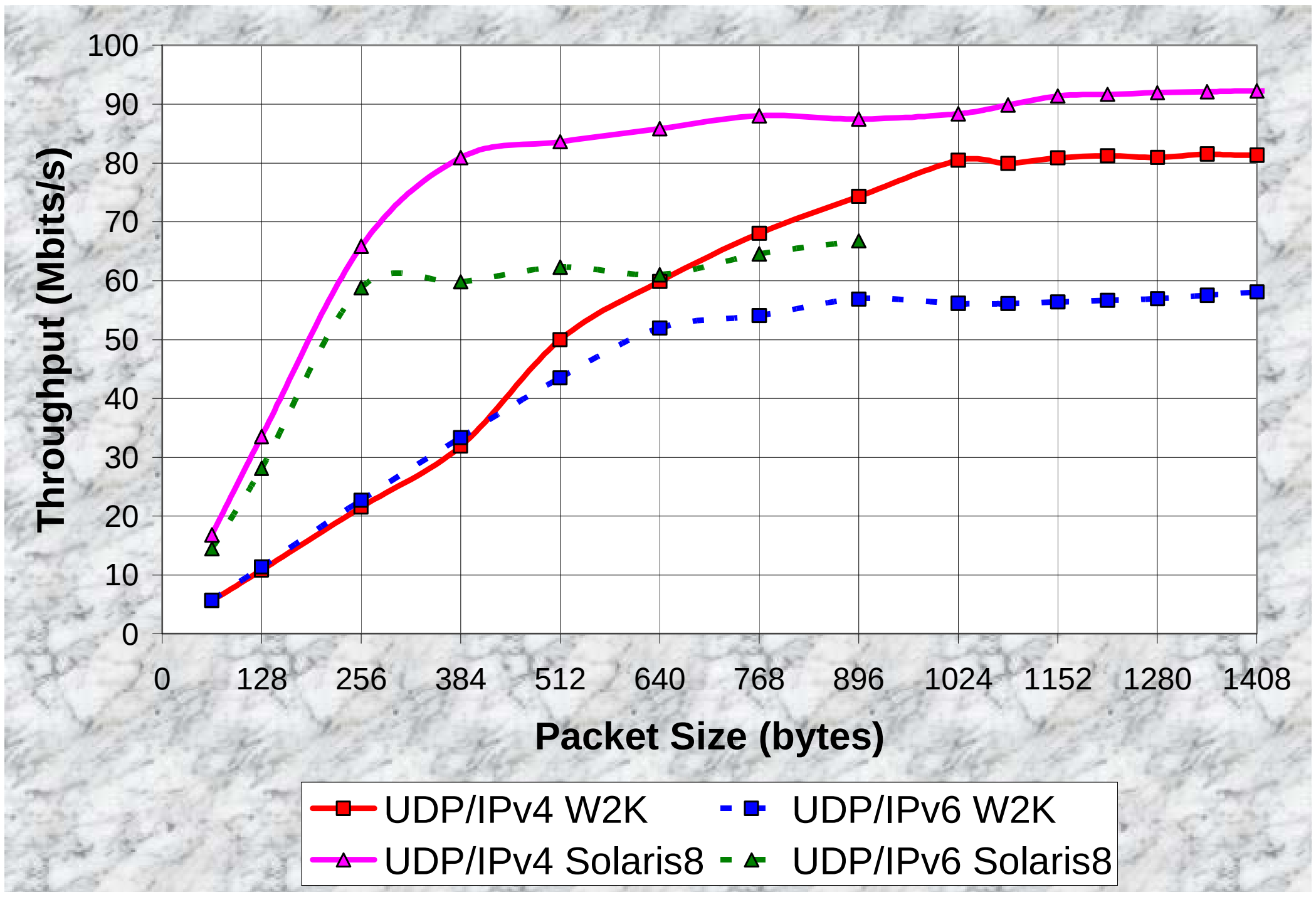

**Figure 25: IBM-Ericsson Test-bed: UDP throughput results for IPv4 & IPv6 over Windows 2000 & Solaris 8.0 with packet size ranging from 64 bytes to 1408 bytes**

The interesting fact about Figure 26's results in CPU utilization is that IPv6 utilizes the CPU about the same as IPv4 for TCP, and in some instances, even less. This at first is counterintuitive, however if the performance numbers regarding TCP's throughput are recalled, they were as much as 250% lower in IPv6 than in IPv4. Therefore, it makes complete sense that a host transmitting at 28 Mbit/s compared to 88 Mbit/s should utilize the CPU less considering all things being equal.

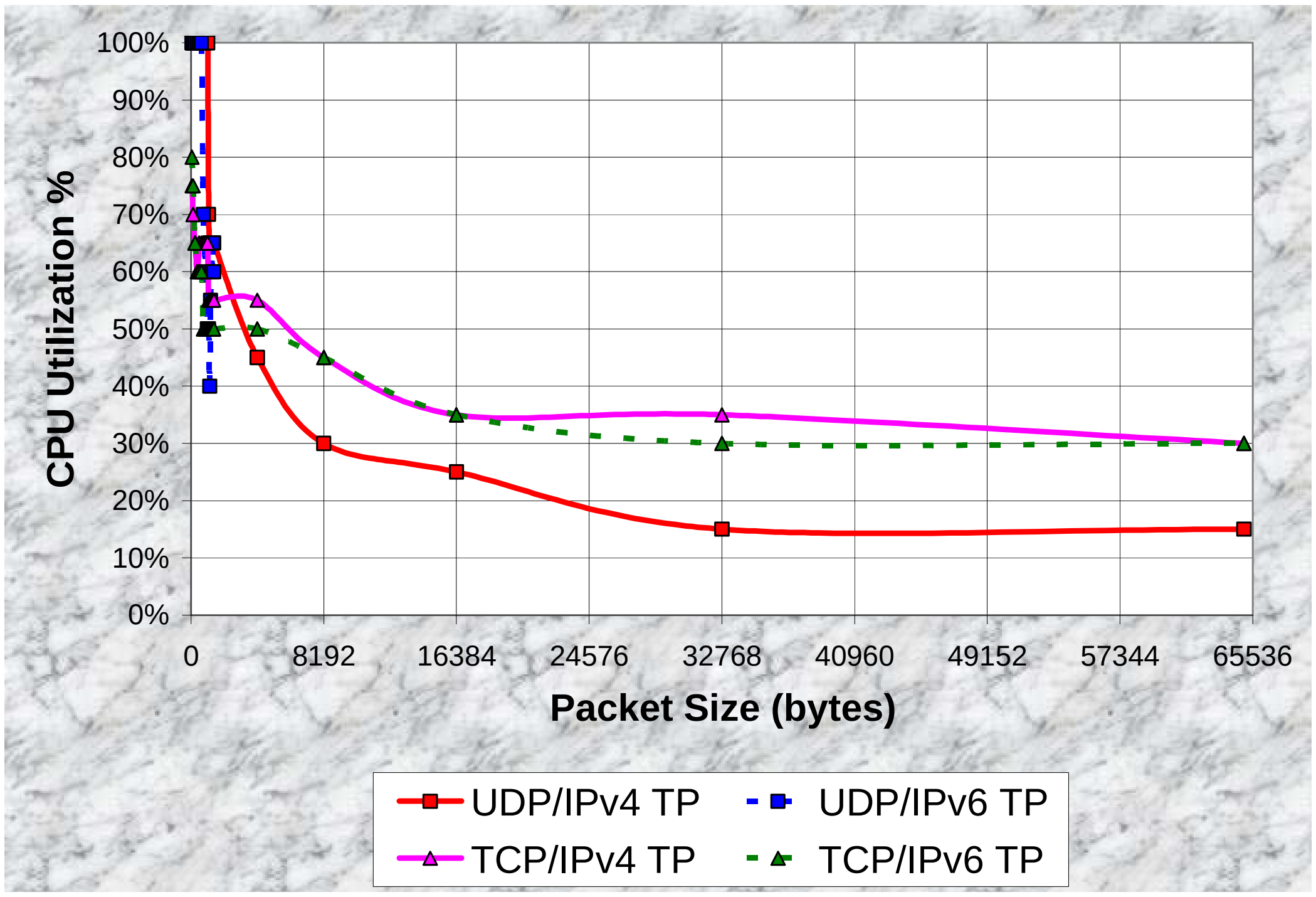

**Figure 26: IBM-Ericsson Test-bed: CPU Utilization results for the throughput experiments in IPv4 and IPv6 running TCP and UDP over Windows 2000 with packet size ranging from 64 bytes to 64 Kbytes**

However, IPv6 is not equivalent to IPv4 and would most likely utilize the CPU more than IPv4, but since there is such a huge difference in the amount of work the end hosts must do, the CPU utilization reflected is nearly identical for both protocols. Furthermore, realize that just as before, the UDP transport protocol consistently requires less CPU utilization compared to the TCP transport protocol, which was to be expected.

#### *4.2.2.2 Latency*

Figure 27 clearly depicts similar performance deficits for IPv6 in the larger packet sizes. Notice how Windows and Solaris offer nearly identical performance.

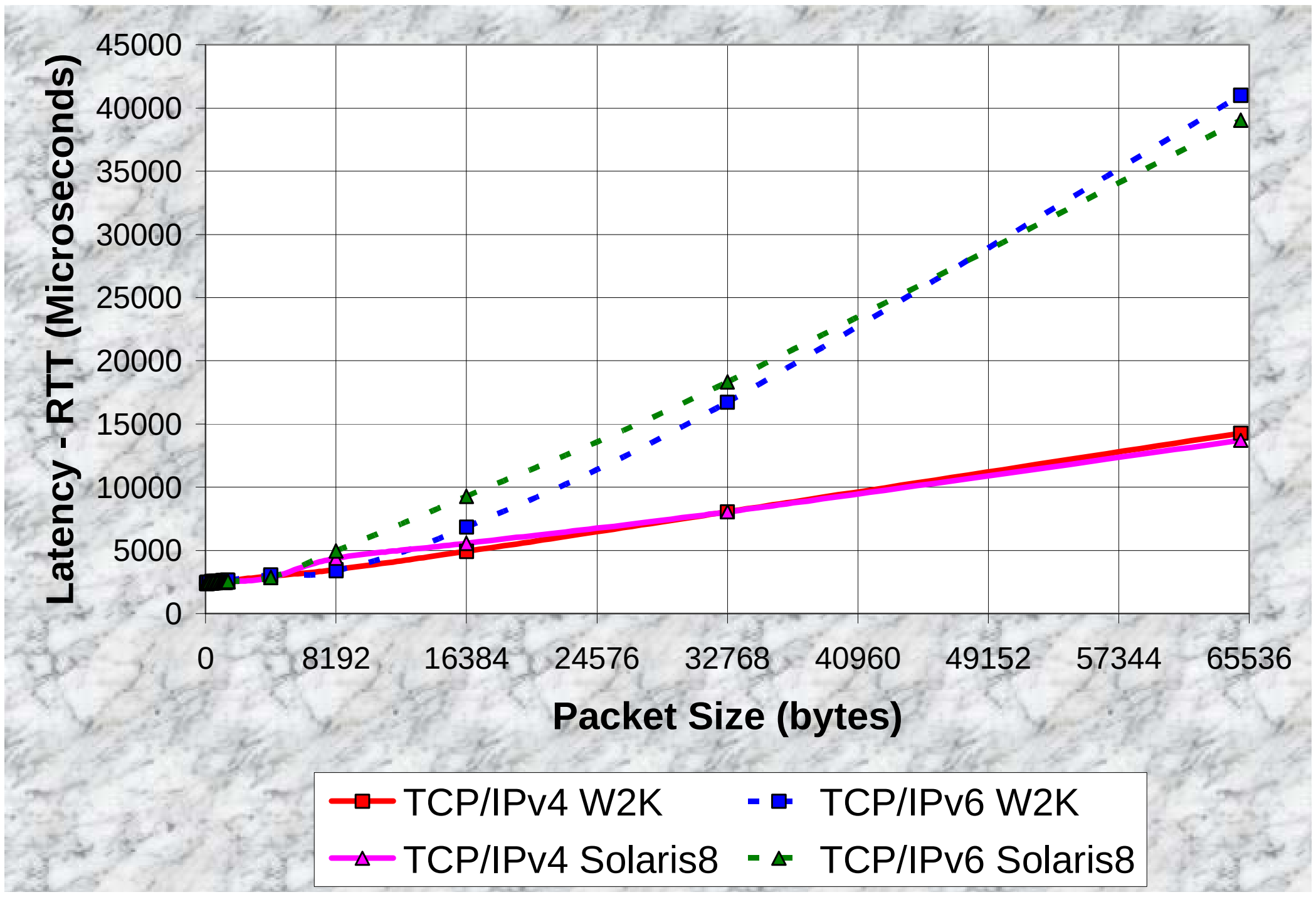

**Figure 27: IBM-Ericsson Test-bed: TCP latency results for IPv4 & IPv6 over Windows 2000 & Solaris 8.0 with packet size ranging from 64 bytes to 64 Kbytes**

However, for 64 Kbyte packets, IPv6 has a latency of about 40,000 μs (40 ms) while IPv4 retains the fairly low 14,000 μs (14 ms) similar to the tests performed. In evaluating the latency performance, it is beginning to make sense why the throughput performance of IPv6 was so bad under the TCP transport protocol.

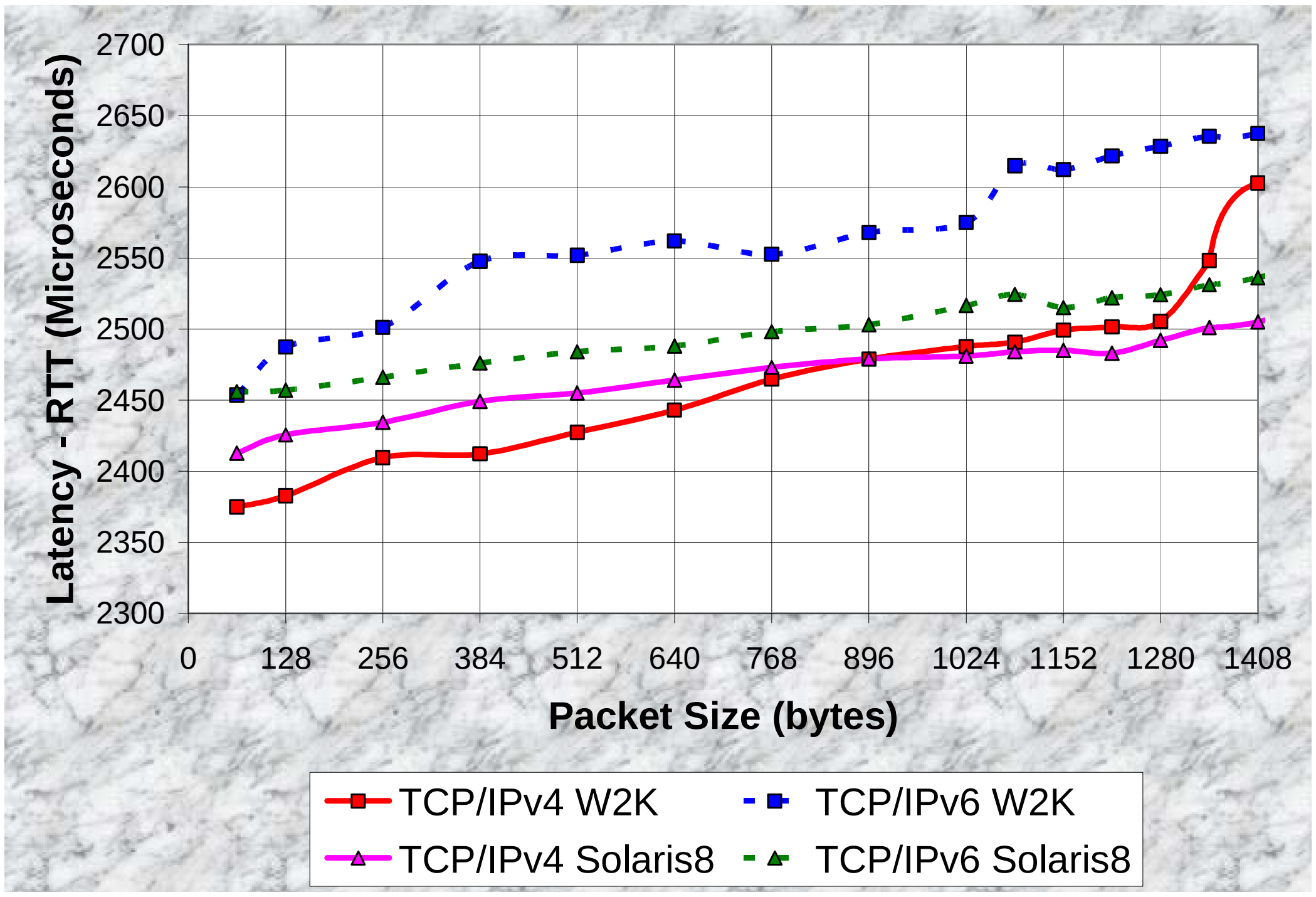

**Figure 28: IBM-Ericsson Test-bed: TCP latency results for IPv4 & IPv6 over Windows 2000 & Solaris 8.0 with packet size ranging from 64 bytes to 1408 bytes**

The interesting fact about Figure 28 is that the latency for the smallest packet size (64 bytes) tested revolved around 2,400 μs (2 ms) instead of about 300 μs (0.3 ms) for the P2P Test-bed. This is obviously the extra overhead that the two routers (IBM and Ericsson) are incurring; according to our results, it is on the order of about 2 ms which the combination of the IBM and Ericsson routers are slowing down every single packet. For the smaller packet sizes of 64 bytes to 1408 bytes,

Windows had a 4% to 5% overhead for the IPv6 network protocol while Solaris only had a very acceptable 1% to 2% overhead.

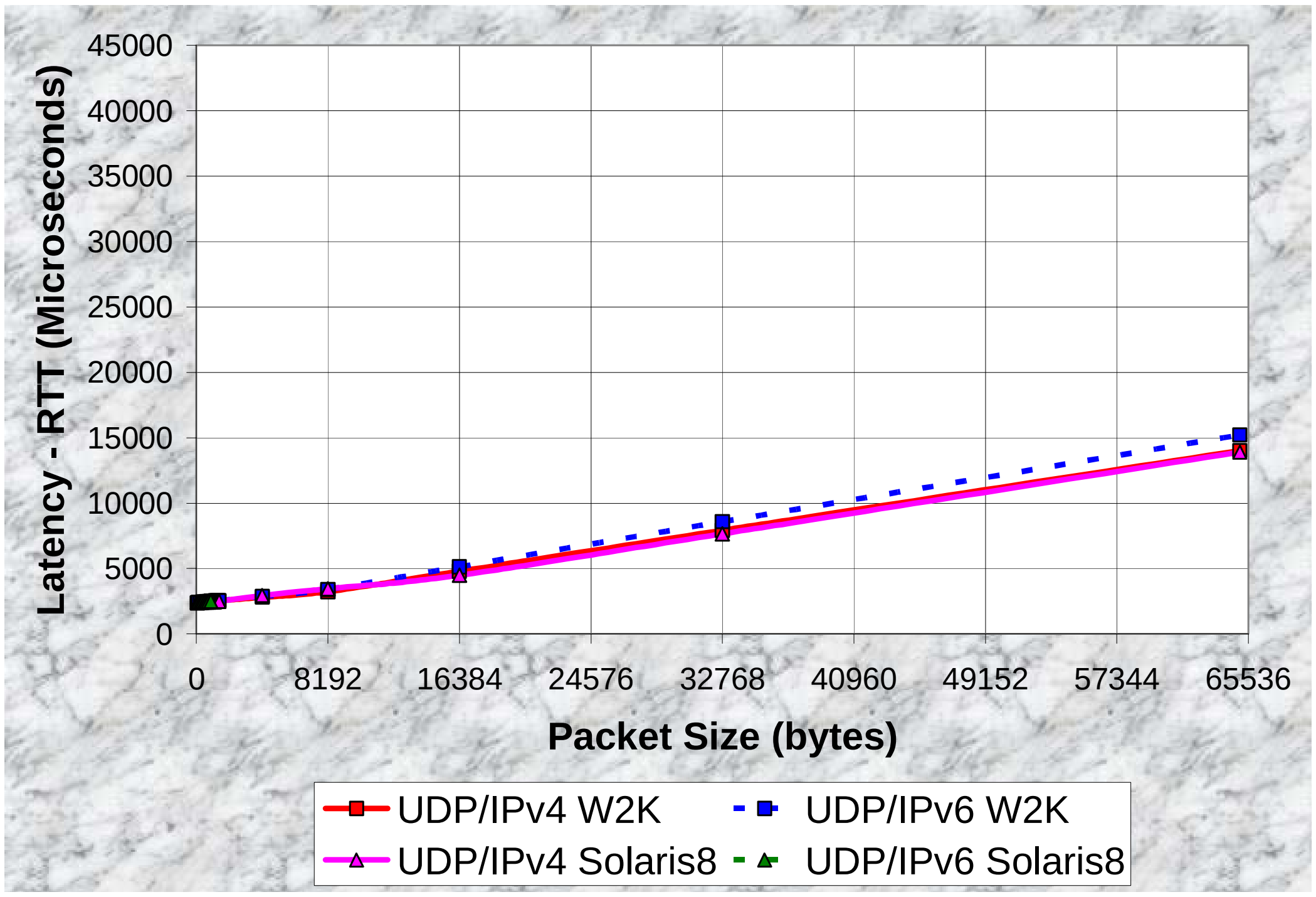

**Figure 29: IBM-Ericsson Test-bed: UDP latency results for IPv4 & IPv6 over Windows 2000 & Solaris 8.0 with packet size ranging from 64 bytes to 64 Kbytes**

In this experiment (Figure 29), there is nothing very interesting since everything is consistent with our previous results. The noteworthy fact is that IPv6 incurs 0% to 8% overhead ranging from smaller packet to larger packets, while Solaris incurs a mere 1% to 4% overhead over the same packet size range.

In Figure 30 below, it appears that IPv6 offers near identical performance at IPv4 under Windows 2000. Under Solaris 8.0, the same comparison has a 1% to 3% overhead which is very acceptable. Notice that for small packet sizes, the round trip time of a packet again revolves around 2,400 μs (2.4 ms) instead of the 300 μs (0.3 ms) of the previous tests performed on the P2P Test-bed; realize that the delay incurred by the routers is again about 2 ms. The interesting point about the results below is the roughness of the plotted line for any of the protocols evaluated. We cannot conclude anything merely by the fact that we have a rough graph; however we can imply that something in the routers is causing the performance of these protocols to be degraded. We will discuss this in more detail in our conclusion.

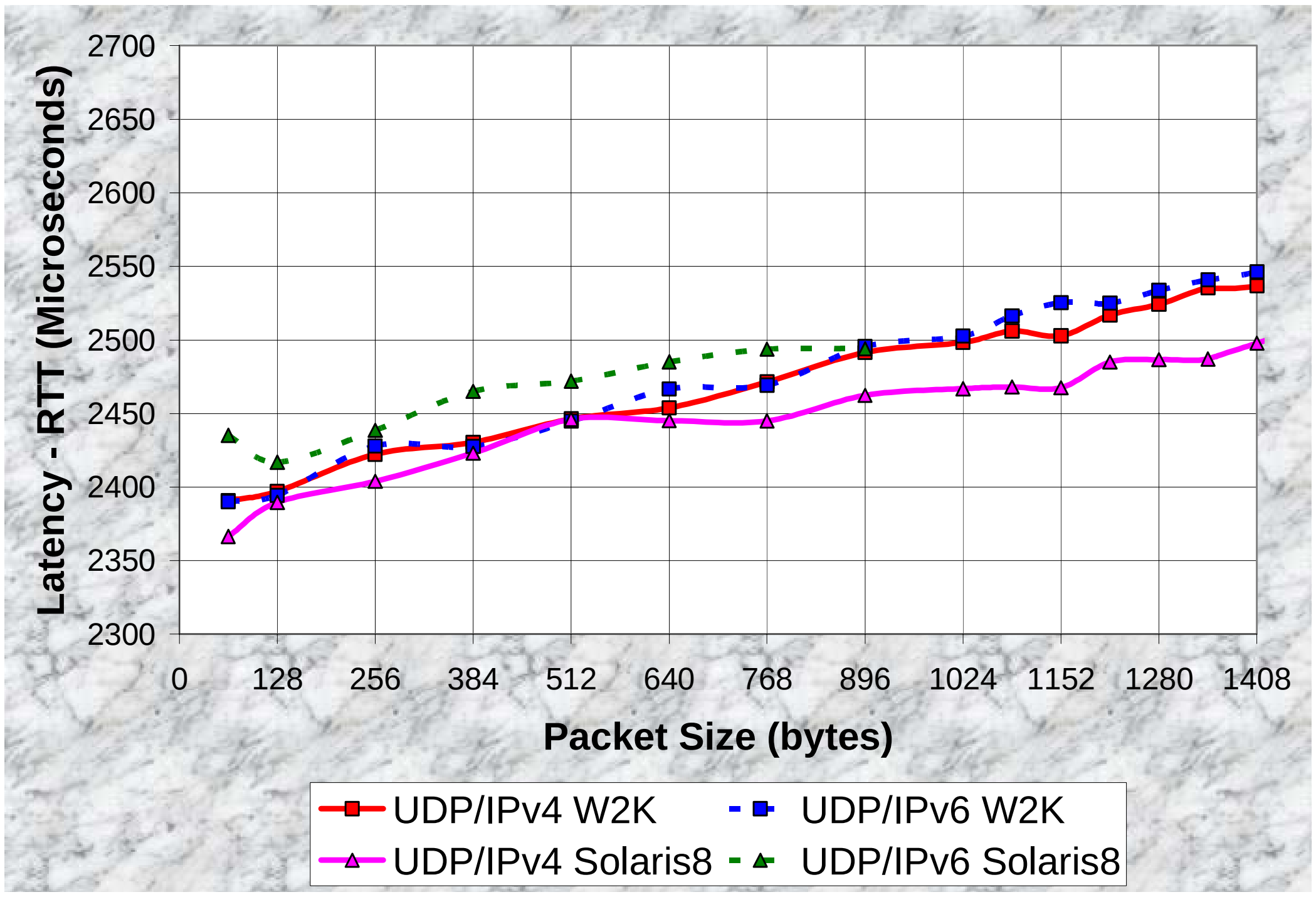

**Figure 30: IBM-Ericsson Test-bed: UDP latency results for IPv4 & IPv6 over Windows 2000 & Solaris 8.0 with packet size ranging from 64 bytes to 1408 bytes**

Figure 31 depicts the CPU utilization which seems very consistent with our previous experiments. The TCP transport protocol requires more CPU cycles than the UDP transport protocol, and IPv6 requires yet even more CPU utilization.

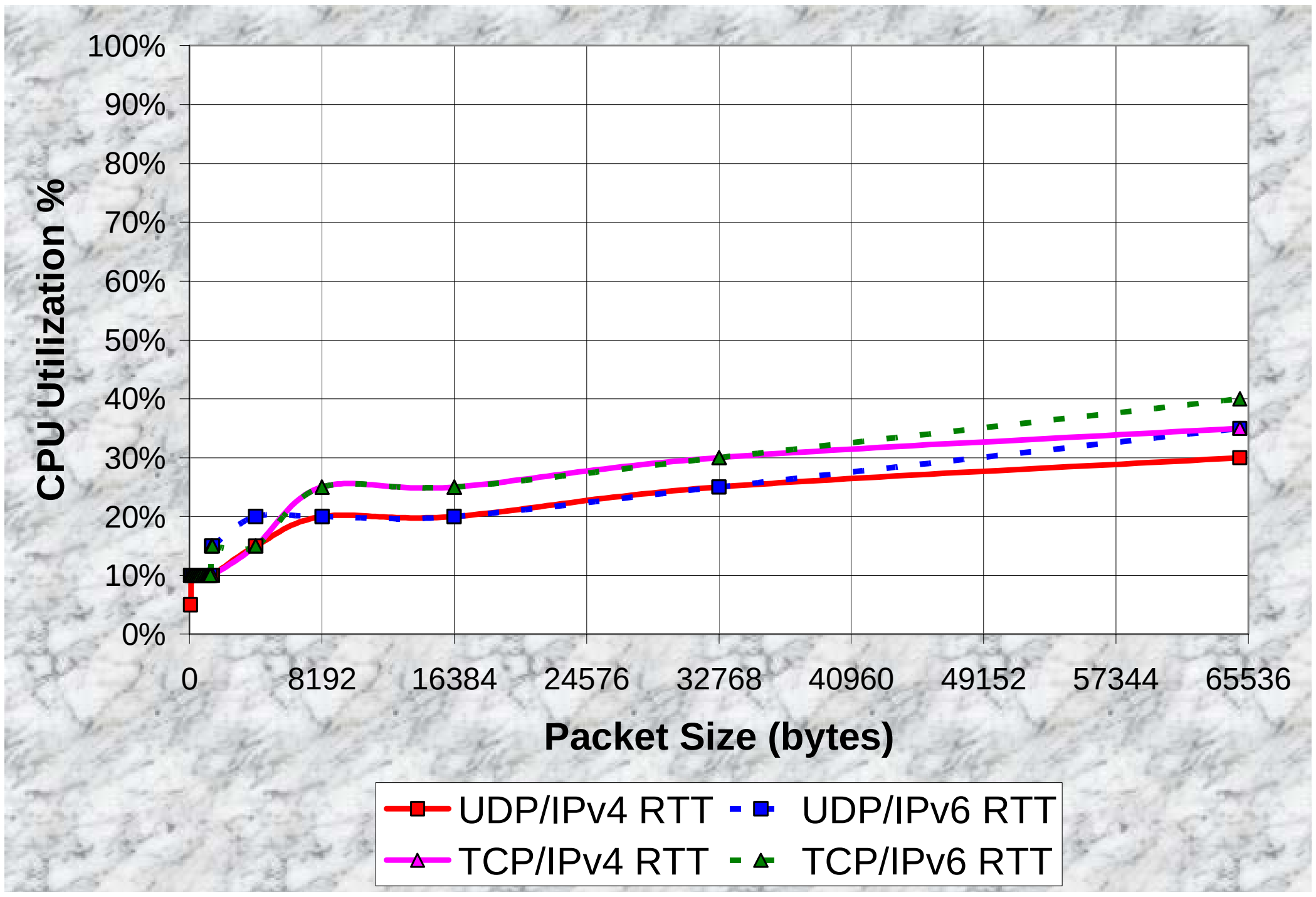

**Figure 31: IBM-Ericsson Test-bed: CPU Utilization results for the latency experiments in IPv4 and IPv6 running TCP and UDP over Windows 2000 with packet size ranging from 64 bytes to 64 Kbytes**

### *4.2.2.3 Socket Creation Time and TCP Connection Time*

For these experiments depicted by Table 7, the socket creation time remained unchanged from the P2P Test-bed since the socket creation time is independent of other entities in the network and therefore it makes sense that the values have not changed. On the other hand, the connection time required to make the connection between the client and server significantly increased from values of 500 μs and 1,000 μs to values like 2,500 μs and 3,000 μs.

When TCP invokes the connect function, the TCP transport protocol deals with very small packets sizes as it negotiates the connection. It therefore makes sense that the connection time be similar to the smallest packet size TCP RTT we observed in the previous section. Notice that Solaris outperforms Windows marginally in the IPv4 connection time, and pretty significantly in the IPv6 connection time.

| OS | IP Version | Transport Protocol | Sock. Cr. Time (μs) | Con. Time (μs) |
|---|---|---|---|---|
| **Solaris 8.0** | IPv4 | TCP | 1622.51 | 2509.86 |
| **Solaris 8.0** | IPv6 | TCP | 1736.45 | 2558.54 |
| **Solaris 8.0** | IPv4 | UDP | 1908.21 | N/A |
| **Solaris 8.0** | IPv6 | UDP | 2041.74 | N/A |
| **Windows 2000** | IPv4 | TCP | 6128.74 | 2608.93 |
| **Windows 2000** | IPv6 | TCP | 8006.51 | 2959.13 |
| **Windows 2000** | IPv4 | UDP | 6002.74 | N/A |
| **Windows 2000** | IPv6 | UDP | 6812.13 | N/A |

**Table 7: IBM-Ericsson Test-bed: TCP and UDP socket creation time and TCP connection time in microseconds for both IPv4 and IPv6 running Windows 2000 and Solaris 8.0**

Similar to our previous results in Table 4, the socket creation time for Windows is much longer than that of Solaris. For Solaris, creating an IPv6 socket adds about a 7% overhead above IPv4. On the other hand, for Windows creating an IPv6 socket adds 14% to 30% overhead on top of IPv4.

#### *4.2.2.4 Number of TCP Connections per Second*

The previous results from the P2P Test-bed found in Table 5 of 430, 404, 147 and 115 for Solaris IPv4, IPv6, Windows IPv4 and IPv6 respectively were way too high for the typical network that would be found in present day networks due to the fact that there were no routers between the client and server and therefore it would not suffer from extra delays which routers ultimately inflict.  The numbers bellow are probably very typical to the number of connection that Solaris and Windows can maintain per second.

| OS | IP Version | Number of Connections |
|---|---|---|
| **Solaris 8.0** | IPv4 | 161 |
| **Solaris 8.0** | IPv6 | 156 |
| **Windows 2000** | IPv4 | 94 |
| **Windows 2000** | IPv6 | 79 |

**Table 8: IBM-Ericsson Test-bed: The number of TCP connections per second for IPv4 and IPv6 running Windows 2000 and Solaris 8.0**

#### *4.2.2.5 Video Client/Server Application*

Similarly to our findings for the TCP throughput experiments, the video client/server application confirms our findings that the IBM-Ericsson Test-bed indeed hinders the performance of the TCP stream, especially since the video application utilizes packet sizes of 64 Kbytes. It should be clear why the CPU utility of the IPv6 network protocol is only 30% compared to the 60% experienced when operated over the P2P Test-bed; the achieved throughput this time was only 26 Mbit/s while it was previously 64 Mbit/s and hence it required a much higher CPU utilization.

| **Video Client/Server Application under Windows 2000** | | | |
|---|---|---|---|
| IP Version | Frame Rates (fps) | Transfer Rates (Mbit/s) | Client CPU Utilization |
| IPv4 | 8.9 | 66.21 | 25% |
| IPv6 | 3.5 | 26.43 | 30% |

**Table 9: IBM-Ericsson Test-bed: Frame rates and transfer rates for the video client/server application for both IPv4 and IPv6 running Windows 2000**

### 4.2.3 IBM Test-bed Performance Results

Due to the very poor performance of IPv6 over the IBM-Ericsson Test-bed in comparison to the P2P Test-bed, we decided to try to isolate the problem by performing some of the experiments again with only one router. Section 4.2.3 and section 4.2.4 do exactly this, and focus mainly on the TCP Latency and throughput experiments since they are the key to the rest of the experiments results. The UDP experiments were not as dramatically affected by the IBM-Ericsson Test-bed, and hence in order to conserve space we will omit the results for the UDP tests. Similarly, we will omit the other 3 experiments (socket creation time and TCP connection time, the number of TCP connections per second, and the video client/server application) since the results can be approximately inferred from the findings in the other experiments. The main purpose of these next two sections is to offer some explanation for the poor performance of IPv6 under the TCP transport protocol.

#### *4.2.3.1 Throughput*

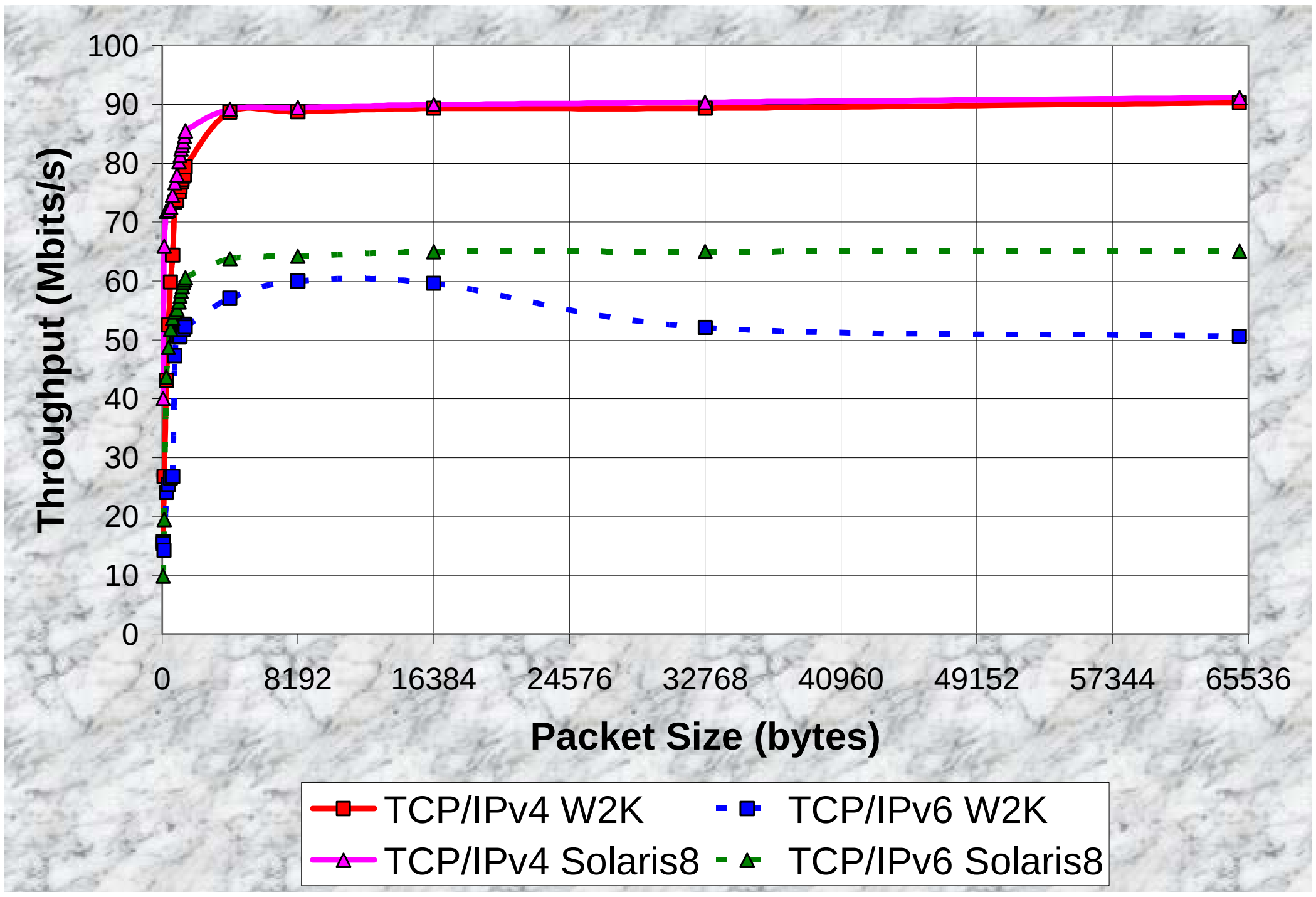

**Figure 32: IBM Test-bed: TCP throughput results for IPv4 & IPv6 over Windows 2000 & Solaris 8.0 with packet size ranging from 64 bytes to 64 Kbytes**

From Figure 32, we can see that the performance of IPv6 is dramatically lower than that of IPv4, however it is much better than the results obtained when we had both routers in the picture.

Figure 33 below clearly shows the consistent behavior we have see throughout which depicts Solaris outperforming Windows and similarly IPv4 outperforming IPv6.

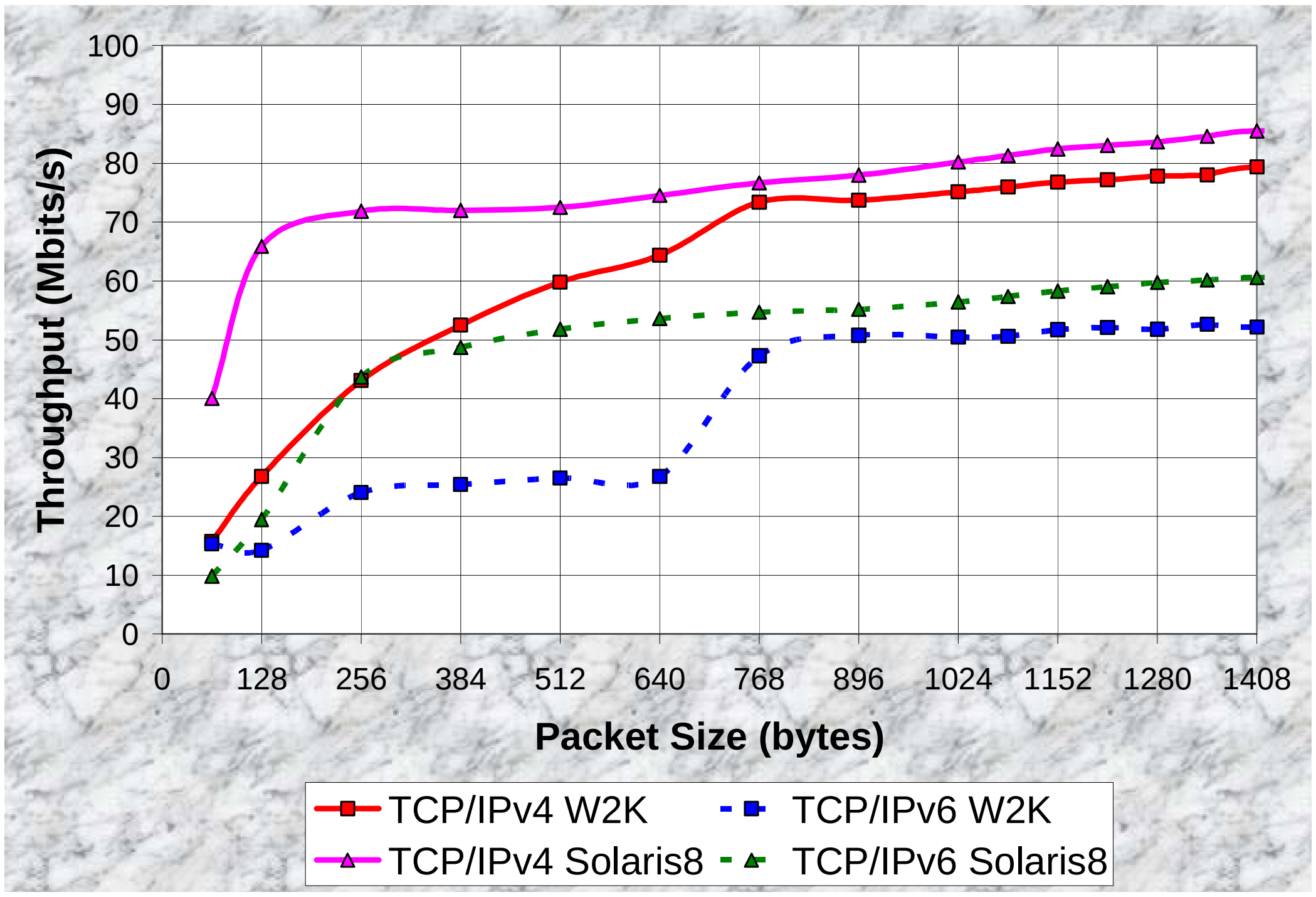

**Figure 33: IBM Test-bed: TCP throughput results for IPv4 & IPv6 over Windows 2000 & Solaris 8.0 with packet size ranging from 64 bytes to 1408 bytes**

### *4.2.3.2 Latency*

For the latency tests in Figure 34, it is quite evident that the RTT experiences similarly larger values for large packets. However, notice that the RTT for a 64 Kbyte IPv6 packet is about 24 ms while it used to be about 40 ms in the IBM-Ericsson Test-bed. Obviously, the IBM router is contributing to the entire problem somewhat, but it is not clear to what degree until we examine the Ericsson Test-bed performance in the next section.

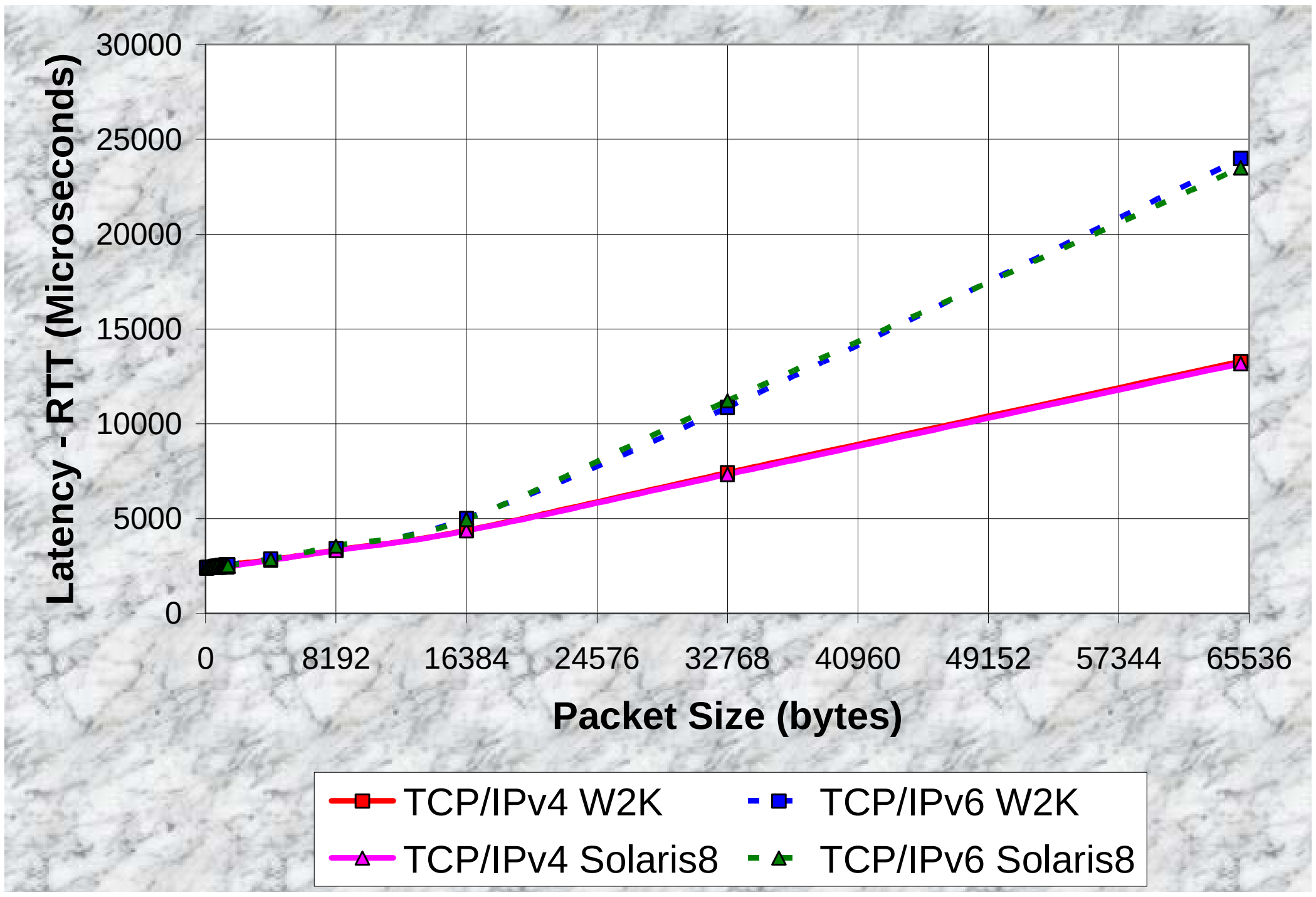

**Figure 34: IBM Test-bed: TCP latency results for IPv4 & IPv6 over Windows 2000 & Solaris 8.0 with packet size ranging from 64 bytes to 64 Kbytes**

It is obvious that Figure 35 is very similar to Figure 28 in which we performed the same experiment but on the IBM-Ericsson Test-bed. Notice the round trip times of the smallest size packets is around 2.4 ms to 2.5 ms which is consistent to the IBM-Ericsson Test-bed RTTs. Since the IBM router seems to be incurring most of the delay in the TCP/IPv6 packets, we must conclude that the Ericsson router incurs very little to no delay on the TCP/IPv6 packets it processes. The next section performs the same two experiments we had in this subsection on the Ericsson Test-bed and we will finally be able to conclude where the performance deficit was all along.

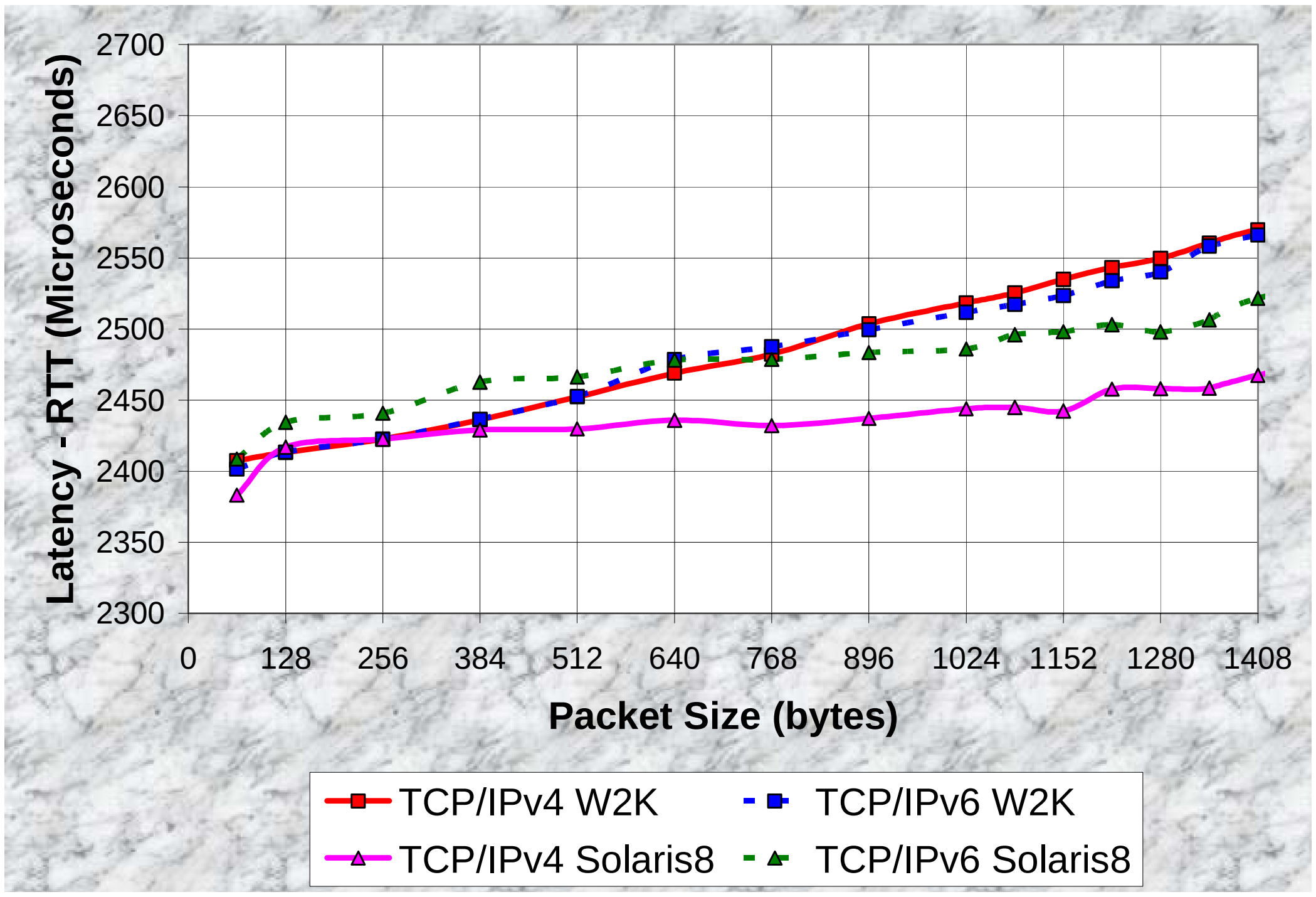

**Figure 35: IBM Test-bed: TCP latency results for IPv4 & IPv6 over Windows 2000 & Solaris 8.0 with packet size ranging from 64 bytes to 1408 bytes**

### 4.2.4 Ericsson Test-bed Performance Results

Due to the very poor performance of IPv6 over the IBM-Ericsson Test-bed and the IBM Test-bed in relation to the P2P Test-bed, we decided to try to isolate the problem by performing the same experiments again, but this time using just the Ericsson router. This section and the previous one (section 4.2.3) do exactly this, and focus mainly on the TCP Latency and throughput experiments since they are the key to the rest of the experiments results. The UDP transport protocol's performance in the IBM-Ericsson Test-bed was to be expected more or less, and therefore in order to conserve space, we will ignore them. The experiments preformed in this section are over the Ericsson Test-bed which omitted the IBM router.

#### *4.2.4.1 Throughput*

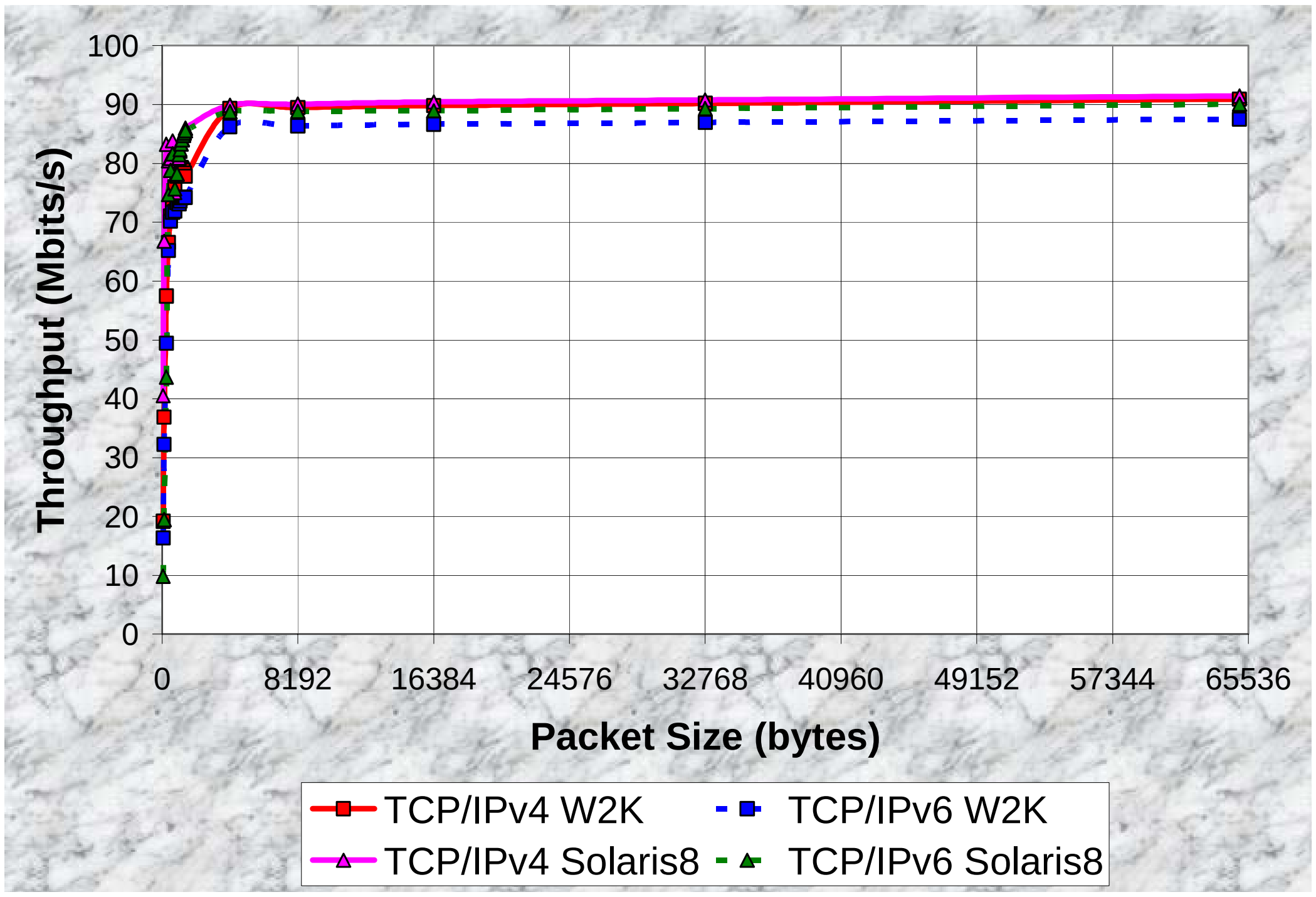

**Figure 36: Ericsson Test-bed: TCP throughput results for IPv4 & IPv6 over Windows 2000 & Solaris 8.0 with packet size ranging from 64 bytes to 64 Kbytes**

Both Figure 36 and Figure 37 confirm that the Ericsson Test-bed has minimal impact in terms of performance overhead of IPv6 compared to IPv4. The results depicted here are rather similar to the ones presented for the P2P Test-bed and therefore it is clear that the Ericsson router handles the TCP/IPv6 packets just as efficiently as the TCP/IPv4 packets. Obviously, there is still the usual overhead of 17% to 1% for the smaller packets to the larger ones, but this was to be expected considering the larger IPv6 header size.

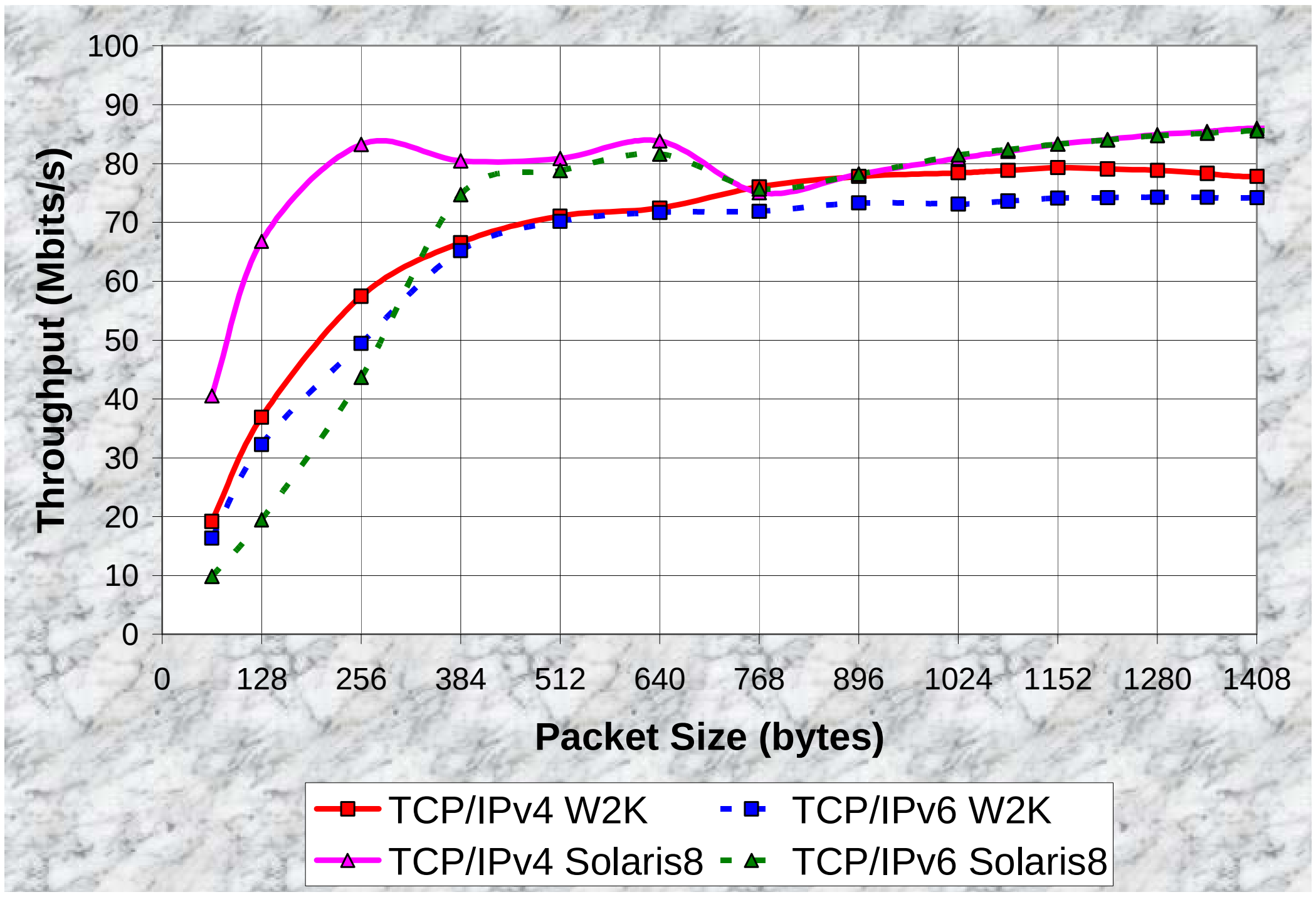

**Figure 37: Ericsson Test-bed: TCP throughput results for IPv4 & IPv6 over Windows 2000 & Solaris 8.0 with packet size ranging from 64 bytes to 1408 bytes**

### *4.2.3.2 Latency*

Figure 38 below clearly shows that the latency incurred on the Ericsson Test-bed is slightly more than that of the P2P Test-bed. It should now be evident that the Ericsson router incurs minimal delays while the IBM router was in the most part responsible for the very poor performance of IPv6 over the IBM-Ericsson Test-bed. There is only a 2% to 5% overhead for the larger packet sizes which is very acceptable.

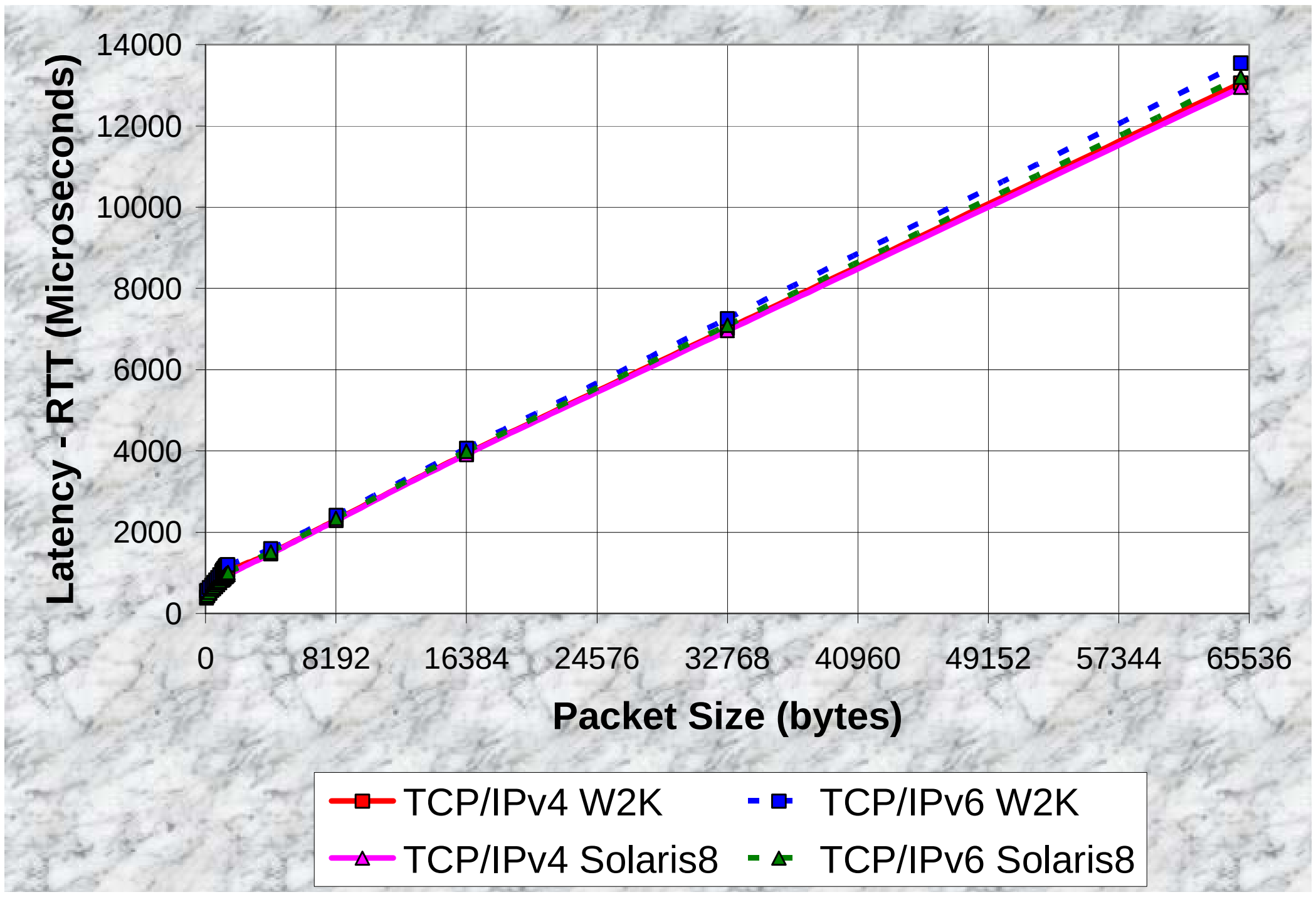

**Figure 38: Ericsson Test-bed: TCP latency results for IPv4 & IPv6 over Windows 2000 & Solaris 8.0 with packet size ranging from 64 bytes to 64 Kbytes**

In Figure 39, the overheads of IPv6 increases to as much as 36% for small packets and as little as 13% for the larger packets under Windows; for Solaris, it is 7% to 5% ranging from the smaller packets to the larger ones. Again, Solaris outperforms Windows in both IPv4 and IPv6 throughout most of the packets size range.

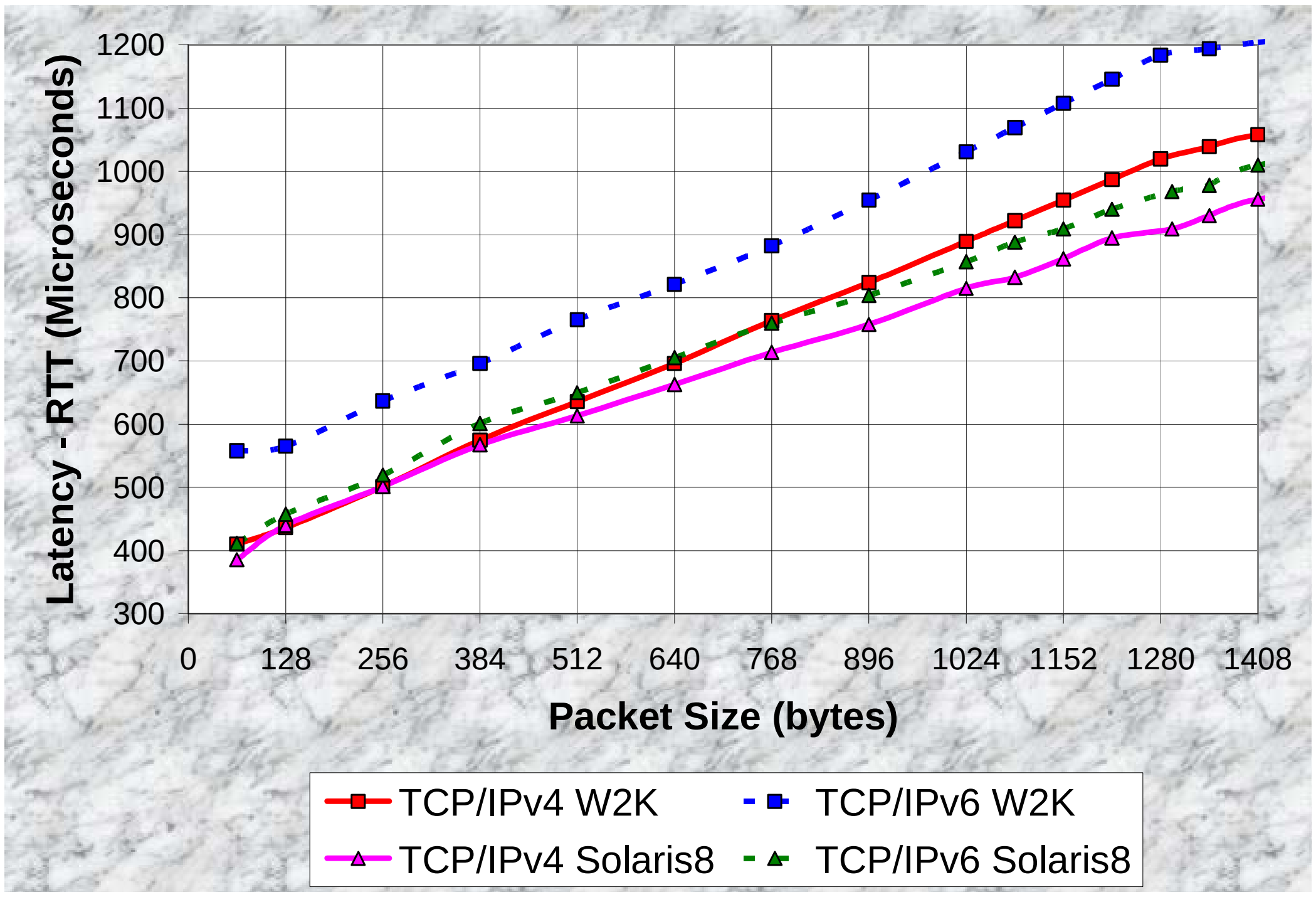

**Figure 39: Ericsson Test-bed: TCP latency results for IPv4 & IPv6 over Windows 2000 & Solaris 8.0 with packet size ranging from 64 bytes to 1408 bytes**

## 4.3 Chapter Conclusion

In this chapter, we presented an unbiased empirical performance evaluation between IPv4 and IPv6 running two different implementations, Windows 2000 and Solaris 8.0 over 4 different test-beds. For Windows 2000, IPv4 consistently outperformed IPv6 by a much larger overhead than anticipated.

Both the P2P Test-bed and the Ericsson Test-bed had reasonable performance overhead of 36% in the worst case for smaller packets and as good as 1% in the best case for larger packets; these best and worst case overhead percentages are all inclusive between Windows and Solaris, UDP and TCP, and the throughput and latency experiments.

However, the IBM Test-bed and the IBM-Ericsson Test-bed both proved to have very bad IPv6 performance in relation to the IPv4 network protocols. Throughout the various experiments, we saw performance as bad as 250% and over 300% worse on the IPv6 network protocol in comparison to its IPv4 counterpart.

It is our belief that the very bad performance which seemed to be cause by the IBM router has nothing to do with the IPv6 protocol and is not the usual performance overhead that will be realized in future networks as they upgrade to IPv6. However something did cause the IBM router to perform very poorly, which leads us to believe that the IPv6 protocol was poorly implemented in the router, or even worst, they implemented much of the processing in software which makes it more inefficient when compared to other routers that implemented the same protocol in hardware.

The IBM router was purchased in 2000 while the Ericsson router was purchased in 2001, almost a year and half later. Since IPv6 is still in its infancy and matures on a daily basis, it might just be a matter of time until most routers will be relatively equivalent and reach performance levels that are close to their theoretical counterparts.

# CHAPTER 5

# IPV4 TO IPV6 TRANSITION MECHANISMS PERFORMANCE EVALUATION

In this chapter, we analyze various transition mechanisms for upgrading from IPv4 to IPv6. This should give a clear indication of what additional performance hits IPv6 will suffer because of a lack of deployment infrastructure.

Since chapter 5 has a similar structure as chapter 4 did, some of the common theoretical sections will not be repeated again in chapter 5. For example the explanation of the performance metrics will not be covered in this chapter and for a full explanation, chapter 4 should be referenced.

Chapter 5 deals with the performance evaluation of two transition mechanisms, namely host-host encapsulation and router-router tunneling. We present the results for the two transition mechanisms set against the results of IPv4 and IPv6 from the previous chapter. Unlike chapter 4, we evaluated the network protocols using both TCP and UDP transport protocols only under the Windows 2000 operating systems.

Since router-router tunneling requires at least two routers for the proper operation of the encapsulation mechanism, we focused our performance evaluation strictly on the IBM-Ericsson Test-bed which involved both routers between the end nodes. For more information regarding the specifics of the IBM-Ericsson test-bed's configuration, please refer to chapter 3 in which the IBM-Ericsson is discussed in detail.

### 5.1 Performance Metrics

Our metrics of evaluation were: throughput, latency, CPU utilization, socket creation time, TCP connection time, the number of TCP connections per second, and the performance of a video application designed by our lab. All the performance measurement software was written in C++.

The majority of the tests were done for a period of about 60 seconds, which netted about 50,000 packets to about 1,000,000 packets, depending on the size of the packets sent and what tests were being completed. The tests dealing with testing the throughput of the UDP transport protocol were limited to 1472 byte datagrams because of a fragmentation bug in the IPv6 protocol stack. All other tests were done using various packet sizes ranging from 64 bytes to 64 Kbytes. Each test was repeated three times in order to rule out any inconsistencies. On occasion when the three different tests were not consistent enough to have a solid conclusion, the experiments were performed several more times until there was enough data to conclude our findings.

For an in-depth analysis of the various performance metrics, please refer to section 4.1 where all 6 performance metrics (throughput, latency, CPU utilization, socket creation time, TCP connection time, the number of TCP connections per second, and the performance of a video application designed by our lab) is described in much detail.

## 5.2 Performance Results

For all figures depicting performance results such as throughput, latency, and CPU utilization, we will use the following consistent conventions as explained in the legends of each figure. The native IPv4/IPv6 network protocol performance results remain consistent with Chapter 4's conventions of being represented by a solid line and dotted line respectively. The native IPv4 and IPv6 protocol is denoted by a square (■), while the transition mechanisms are denoted by a triangle (▲). The host-host encapsulation is denoted by IPv4(IPv6) and the router-router tunneling is denoted by IPv6 Tunneling within the legends of the figures. The x-axis is the packet size in the corresponding experiment, while the y-axis represents the measured metric. For each test we have two figures: one represents the large global view with packet sizes ranging from 64 bytes to 64 Kbytes, while the other represents only a small part of the bigger graph displaying the results for packet size between 64 bytes and 1408 bytes.

### 5.2.1 Throughput

As Figure 40 indicates, it can be clearly seen that each layer of complexity adds additional overhead over the entire packet size spectrum.

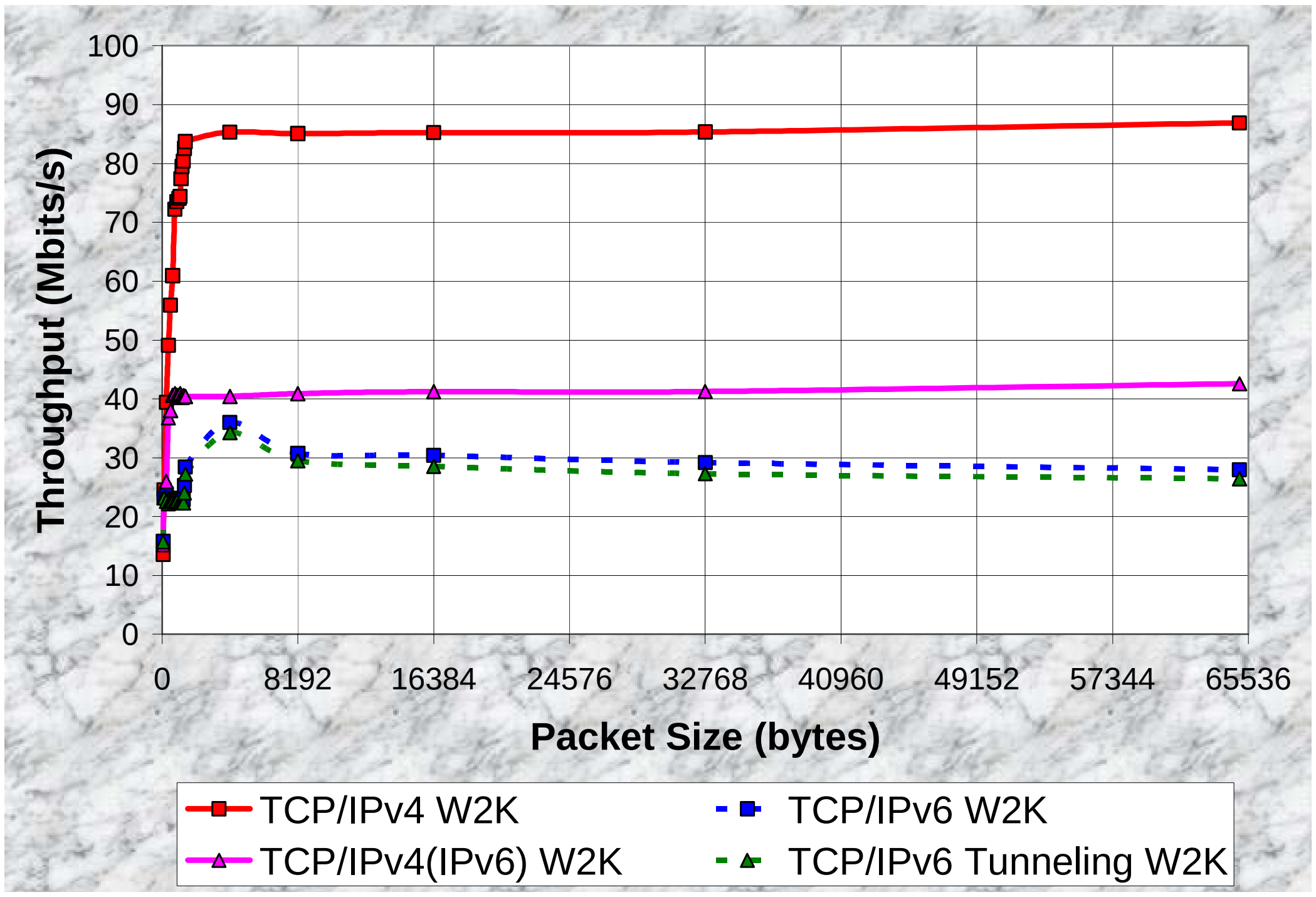

**Figure 40: IBM-Ericsson Test-bed: TCP throughput results for IPv4, IPv6, and IPv4-IPv6 transition mechanisms under Windows 2000 with packet size ranging from 64 bytes to 64 Kbytes**

There are two interesting facts that can be concluded from the figure above. First of all, the router-to-router tunneling seems to have very little overhead on top of the IPv6 protocol stack. Specifically, it incurs about 1% to 7% overhead which is very acceptable. On the other hand, the host-to-host encapsulation seems to perform pretty badly when compared to IPv4. It incurs as much as 110% overhead on top of IPv4 for larger packet sizes. The throughput achieved with host-to-host encapsulation is higher than that achieved with the native IPv6 network protocol. This was to be expected since the native IPv6 network protocol has its poor performance due to the IBM router.

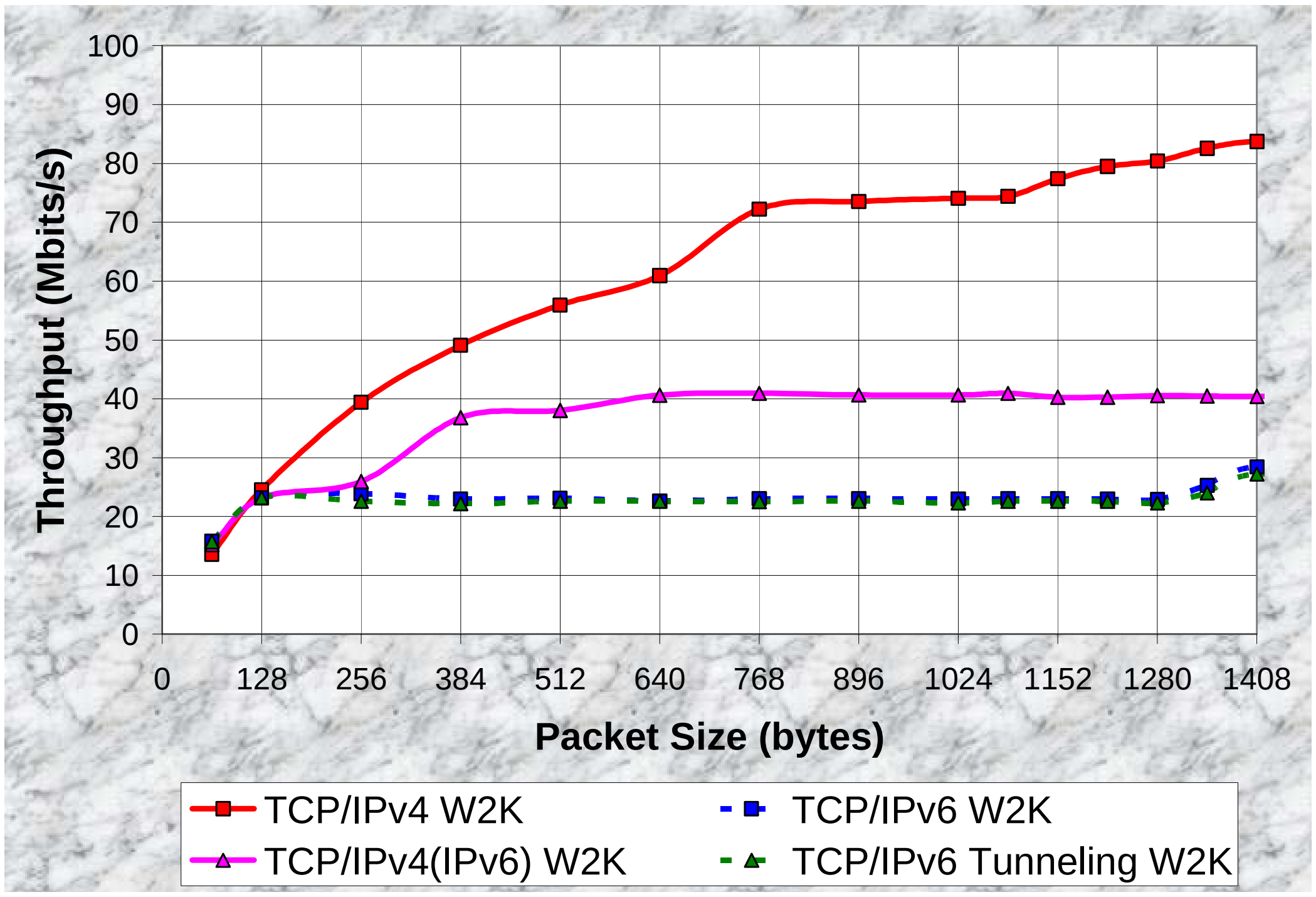

**Figure 41: IBM-Ericsson Test-bed: TCP throughput results for IPv4, IPv6, and IPv4-IPv6 transition mechanisms under Windows 2000 with packet size ranging from 64 bytes to 1408 bytes**

The host-to-host encapsulation should enjoy the higher throughput of the IPv4 network protocol, but its problem is the extra time the sending and receiving host takes to encapsulate and de-capsulate the IPv6 packet. It should be obvious that most of the overhead in the host-to-host encapsulation is due to the host's inability to process the encapsulation mechanism fast enough. This is most likely due to a software implementation in the protocol stack.

As Figure 42 indicates, the only experiment that was performed above the Ethernet MTU size was the UDP/IPv4 experiment. As far as the host is concerned, the other three experiments (IPv6, IPv4(IPv6), and IPv6 tunneling) are all part of the IPv6 stack, and therefore all have the same limitation.

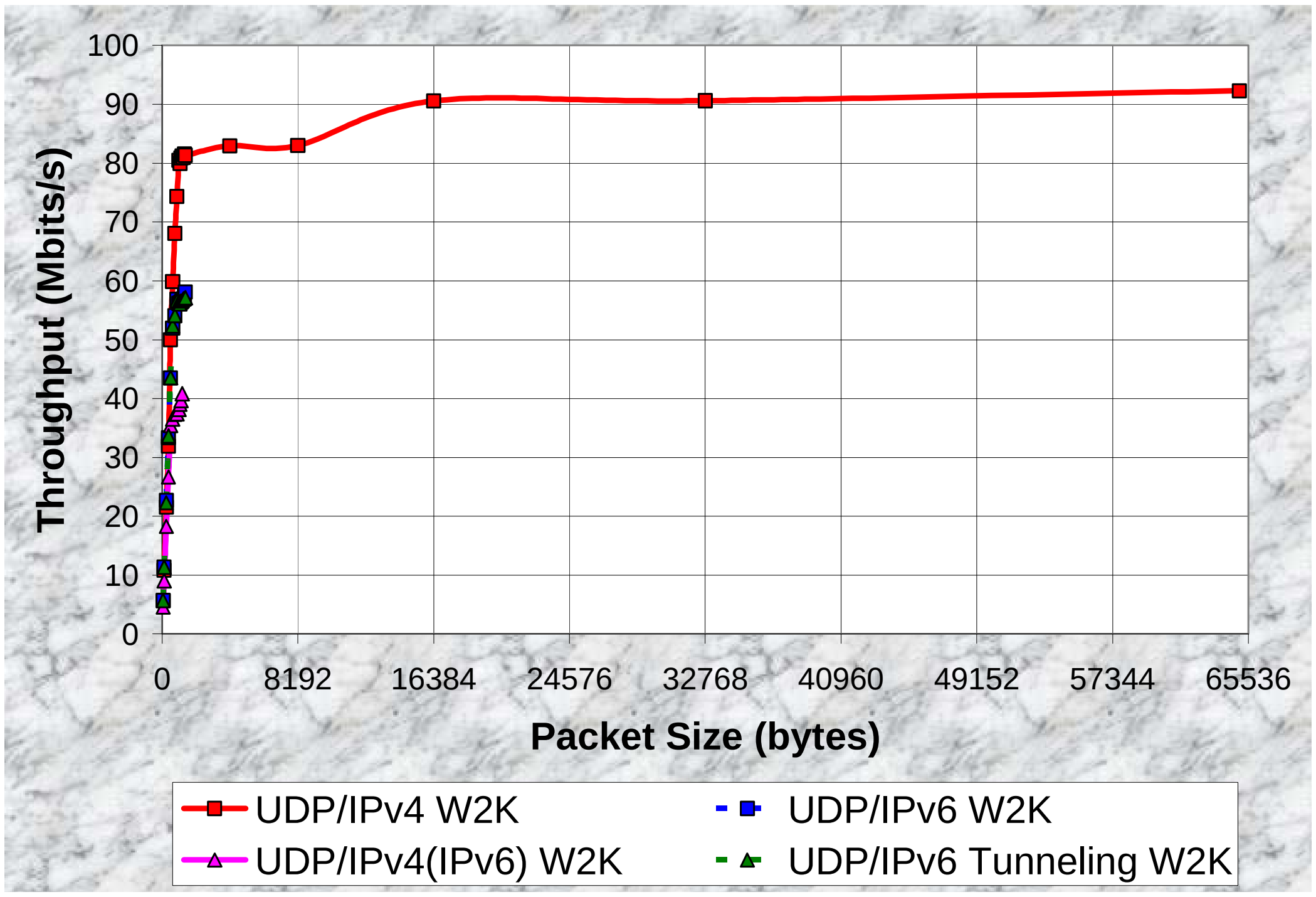

**Figure 42: IBM-Ericsson Test-bed: UDP throughput results for IPv4, IPv6, and IPv4-IPv6 transition mechanisms under Windows 2000 with packet size ranging from 64 bytes to 64 Kbytes**

As it can be depicted in Figure 43, the IPv4(IPv6) experiment experienced the bug in the IPv6 network protocol stack at a packet size of 1280 bytes. This is most likely due to the encapsulation of the IPv6 packet inside an IPv4 packet in which an extra level of header information was wasting valuable space inside the data payload of the IPv4 packet.

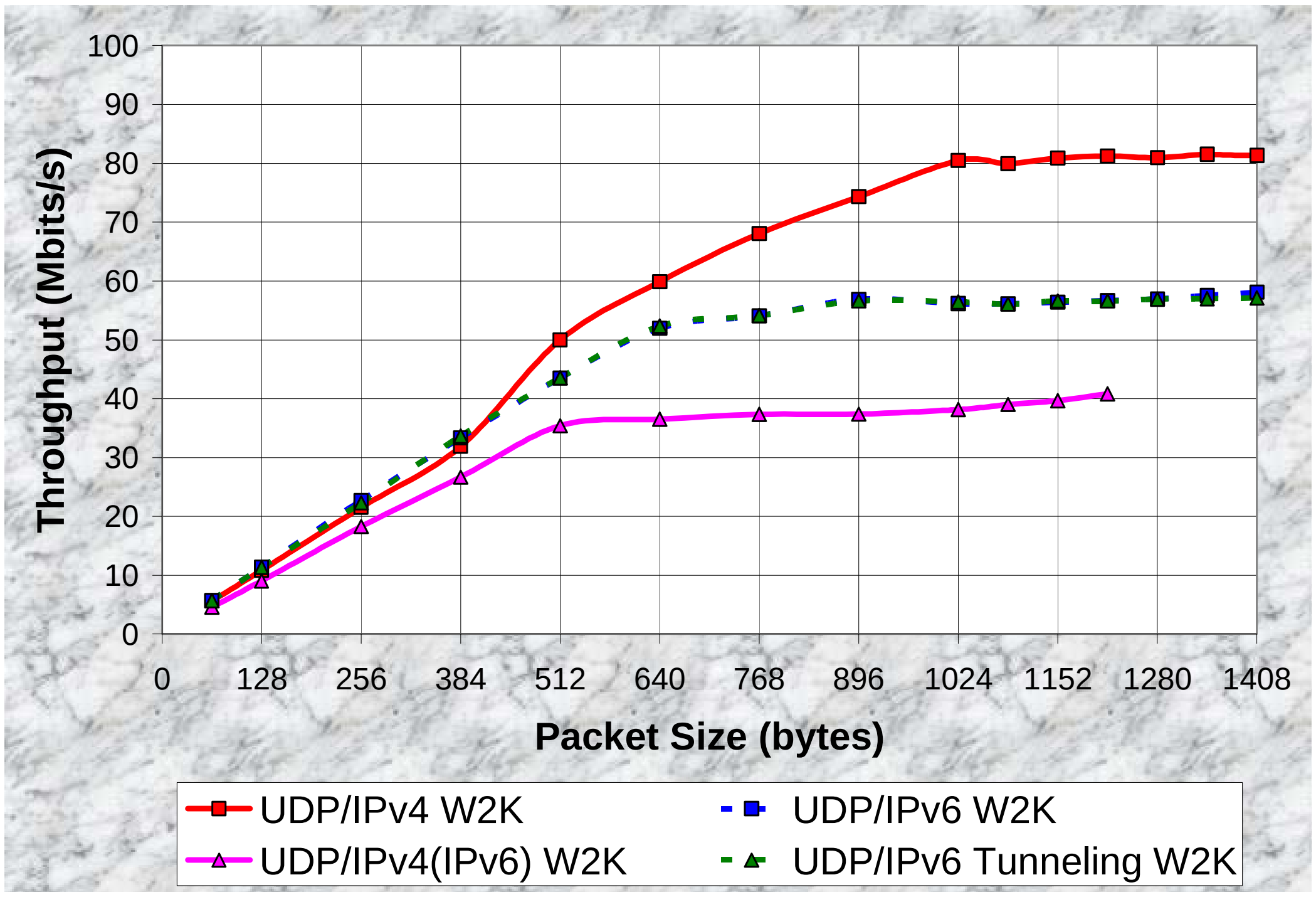

**Figure 43: IBM-Ericsson Test-bed: UDP throughput results for IPv4, IPv6, and IPv4-IPv6 transition mechanisms under Windows 2000 with packet size ranging from 64 bytes to 1408 bytes**

As for the CPU utilization depicted in Figure 44 and Figure 45, it is clear that the host-to-host encapsulation incurred the most CPU overhead. This was expected since the end host had to encapsulate and de-capsulate every single packet that was transmitted or received. We are positive that the router experiences similar behaviors in the CPU load for the router-to-router tunneling, however we have no way to verify our assumptions. Also, if it seems odd that the IPv4 protocol, the IPv6 protocol, and the router-to-router tunneling all incurred relatively similar CPU utilization, it really is not. The throughput rates were different (much higher for IPv4 than IPv6) among the various experiments and therefore it is justifiable that IPv4 has a CPU utilization as high as that of IPv6.

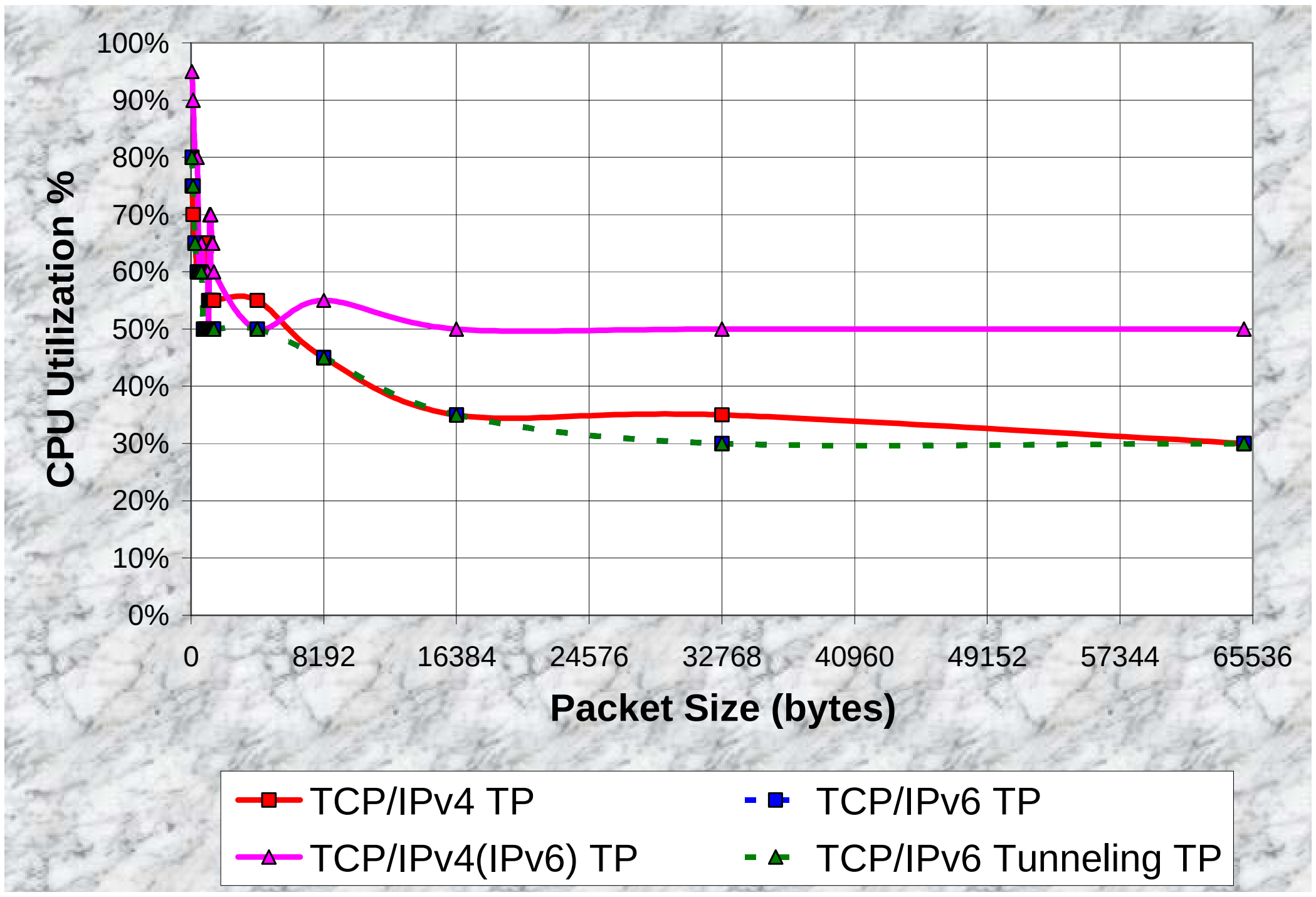

**Figure 44: IBM-Ericsson Test-bed: CPU utilization for TCP throughput results for IPv4, IPv6, and IPv4-IPv6 transition mechanisms under Windows 2000 with packet size ranging from 64 bytes to 64 Kbytes**

Figure 45 obviously look incomplete because we could not perform the experiments that were greater than the Ethernet MTU size for the IPv6 protocol stack. Once again, the UDP shows consistently that it requires less CPU utilization in order to achieve similar throughput performance.

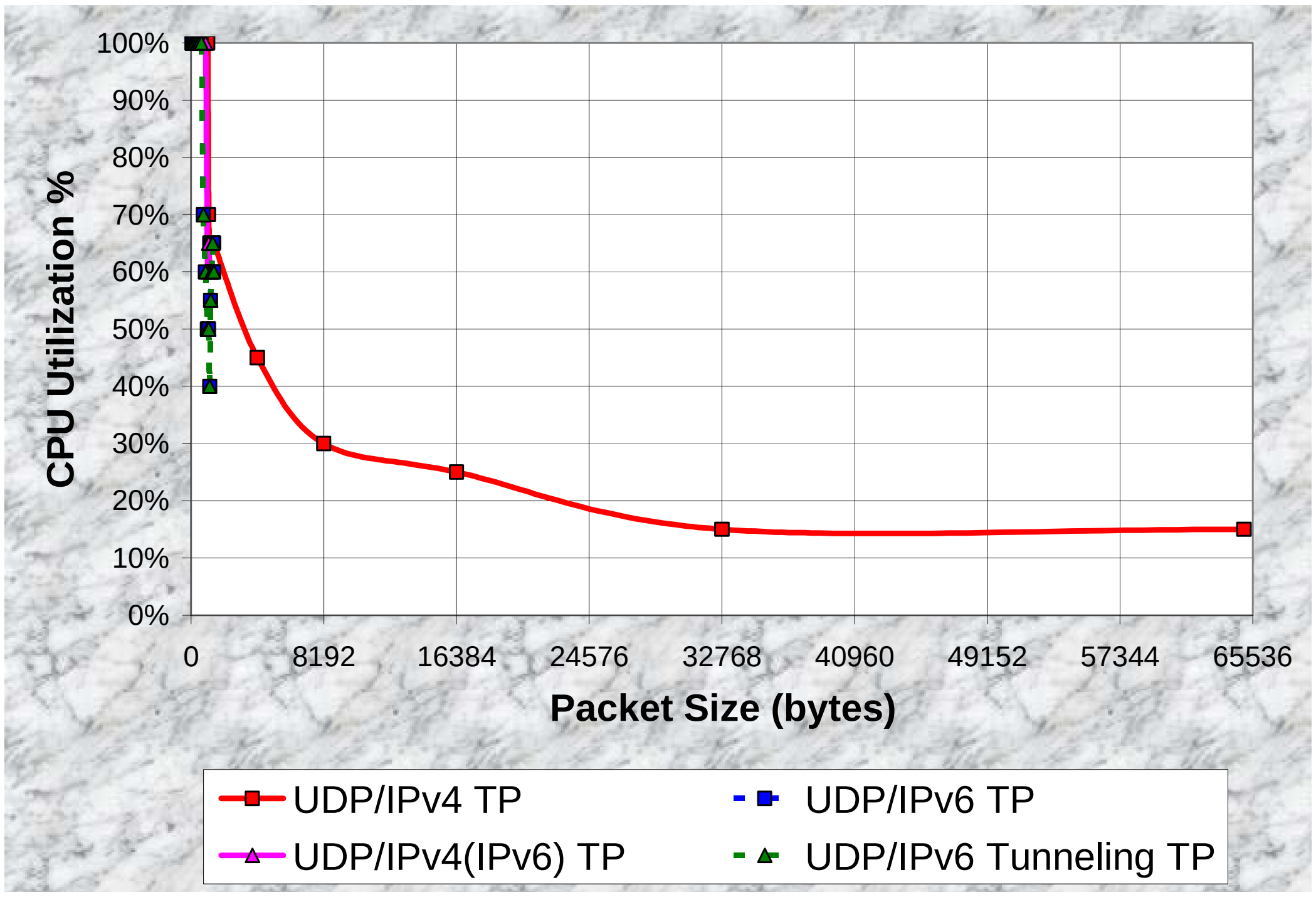

**Figure 45: IBM-Ericsson Test-bed: CPU utilization for UDP throughput results for IPv4, IPv6, and IPv4-IPv6 transition mechanisms under Windows 2000 with packet size ranging from 64 bytes to 64 Kbytes**

### 5.2.2 Latency

As Figure 46 indicates, the RTT for the host-to-host encapsulation was as 30 ms for 64 Kbyte packets compared to about 15 ms for the IPv4 network protocol. Both IPv6 and the router-to-router tunneling experienced similar trends and had RTTs as high as 42 ms for 64 Kbyte packets. The overheads are very clear to be very high for either transition mechanisms; however, they might be cause in part because of the poor implementation of the IPv6 network protocol in the IBM router.

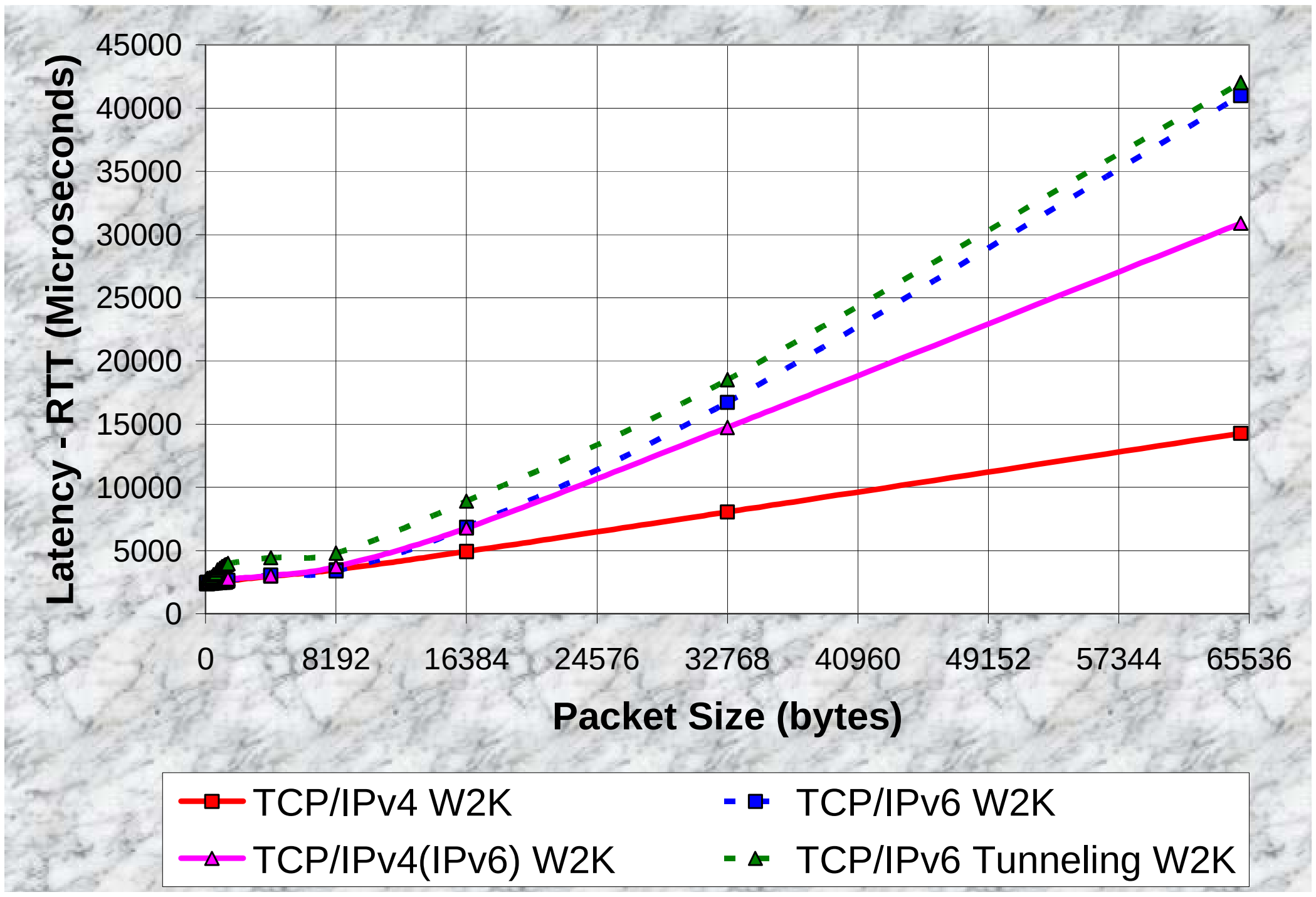

**Figure 46: IBM-Ericsson Test-bed: TCP latency results for IPv4, IPv6, and IPv4-IPv6 transition mechanisms under Windows 2000 with packet size ranging from 64 bytes to 64 Kbytes**

As Figure 47 indicates, IPv6 tunneling incurs a pretty heavy performance overhead on top of all the other experiments. Our best assumption is that when the packet sizes are too small, the routers still take a minimum amount of time in order to process each encapsulation and de-capuslation, and therefore we see the very large increase in RTT's for small packet sizes. Obviously, the difference is amortized as the packet size get larger and eventually the IPv6 tunneling curve follws the native IPv6 curve very closely.

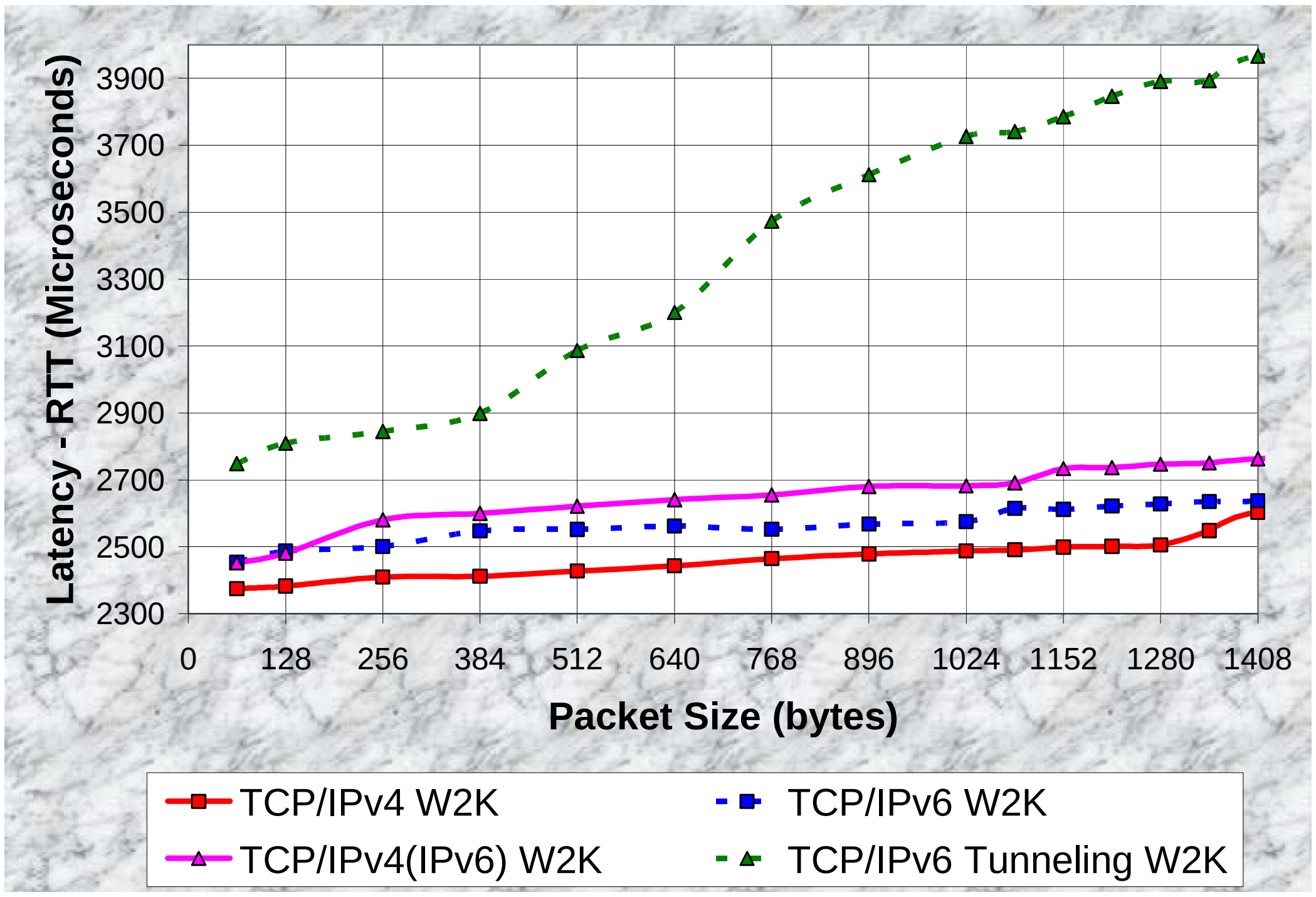

**Figure 47: IBM-Ericsson Test-bed: TCP latency results for IPv4, IPv6, and IPv4-IPv6 transition mechanisms under Windows 2000 with packet size ranging from 64 bytes to 1408 bytes**

As for the UDP latency experiments depicted in Figure 48 and Figure 49, nothing very interesting is found, except that all the experiments perform almost identically, except for some minimal expected overhead incurred by the transition mechanisms. The host-to-host encapsulation performed nearly identical to the native IPv6 network protocol, and the router-to-router tunneling had a small 5% tp 3% overhead above native IPv6 for packet sizes ranging from smaller ones to larger ones.

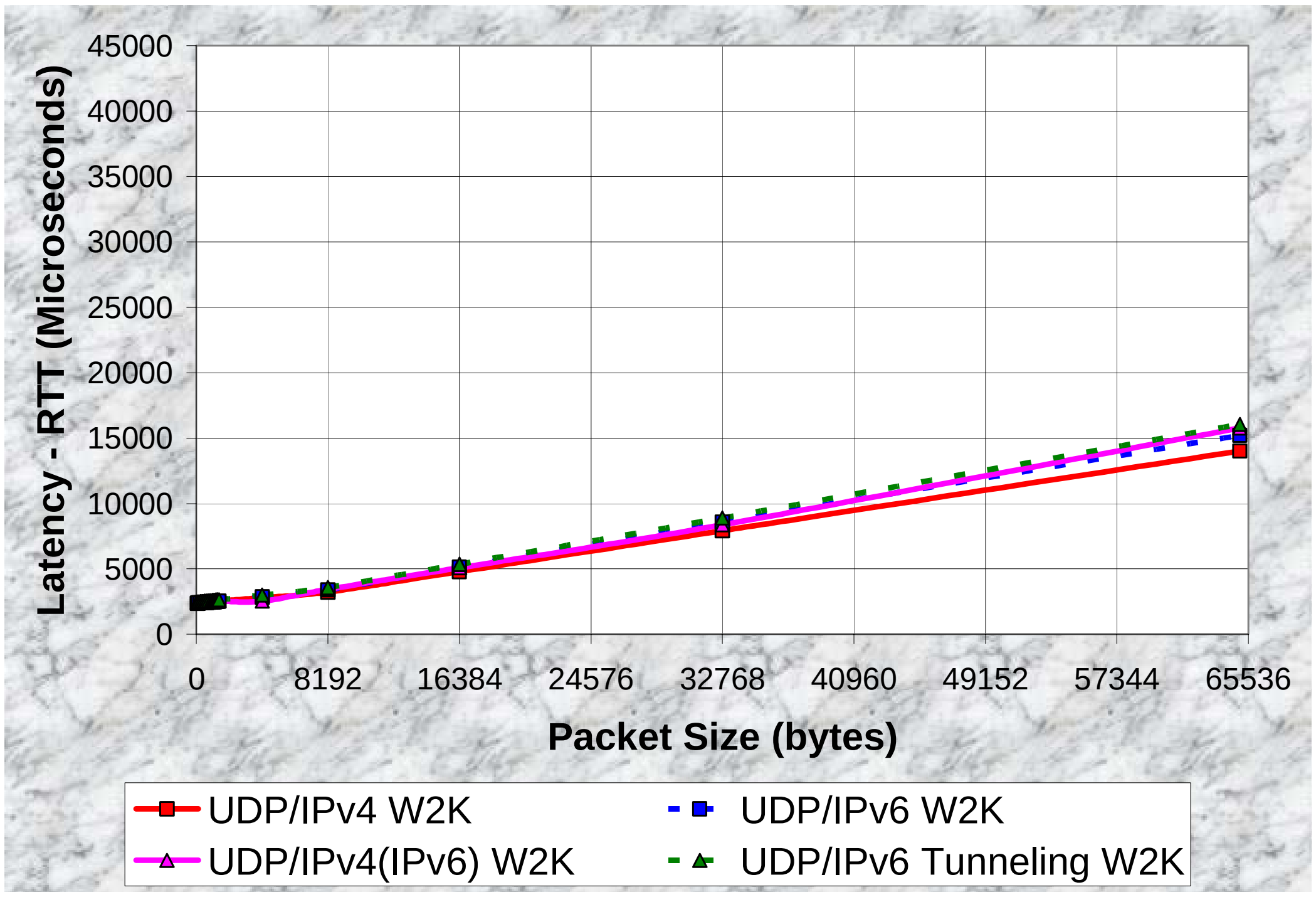

**Figure 48: IBM-Ericsson Test-bed: UDP latency results for IPv4, IPv6, and IPv4-IPv6 transition mechanisms under Windows 2000 with packet size ranging from 64 bytes to 64 Kbytes**

Figure 49 clearly depicts the above statements even more since the scale represented in the figure is in a way that it is much easier to see and understand the performance overhead of 5% to 3% overhead that IPv6 tunneling incurs above native IPv6. It can also be seen how the other three experiments all performed relatively the same. For the UDP transport protocol, all the transition mechanisms seem very promising since they incur relatively very little overhead. However, the TCP transport protocol, as we showed earlier, needs to improve significantly before IPv6 will reach a level of performance similar to that of IPv4 network protocol.

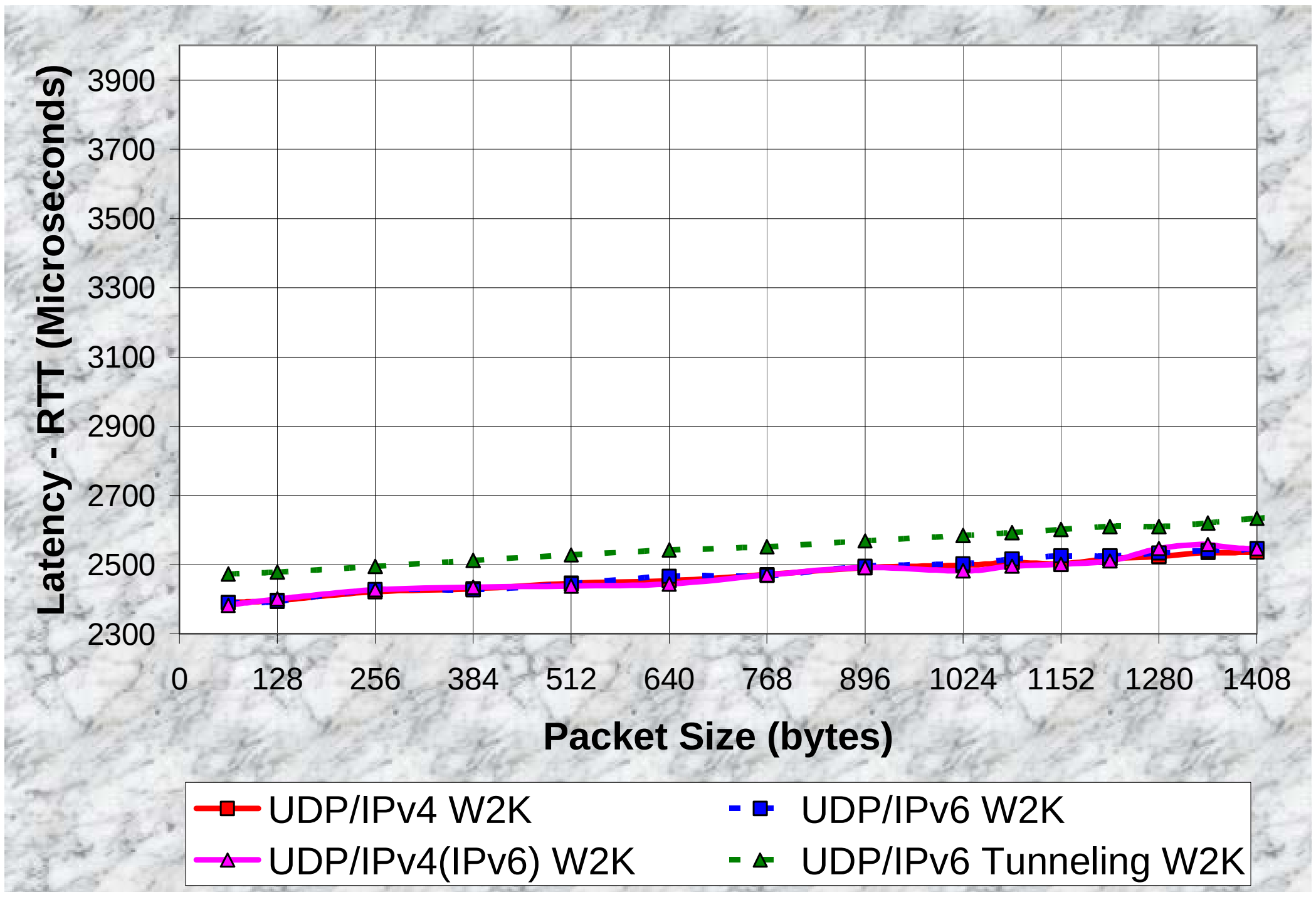

**Figure 49: IBM-Ericsson Test-bed: UDP latency results for IPv4, IPv6, and IPv4-IPv6 transition mechanisms under Windows 2000 with packet size ranging from 64 bytes to 1408 bytes**

The last experiment we had was the CPU utilization for both TCP and UDP as can be seen in Figure 50 and Figure 51 respectively. It is rather clear that host-to-host encapsulation required the most CPU utilization due to the fact that the host had to do much more work for every packet sent and received. It virtually had to encapsulate and de-capsulate every packet before transmitting it or before understanding the data payload of a received packet.

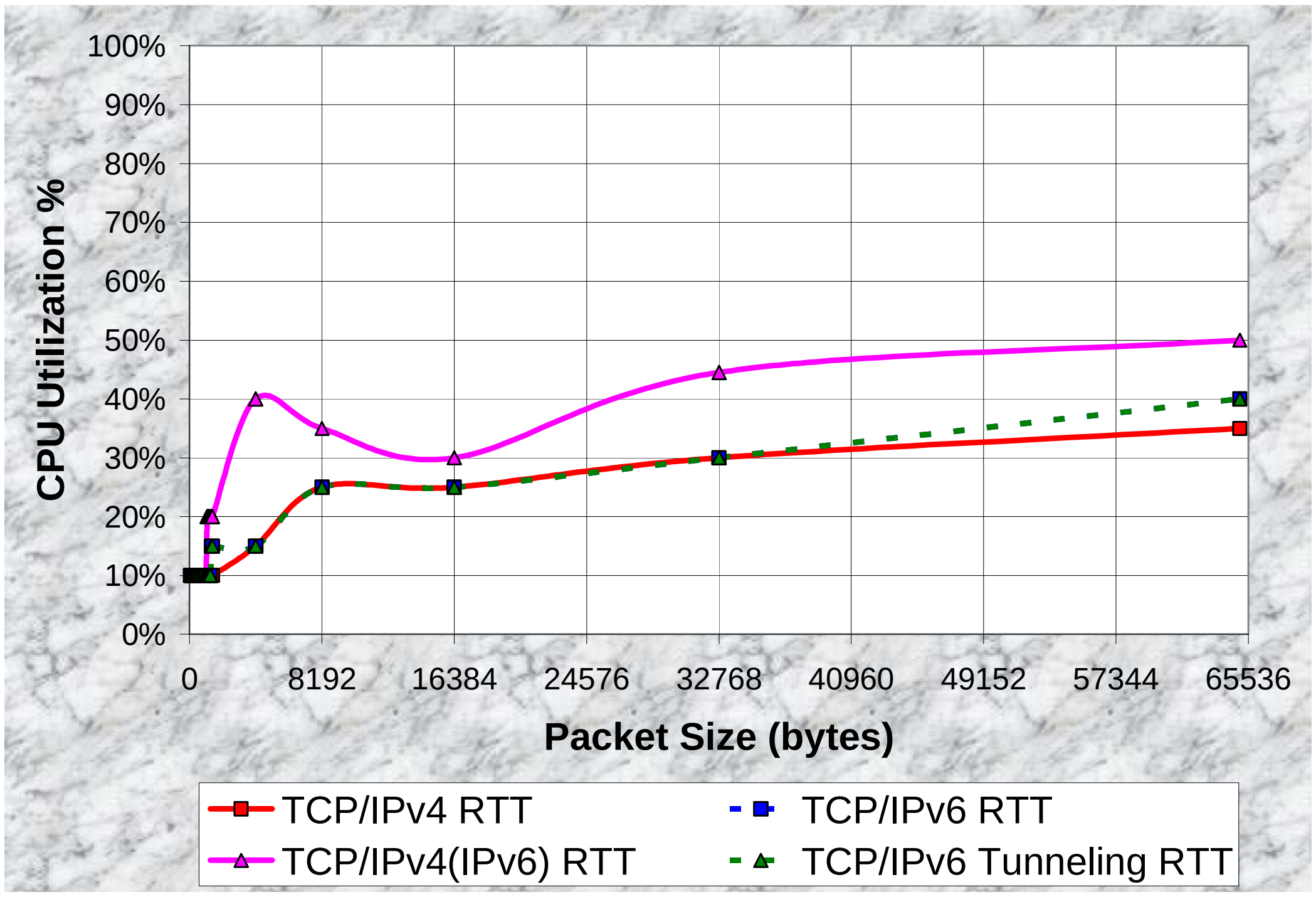

**Figure 50: IBM-Ericsson Test-bed: CPU utilization for TCP latency results for IPv4, IPv6, and IPv4-IPv6 transition mechanisms under Windows 2000 with packet size ranging from 64 bytes to 64 Kbytes**

A similar trend is experienced in Figure 51 in which the IPv4(IPv6) (host-to-host encapsulation) incurs the most CPU utilization overhead for the UDP transport protocol. The other three experiments are all relatively using the same CPU utilization. Our findings make much sense and therefore emphasize the need to simplify the network protocols as much as possible without loosing any of its functionality.

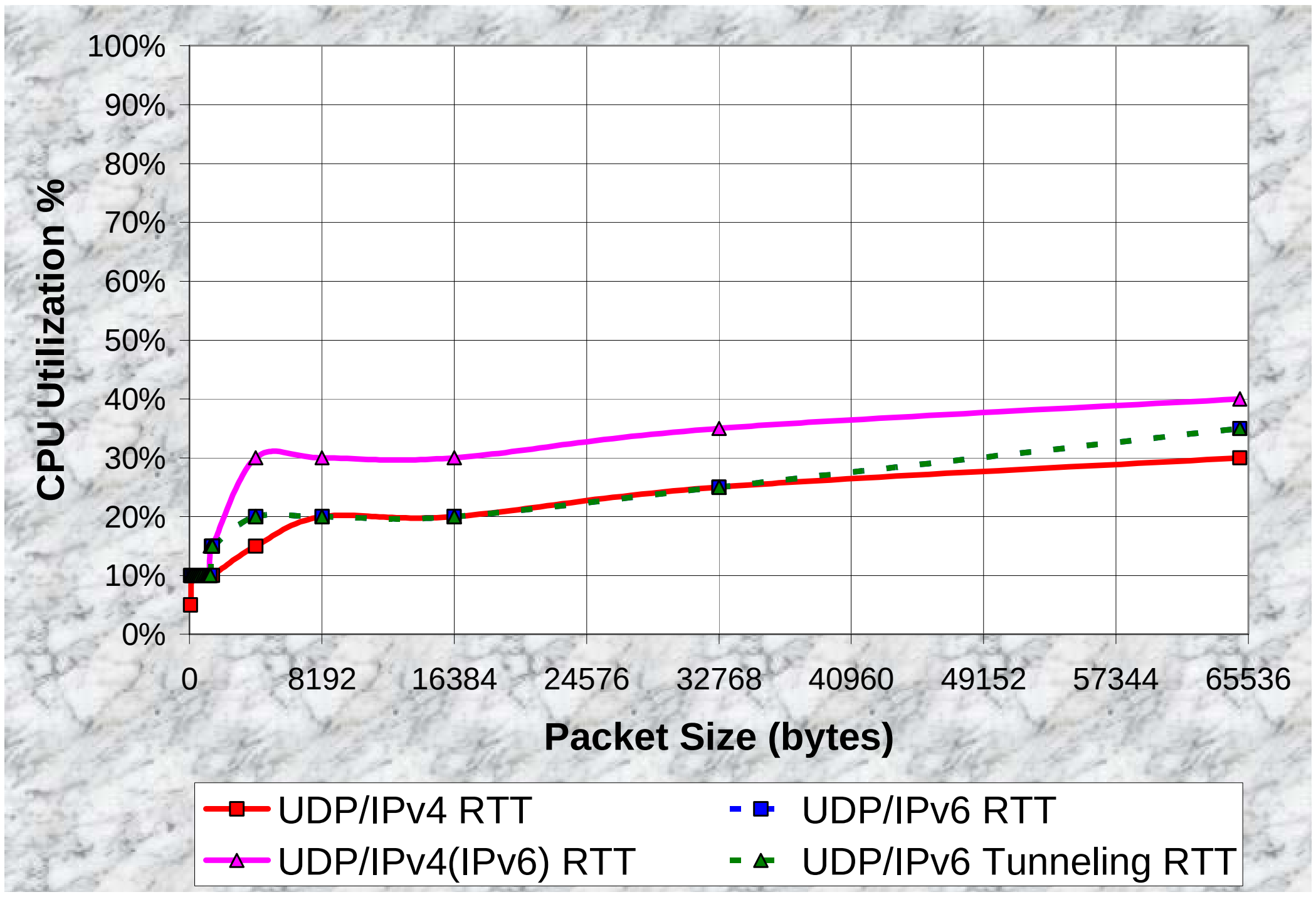

**Figure 51: IBM-Ericsson Test-bed: CPU utilization for UDP latency results for IPv4, IPv6, and IPv4-IPv6 transition mechanisms under Windows 2000 with packet size ranging from 64 bytes to 64 Kbytes**

### 5.2.3 Socket Creation Time and TCP Connection Time

The socket creation time remained relatively the same for both IPv4 and IPv6 under both TCP and UDP as it was in chapter 4 in our previous evaluation. However, the connection time varies from each experiment we ran to another. Without any surprise, the best performer was native IPv4 with the next best candidate being the host-to-host encapsulation (IPv4(IPv6)). The next best was the native IPv6 network protocol and the worst was the router-to-router tunneling. The good part is that the biggest difference between two experiments is 25% while the smallest difference is 6%. Obviously, the UDP connection time was irrelevant since UDP is a connection-less protocol. Since UDP does not have a connection mechanism such as TCP, we cannot measure the connection time.

| OS | IP Version | Transport Protocol | Sock. Cr. Time (μs) | Con. Time (μs) |
|---|---|---|---|---|
| **Windows 2000** | IPv4 | TCP | 6128.74 | 2608.93 |
| **Windows 2000** | IPv6 | TCP | 8006.51 | 2959.13 |
| **Windows 2000** | IPv4(IPv6) | TCP | 8006.51 | 2784.42 |
| **Windows 2000** | IPv6 Tunneling | TCP | 8006.51 | 3261.58 |
| **Windows 2000** | IPv4 | UDP | 6002.74 | N/A |
| **Windows 2000** | IPv6 | UDP | 6812.13 | N/A |
| **Windows 2000** | IPv4(IPv6) | UDP | 6812.13 | N/A |
| **Windows 2000** | IPv6 Tunneling | UDP | 6812.13 | N/A |

**Table 10: IBM-Ericsson Test-bed: TCP and UDP socket creation time and TCP connection time in microseconds for both IPv4, IPv6, and the transition mechanisms running Windows 2000**

### 5.2.4 Number of TCP Connections per Second

The results from Table 11 clearly shows that the order of the contestants is really the same as it was in the previous subsection with the connection times for the TCP transport protocol. This should be no surprise since these experiments heavily rely on the time it takes to setup a socket and the time it takes to perform a connect operation. The good part of the results depicted below is that the IPv6 tunneling does not seem to incur a much higher overhead on top of the native IPv6 network protocol.

| OS | IP Version | Number of Connections |
|---|---|---|
| **Windows 2000** | IPv4 | 94 |
| **Windows 2000** | IPv6 | 79 |
| **Windows 2000** | IPv4(IPv6) | 80 |
| **Windows 2000** | IPv6 Tunneling | 76 |

**Table 11: IBM-Ericsson Test-bed: The number of TCP connections per second over IPv4, IPv6, and the transition mechanisms running Windows 2000**

### 5.2.5 Video Client/Server Application

The results of this experiment are consistent with those obtained in earlier experiments, namely the TCP throughput experiments. With IPv4, the program netted about 66 Mbit/s and about 8.9 frames per second. Under IPv6, the transfer rates dropped to a mere 26 Mbit/s and about 3.5 frames per second; the IPv6 tunneling mechanism yielded very similar performance. Finally, the host-to-host encapsulation netted better results than the native IPv6 network protocol, however that was expected since once the packet is converted into an IPv4 packet, it enjoys the higher throughput and lower latency of the IPv4 network protocol. However, the host-to-host encapsulation does incur a higher CPU utilization overhead which is also consistent with our other experiments.

| Video Client/Server Application under Windows 2000 | | | |
|---|---|---|---|
| IP Version | Frame Rates (fps) | Transfer Rates (Mbit/s) | Client CPU Utilization |
| IPv4 | 8.9 | 66.21 | 25% |
| IPv6 | 3.5 | 26.43 | 30% |
| IPv4(IPv6) | 5.1 | 38.16 | 40% |
| IPv6 Tunneling | 3.4 | 25.23 | 30% |

**Table 12: IBM-Ericsson Test-bed: Frame rates and transfer rates for the video client/server application for IPv4, IPv6 and the transition mechanisms running Windows 2000**

Notice that all the results are somewhat lower than the TCP throughput experiments would suggest; the majority of the slower throughput results can be attributed to processing overhead for the client to render the high quality uncompressed video.

## 5.3 Chapter Conclusion

In chapter 5, we presented an extension to chapter 4 in which we performed an unbiased empirical performance evaluation between IPv4, IPv6, and host-to-host encapsulation (IPv4(IPv6)) and router-to-router tunneling (IPv6 tunneling) running over Windows 2000. For every level of complexity (IPv4, IPv6, tunneling), the performance overhead was always increasing. As IPv6 is still maturing, perhaps it is just a matter of time until IPv6’s actual performance will finally reflect its theoretical counterpart.

# CHAPTER 6

# CONCLUSION & FUTURE WORK

Trough our work, we presented an unbiased empirical performance evaluation between IPv4 and IPv6 running two different implementations, Windows 2000 and Solaris 8.0. Furthermore, we also analyzed various transition mechanisms in order to see the extent of the performance degradation during the process of upgrading from IPv4 to IPv6. We came to the conclusion that the IPv6 protocol stack needs much improvement in order to reduce the overhead it presently incurs over IPv4.

Since IPv6 is still maturing, perhaps it is just a matter of time until its performance will finally reflect its theoretical counterpart. Perhaps it was just an implementation issue with the IBM router, but nevertheless, it states that standardization of the hardware and the IPv6 protocol stack is far from being finalized. In an ideal case in which IPv6 has had enough time to prove itself, there should not be the kind of discrepancy between hardware of different vendors as we found between the IBM router and the Ericsson router.

We must admit that the toughest part of our work was in configuring the routers. The IBM router was not that bad once one had a little bit of time to get accustomed to it. On the other hand, the Ericsson router has an interface designed by someone who clearly never iontended to configure the router himself. It is very cumbersome and has many bugs with poor documentation and user feedback. When we were configuring the Ericsson router, we usually managed to lock up the router once every couple of hours due to invalid inputs. Since we had such a hard time in configuring the Ericsson router, we created Appendix C that clearly describes the commands issued to the router to configure it to the state in which the router was ready to perform the kind of experiments we executed in this work.

In the near future, we plan on using RSVP as a means to make reservations in the system and see how IPv6 performs again versus IPv4. IPv6 also supports prioritizing packets, which might be an easy way to offer a lighter version of QoS without specifying any requirements. Specifically, we can utilize the flow label field in the IPv6 header in order to specify specific requirements in how a particular class of packets are to be handled.

The real value of our work lies in the potential of IPv6 and not necessarily in the performance overhead that we showed to be rather high. According to our evaluation, IPv6 has a performance deficit when utilizing traditional data streams, but as multimedia content is becoming more abundant in the Internet, only an in-depth evaluation of new emerging applications will net the real performance gain of IPv6 using various QoS features not previously supported by IPv4.

# APENDIX A: GLOSSARY

1. **QoS** – Quality of Service; a networking term that specifies a guaranteed throughput level.
2. **RSVP** – Resource reSerVation Protocol; a new Internet protocol being developed to enable the Internet to support a specified QoS. Using RSVP, an application will be able to reserve resources along a route from source to destination. RSVP-enabled routers will then schedule and prioritize packets to fulfill the QoS. RSVP is a chief component of a new type of Internet being developed, known broadly as an integrated services Internet. The general idea is to enhance the Internet to support transmission of real-time data.
3. **IPv4** – Internet Protocol version 4; current version of IP, which was finally revised in 1981; it has a 32 bit address looking like 255.255.255.255, and it supports up to 4,294,967,296 addresses.
4. **IPv6** – Internet Protocol version 6; IPv6 is designed as an evolutionary upgrade to the Internet Protocol and will, in fact, coexist with the older IPv4 for some time. IPv6 is designed to allow the Internet to grow steadily, both in terms of the number of hosts connected and the total amount of data traffic transmitted; it has a 128 bit address represented by the hexadecimal notation separated by colons

1234:5678:90AB:CDEF:FFFF:FFFF:FFFF:FFFF, and will support up to 340,282,366,920938,463,463,374,607,431,768,211,456 ($3.4x10^{38}$) unique addresses.

5. **IP** – Internet Protocol; specifies the format of packets, also called datagrams, and the addressing scheme.

6. **TCP** – Transmission Control Protocol; TCP, a connection oriented protocol, enables two hosts to establish a connection and exchange streams of data. TCP guarantees delivery of data and also guarantees that packets will be delivered in the same order in which they were sent.

7. **UDP** – User Datagram Protocol, a connectionless protocol that, like TCP, runs on top of IP networks. Unlike TCP/IP, UDP/IP provides very few error recovery services, offering instead a direct way to send and receive datagrams over an IP network. It's used primarily for broadcasting messages over a network.

8. **Connectionless** – Refers to network protocols in which a host can send a message without establishing a connection with the recipient. That is, the host simply puts the message onto the network with the destination address and hopes that it arrives.

9. **Connection-oriented** – require a channel to be established between the sender and receiver before any messages are transmitted.

10. **ATM** – Asynchronous Transfer Mode; a network technology based on transferring data in cells or packets of a fixed size. The cell used with ATM is relatively small compared to units used with older technologies. The small, constant cell size allows ATM equipment to transmit video, audio, and computer data over the same network, and assure that no single type of data dominates the line. One of the biggest advantages of ATM over competing technologies such as Frame Relay and Fast Ethernet, is that it supports QoS levels. This allows ATM providers to guarantee to their customers that end-to-end latency will not exceed a specified level.
11. **Unicast** – sending a message to a specific recipient on a network.
12. **Multicast** – sending a message to a select group on a network.
13. **Broadcasting** – sending a message to everyone connected to a network.
14. **Bandwidth** – The amount of data that can be transmitted in a fixed amount of time. For digital devices, the bandwidth is usually expressed in bits per second(bps) or bytes per second. For analog devices, the bandwidth is expressed in cycles per second, or Hertz (Hz).
15. **Throughput** – The speed with which data can be transmitted from one device to another. Data rates are often measured in megabits

(million bits) or megabytes (million bytes) per second. These are usually abbreviated as Mbps and MBps, respectively.

16. **Latency** – the amount of time it takes a packet to travel from source to destination.

19. **Synchronous** – processes where data is transmitted at regular intervals; most rigid.

20. **Asynchronous** – processes where data can be transmitted intermittently rather than in a steady stream; each party would be required to wait a specified interval before transmitting; most lenient.

21. **Isynchronous** – processes where data must be delivered within certain time constraints; in between synchronous and asynchronous for rigidity.

22. **Network** – a collection of interconnected autonomous computers.

23. **Interconnected** - must be able to share information.

24. **Autonomous** - must be able to operate independent of others.

# APPENDIX B: SOURCE CODE

In this appendix, a brief summary of the more important source code is presented. Complete source code would have been much too large, and therefore we emphasize merely the socket creation of both IPv4 and IPv6 in both Windows 2000 and Solaris 8.0. Because Windows 2000 only provides a millisecond resolution timer, we developed a microsecond resolution timer depicted in section 8.3. All the source code in this chapter will be italicized and reduced in font size compared to regular text. The comments will follow the C++ convention of "//" preceding the commented text and will appear bold.

## B.1 IPv4 and IPv6 in Windows 2000

Bellow is some sample source code which depicts how to build IPv4 and IPv6 sockets, set the transport protocol between UDP and TCP, establish a connection with a server, send and receive messages, and close the sockets; the source code is for the Windows 2000 platform using the Winsock2 API. The source code below is the client side of the client/server duo, however since all the basic components exist in the client, a server program is a trivial exercise since all the information needed is contained in the client software. The source code is commented throughout and hence should be self explanatory.

```
//client side of IPv4/IPv6 application
#include <iostream.h>
#include <winsock2.h>
#include <ws2tcpip.h>
#include <tpipv6.h>
#include <windows.h>

int main(int argc, char **argv)
{
	char* sendString;			//used for sending buffer
	char* recvString;			//used for receiving buffer
	char error_buf[100];			//used for outputing error messages
	char AddrName[NI_MAXHOST];
	char *Server = "141.217.17.27";		//ip address of server
	int Family = PF_UNSPEC;			//PF_INET for IPv4
						//PF_INET6 for IPv6
						//PF_UNSPEC for unknown network protocol
	int SocketType = SOCK_DGRAM;			//SOCK_STREAM for TCP transport protocol
							//SOCK_DGRAM for UDP transport protocol
	char *Port = "2806";				//port to connect with server
	int RetVal, AddrLen;
	int packet_size=65535;				//packet size
	int BUFFER_SIZE=60000;				//buffer size
	DWORD dwSockSize = BUFFER_SIZE;			//used for setting the buffer size

	//data structures needed for socket setup
	WSADATA wsaData;
	ADDRINFO Hints, *AddrInfo, *AI;
	SOCKET ConnSocket;
	struct sockaddr_storage Addr;

	// Ask for Winsock version 2.2.
	if ((RetVal = WSAStartup(MAKEWORD(2, 2), &wsaData)) != 0)
	{
		fprintf(stderr, "WSAStartup failed with error %d: %s\n", RetVal, DecodeError(RetVal));
		WSACleanup();
		return -1;
	}

	// By not setting the AI_PASSIVE flag in the hints to getaddrinfo, we're indicating that we
	//intend to use the resulting address(es) to connect to a service.  This means that when the
	//Server parameter is NULL, getaddrinfo will return one entry per allowed protocol family
	//containing the loopback address for that family.
	memset(&Hints, 0, sizeof(Hints));
	Hints.ai_family = Family;
	Hints.ai_socktype = SocketType;
	RetVal = getaddrinfo(Server, Port, &Hints, &AddrInfo);
	if (RetVal != 0)
	{
		fprintf(stderr, "Cannot resolve address [%s] and port [%s], error %d: %s\n", Server, Port,
			RetVal, gai_strerror(RetVal));
		WSACleanup();
		return -1;
	}

	// Try each address getaddrinfo returned, until we find one that we can successfully connect
	for (AI = AddrInfo; AI != NULL; AI = AI->ai_next)
	{
		// Open a socket with the correct address family for this address.
		ConnSocket = socket(AI->ai_family, AI->ai_socktype, AI->ai_protocol);
		 if (ConnSocket == INVALID_SOCKET)
		{
			fprintf(stderr,"Error Opening socket, error %d: %s\n", WSAGetLastError(),
				DecodeError(WSAGetLastError()));
			return -1;
```

```
        }

        // Notice that nothing in this code is specific to whether we are using UDP or TCP.
        // When connect() is called on a datagram socket, it does not actually establish the
        //connection as a stream (TCP) socket would. Instead, TCP/IP establishes the remote
        //half of the (LocalIPAddress, LocalPort, RemoteIP, RemotePort) mapping.  This
        //enables us to use send() and recv() on datagram sockets, instead of recvfrom() and
        //sendto().
        printf("Attempting to connect to: %s\n", Server ? Server : "localhost");

        if (connect(ConnSocket, AI->ai_addr, AI->ai_addrlen) != SOCKET_ERROR)
                break;

        RetVal = WSAGetLastError();
        if (getnameinfo(AI->ai_addr, AI->ai_addrlen, AddrName, sizeof(AddrName), NULL, 0,
                NI_NUMERICHOST) != 0)
        strcpy(AddrName, "<unknown>");
        fprintf(stderr, "connect() to %s failed with error %d: %s\n", AddrName, RetVal,
                DecodeError(i));
 } //end of for loop

if (AI == NULL)
{
        fprintf(stderr, "Fatal error: unable to connect to the server.\n");
        WSACleanup();
        return -1;
}

// This demonstrates how to determine to where a socket is connected.
AddrLen = sizeof(Addr);
if (getpeername(ConnSocket, (LPSOCKADDR)&Addr, &AddrLen) == SOCKET_ERROR)
{
        fprintf(stderr, "getpeername() failed with error %d: %s\n", WSAGetLastError(),
                DecodeError(WSAGetLastError()));
}
else
{
        if (getnameinfo((LPSOCKADDR)&Addr, AddrLen, AddrName, sizeof(AddrName), NULL, 0,
                NI_NUMERICHOST) != 0)
        strcpy(AddrName, "<unknown>");

        printf("Connected to %s, port %d, protocol %s, protocol family %s\n",
                AddrName, ntohs(SS_PORT(&Addr)),
                (AI->ai_socktype == SOCK_STREAM) ? "TCP" : "UDP",
                (AI->ai_family == PF_INET) ? "PF_INET" : "PF_INET6");
}

// We are done with the address info chain, so we can free it.
freeaddrinfo(AddrInfo);

// Find out what local address and port the system picked for us.
AddrLen = sizeof(Addr);
if (getsockname(ConnSocket, (LPSOCKADDR)&Addr, &AddrLen) == SOCKET_ERROR)
{
        fprintf(stderr, "getsockname() failed with error %d: %s\n",
                        WSAGetLastError(), DecodeError(WSAGetLastError()));
}
else
{
        if (getnameinfo((LPSOCKADDR)&Addr, AddrLen, AddrName, sizeof(AddrName), NULL, 0,
                NI_NUMERICHOST) != 0)
        strcpy(AddrName, "<unknown>");
        printf("Using local address %s, port %d\n", AddrName, ntohs(SS_PORT(&Addr)));
}
```

```
        //set the send buffer to the specified value
        setsockopt(ConnSocket, SOL_SOCKET, SO_SNDBUF, (const char*)&dwSockSize,
                sizeof(dwSockSize));
        //set the receive buffer to the specified value
        setsockopt(ConnSocket, SOL_SOCKET, SO_RCVBUF, (const char*)&dwSockSize,
                sizeof(dwSockSize));

        //send message contained in sendString
        RetVal = send(ConnSocket, sendString, packet_size, 0);
        if (RetVal == SOCKET_ERROR)
        {
                sprintf(error_buf, "Windows Sockets error %d: Couldn't connect socket.",
                        WSAGetLastError() );
                cout<<"ERROR: send()..."<<error_buf<<endl;
                return -1;
        }

        //receive message into recvString
        RetVal = recv(ConnSocket, recvString, packet_size, 0);
        if (RetVal == SOCKET_ERROR)
        {
                sprintf(error_buf, "Windows Sockets error %d: Couldn't connect socket.",
                        WSAGetLastError() );
                cout<<"ERROR: recv()..."<<error_buf<<endl;
                return -1;
        }

        //close socket
        printf("Server closed connection\n");
        shutdown(ConnSocket, SD_SEND);
        closesocket(ConnSocket);
        WSACleanup();
        return 0;
}
```

**Figure 52: Windows 2000 IPv4/IPv6 source code**

## B.2 IPv4 and IPv6 in Solaris 8.0

Similarly as in section B.1, this section includes some sample source code which depicts how to build IPv4 and IPv6 sockets, set the transport protocol between UDP and TCP, establish a connection with a server, send and receive messages, and close the sockets; the source code is for the Solaris 8.0 operating system. The source code below is the client side of the client/server duo, however since all the basic components exist in the client, a server program is a trivial exercise since all the information needed is contained in the client software.

```
#include <stdio.h>              /* NULL, O_RDWR */
#include <sys/types.h>          /* u_short, u_long */
#include <sys/socket.h>         /* Family, SocketType */
#include <netinet/in.h>         /* sockaddr_in */
#include <netdb.h>              /* /etc/hosts */
#include <arpa/inet.h>
#include <unistd.h>
#include <string.h>              /* memset() */
#include <sys/time.h>
#include <signal.h>
#include <stdio.h>
#include <sys/errno.h>
#include <sys/filio.h>
#include <sys/ioctl.h>
#include <netinet/tcp.h>

main (argc, argv)
char *argv[];
{
        char MESSAGE[65535];        //send buffer
        char rMESSAGE[65535];       //receive buffer
        int SERV_PORT=2806;         //server port
        int Family = AF_INET6;      //AF_INET for IPv4
                                    //AF_INET6 for IPv6
        int SocketType = SocketType;//SOCK_DGRAM for UDP
                                    //SOCK_STREAM for TCP
        int sd, sndbytes, rcvbytes, rc;
        struct sockaddr_in6 servaddr; //ipv6 address structure
        struct hostent *hp, *getipnodebyname(); /* /etc/ipnode lookup */
        int cnt;
        int rcvBytes=0;
        int sndBytes=0;
        int datasize = 65535;
        int IPver=4;                //4 for IPv4
                                    //6 for IPv6
        struct sockaddr_in servaddr4; //ipv4 address structure
        int dwSockSize=datasize;

        //IPv4 network protocol stack
        if(IPver==4)
        {
                hp = gethostbyname("141.217.17.27");

                /* create socket */
                if( ( sd = socket(Family, SocketType, 0) ) < 0 )
                {
                        perror("socket failed");
                        exit(1);
                }

                /* fill in socket details */
                bzero(&servaddr4, sizeof(servaddr4));
                servaddr4.sin_family = Family;
                servaddr4.sin_port = htons(SERV_PORT);
                memcpy( &servaddr4.sin_addr.s_addr, hp->h_addr, hp->h_length);

                //allocate send and receive buffers
                rc = setsockopt(sd, SOL_SOCKET, SO_SNDBUF, (const char*)&dwSockSize,
                        sizeof(dwSockSize));
                rc = setsockopt(sd, SOL_SOCKET, SO_RCVBUF, (const char*)&dwSockSize,
                        sizeof(dwSockSize));

                /* connect to server */
                if( connect( sd, (struct sockaddr *) &servaddr4, sizeof(servaddr4)) < 0 )
                {
```

```
                perror("connect failed");
                exit(0);
        }
}
//IPv4 network protocol stack
else if(IPver==6)
{
        hp = getipnodebyname( "8:8:8:8:8:8:8:2", Family, AI_DEFAULT );

        /* create socket */
        if( ( sd = socket(Family, SocketType, 0) ) < 0 )
        {
                perror("socket failed");
                exit(1);
        }

        /* fill in socket details */
        bzero(&servaddr, sizeof(servaddr));
        servaddr.sin6_family = Family;
        servaddr.sin6_port = htons(SERV_PORT);
        memcpy( &servaddr.sin6_addr, hp->h_addr, hp->h_length);

        //allocate send and receive buffers
        rc = setsockopt(sd, SOL_SOCKET, SO_SNDBUF, (const char*)&dwSockSize,
                sizeof(dwSockSize));
        rc = setsockopt(sd, SOL_SOCKET, SO_RCVBUF, (const char*)&dwSockSize,
                sizeof(dwSockSize));

        /* connect to server */
        if( connect( sd, (struct sockaddr *) &servaddr, sizeof(servaddr)) < 0 )
        {
                perror("connect failed");
                exit(0);
        }
}
else
{
        printf("\nWRONG IP version...\n");
        exit(1);
}

//send contents in MESSAGE
sndBytes = send(sd, MESSAGE, datasize, 0);
if(sndBytes<0)
{
        perror("send failed");
        close(sd);
        exit(1);
}

//receive data into rMESSAGE
rcvBytes = recv(sd, &rMESSAGE, datasize, 0);
if(rcvBytes<0)
{
        perror("recv failed");
        close(sd);
        exit(1);
}

//close socket sd
close(sd);
exit(0);
}
```

**Figure 53: Solaris 8.0 IPv4/IPv6 source code**

## B.3 Microsecond Timer Granularity in Windows 2000

When it comes to retrieving time and creating timers within the Microsoft Windows platforms, the best we can do with predefined functions is millisecond (ms) granularity. In practice, this ms granularity is even worse as it proves to actually be accurate to 10 ms intervals. For most applications, this is more than accurate enough; however, for high demanding applications such as real time data streaming, which is very stringent upon time requirements, a more accurate timer is need rather than the conventional SetTimer() supplied by “time.h”.

Bellow is the sample code for obtaining a finer granularity for a timer under Windows. Take note that the code below only returns an unsigned long, which is an integer and represents the number of microseconds since the machine was started. An additional observation would include the constant variable freq, the timer frequency, which in my case was set to 1.193180. This number was derived from Microsoft Corporation, as every processor is different and has a different freq number. Although this would make the timer unreliable when taken to a different machine, there are only a few possible frequency numbers, and thus it can be tested to adapt the frequency on the fly by testing the validity of the new timer against the supplied SetTimer(). Comments are denoted in Figure 54 as bold characters.

```
////////////////////////////////////////////////////////////////////////////////////////////////////////////////////////////////
//                                                                                          //
// PROCEDURE:                                                                               //
//                  MyTimeGetTime()                                                         //
//                                                                                          //
// PARAMETERS:                                                                              //
//                  N/A                                                                     //
//                                                                                          //
// DESCRIPTION:                                                                             //
//                  Gets the system time to an accuracy of microseconds.                    //
//                                                                                          //
// RETURNS:                                                                                 //
//                  Returns the system time to an accuracy of microseconds.                 //
//                                                                                          //
////////////////////////////////////////////////////////////////////////////////////////////////////////////////////////////////

unsigned long MyTimeGetTime()
{
        static unsigned long freq=1.193180;                    // timer frequency
        LARGE_INTEGER curtime;

        if (!freq)
        {       // determine timer frequency
                QueryPerformanceFrequency(&curtime);
                if (curtime.HighPart)
                {                                              // timer is too fast
                        printf("Timer too fast\n");
                        freq = 1;
                }                                              // timer is too fast
                else
                        freq = curtime.LowPart / constTimer;   // i.e., ticks per ms
                printf("freq = %ld\n", freq);
        }                                                      //    determine    timer
        frequency
        QueryPerformanceCounter(&curtime);
        return curtime.LowPart / freq;
}
```

**Figure 54: MyTimeGetTime() – Retrieves microseconds since boot-up under Windows 2000**

# APPENDIX C: ROUTER CONFIGURATIONS

This section is supposed to cover the router configuration commands performed on the Ericsson AXI 462 and the IBM 2216 Nways Multiaccess Connector Model 400 routers. Due to the very well built interface and documentation on the IBM router, we will not go through any details of the IBM router. On the other hand, the Ericsson router proved to be our biggest challenge in terms of configuring it. We compiled a list of commands which if executed in the stated order, will configure the router from scratch ready to be utilized as our IBM-Ericsson Test-bed depicted in Chapter 3. They are depicted in italics and smaller font than normal text. The commands are to be executed at the router console once the user has logged on. “AXI462 %” denotes the console prompt; comments are indicated by “//” preceding the particular text on the respective line, similar to the C++ standard. Comments are also denoted by being bold.

***//log on to the “paxtcl0” system***
*>paxtcl0*
***//enable routing process 1***
*AXI462 % ip routing 1*
*AXI462 % ip enable 3*
***//enable interface lan 3***
*AXI462 % interface lan 3*
***//set the address variable to 3:3:3:3:3:3:3:1 and give it the name LAN3_IPv6***
*AXI462 % ip access LAN3_IPv6 -local 3:3:3:3:3:3:3:1 -prefix 64*
***//write the information from the previous step and assign it to interface lan 3***
*AXI462 % ip laninterface*
***//set the RIP process for IPv6 with the appropriate flags***
*AXI462 % ip ripng -sendhost 1 -sendprf 1 -senddef 1 -sendagg 1 -sendstat 1 -nexthop 1*
***//enable RIP protocol for IPv6 for interface lan 3 and write setting to memory***

*AXI462 % start ripng*
***//enable interface lan 4***
*AXI462 % interface lan 4*
***//set the address variable to 4:4:4:4:4:4:4:1 and give it the name LAN4_IPv6***
*AXI462 % ip access LAN4_Ipv6 -local 4:4:4:4:4:4:4:1 -prefix 64*
***//write the information from the previous step and assign it to interface lan 4***
*AXI462 % ip laninterface*
***//set the RIP process for IPv6 with the appropriate flags***
*AXI462 % ip ripng -sendhost 1 -sendprf 1 -senddef 1 -sendagg 1 -sendstat 1 -nexthop 1*
***//enable RIP protocol for IPv6 for interface lan 4 and write setting to memory***
*AXI462 % start ripng*
***//use interface lan 3, notice that since it is already enable, we just use it***
*AXI462 % use interface lan 3*
***//set the address variable to 10.0.0.1 and give it the name LAN3_IPv4***
*AXI462 % ip access LAN3_Ipv4 -local 10.0.0.1 -prefix 8*
***//write the information from the previous step and assign it to interface lan 3***
***//as a secondary address since the IPv6 is the primary address***
*AXI462 % ip laninterface -secondary*
***//set the RIP2 process with the appropriate flags***
*AXI462 % ip rip*
***//enable RIP2 protocol for interface lan 3 and write setting to memory***
*AXI462 % start rip*
***//use interface lan 4, notice that since it is already enable, we just use it***
*AXI462 % use interface lan 4*
***//set the address variable to 141.217.17.49 and give it the name LAN4_IPv4***
*AXI462 % ip access LAN4_Ipv4 -local 141.217.17.49 -prefix 24*
***//write the information from the previous step and assign it to interface lan 3***
***//as a secondary address since the IPv6 is the primary address***
*AXI462 % ip laninterface -secondary*
***//set the RIP2 process with the appropriate flags***
*AXI462 % ip rip*
***//enable RIP2 protocol for interface lan 3 and write setting to memory***
*AXI462 % start rip*
***//use interface lan 3, notice that since it is already enable, we just use it***
*AXI462 % use interface lan 3*
***//set the tunnel IPv6 start point and end point on lan 3 and call it TUNNEL***
*AXI462 % ip access TUNNEL -local 3:3:3:3:3:3:3:1 -peer 3:3:3:3:3:3:3:3*
***//set the tunnel IPv4 end point and start point on lan 3 and write setting to memory***
*AXI462 % ip 4tunnel 10.0.0.3 -src 10.0.0.1*
***//set the RIP process for IPv6 tunnel with the appropriate flags***
*AXI462 % ip ripng -sendhost 1 -sendprf 1 -senddef 1 -sendagg 1 -sendstat 1 -nexthop 1*
***//enable RIP protocol for IPv6 for interface lan 4 and write setting to memory***
*AXI462 % start ripng*

**Figure 55: Ericsson AXI 462 router configuration commands**

# BIBLIOGRAPHY

[1] Information Sciences Institute, University of Southern California, "Internet Protocol," Request for Comments 791, Internet Engineering Task Force, September 1981

[2] S. Bradner, A. Mankin, "IP: Next Generation (IPng) White Paper Solicitation," Request for Comments 1550, Internet Engineering Task Force, December 1993

[3] R. Gilligan, E. Nordmark, "Transition Mechanisms for IPv6 Hosts and Routers," Request for Comments 1933, Internet Engineering Task Force, April 1996

[4] P. Srisuresh, M. Holdrege, "IP Network Address Translator (NAT) Terminology and Considerations," Request for Comments 2663, Internet Engineering Task Force, August 1999

[5] C. Huitema, "The H Ratio for Address Assignment Efficiency," Request for Comments 1715, Internet Engineering Task Force, November 1994

[6] Microsoft Corporation, "Microsoft IPv6 Technology Preview for Windows 2000," December 12, 2000,

http://www.microsoft.com/windows2000/technologies/communications/ipv6/default.asp

[7] A. S. Tanenbaum, Computer Networks, Third Edition, Prentice Hall Inc., 1996, pp. 686

[8] S. Deering, R. Hinden, "Internet Protocol, Version 6 (IPv6) Specification," Request for Comments 1883, Internet Engineering Task Force, December 1995

[9] A. S. Tanenbaum, Computer Networks, Third Edition, Prentice Hall Inc., 1996, pp. 413-449

[10] William Stallings. High Speed Networks, TCP/IP and ATM Design Principles. Pages 444 - 457.

[11] RSVP for the Multimedia Party. Packet Magazine Archives, Third Quarter 1995.

[12] Seaman, Mick and Klessig, Bob, 3 Com Corp. Going the Distance with QoS, Data Communication, February 1999. Pages 120.

[13] A. S. Tanenbaum, Computer Networks, Third Edition, Prentice Hall Inc., 1996, pp. 28-35

[14] A. S. Tanenbaum, Computer Networks, Third Edition, Prentice Hall Inc., 1996, pp. 35-44

[15] R. Braden, Ed., L. Zhang, S. Berson, S. Herzog, S. Jamin, "Resource ReSerVation Protocol (RSVP) -- Version 1 Functional Specification,"

Request for Comments 2205, Internet Engineering Task Force, September 1997

[16] R. Braden, D. Clark, S. Shenker, “Integrated Services in the Internet Architecture: an Overview,” Request for Comments 1633, Internet Engineering Task Force, June 1994

[17] M. Seaman, A. Smith, E. Crawley, J. Wroclawski, “Integrated Service Mappings on IEEE 802 Networks,” Request for Comments 2815, Internet Engineering Task Force, May 2000

[18] K. Nichols, S. Blake, F. Baker, D. Black, “Definition of the Differentiated Services Field (DS Field) in the IPv4 and IPv6 Headers,” Request for Comments 2474, Internet Engineering Task Force, December 1998

[19] Andrew S. Tanenbaum. Computer Networks, 3rd Edition. Pages 379 - 384.

[20] Hedrick C. “Routing Information Protocol”, Request for Comments 1058, Internet Engineering Task Force, June 1988

[22] Marcus A. Goncalves, Kitty Niles. “IPv6 Networks”, McGraw-Hill, 1998.

[23] S. Deering, R. Hinden, “IP Version 6 Addressing Architecture,” Request for Comments 1884, Internet Engineering Task Force, December 1995.

[24] S. Thomson, T. Narten, "IPv6 Stateless Address Autoconfiguration," Request for Comments 1971, Internet Engineering Task Force, August 1996.

[25] T. Narten, E. Nordmark, W. Simpson, "Neighbor Discovery for IP Version 6 (IPv6)," Request for Comments 1970, Internet Engineering Task Force, August 1996.

[26] C. Huitema, "IPv6, The New Internet Protocol, Second Edition," Prentice Hall Inc., 1997, pp. 197-221.

[27] G. Tsirtsis, P. Srisuresh, "Network Address Translation - Protocol Translation (NAT-PT),“ Request for Comments 2766, Internet Engineering Task Force, February 2000.

[28] A. Durand, P. Fasano, I. Guardini, D. Lento, "IPv6 Tunnel Broker, " Request for Comments 3053, Internet Engineering Task Force, January 2001.

[29] B. Carpenter, K. Moore, "Connection of IPv6 Domains via IPv4 Clouds," Request for Comments 3056, Internet Engineering Task Force, February 2001.

[30] A. Conta, S. Deering, "Generic Packet Tunneling in IPv6 Specification," Request for Comments 2473, Internet Engineering Task Force, December 1998

[31] B. Carpenter, C. Jung, “Transmission of IPv6 over IPv4 Domains without Explicit Tunnels,” Request for Comments 2529, Internet Engineering Task Force, March 1999.

[32] Fiuczynski, Marc E et. Al. “The Design and Implementation of an IPv6/IPv4 Network Adress and Protocol Translator”. 1998.

[33] Karuppiah, Ettikan Kandasamy, et al. “Application Performance Analysis in Transition Mechanism from IPv4 to IPv6”. Research & Business Development Department, Faculty of Information Technology, Multimedia University (MMU), Jalan Multimedia, June 2001.

[34] Draves, Richard P., et al. “Implementing IPv6 for Windows NT”. Proceedings of the 2nd USENIX Windows NT Symposium, Seattle, WA, August 3-4, 1998.

[35] Seiji Ariga, Kengo Nagahashi, Asaki Minami, Hiroshi Esaki, Jun Murai. “Performance Evaluation of Data Transmission Using IPSec over IPv6 Networks”, INET 2000 Proceedings, Japan, July 18th, 2000

[36] Peter Ping Xie. “Network Protocol Performance Evaluation of IPv6 for Windows NT”, Master Thesis, California Polytechnic State University, San Luis Obispo, June 1999.

[37] Ettikan Kandasamy Karuppiah. “IPv6 Dual Stack Transition Technique Performance Analysis: KAME on FreeBSD as the Case”,

Faculty of Information Technology, Multimedia University (MMU), Jalan Multimedia, October 2000

# ABSTRACT

AN EMPIRICAL ANALYSIS OF
INTERNET PROTOCOL VERSION 6 (IPV6)

by

IOAN RAICU

MAY 2002

Advisor: Dr. Sherali Zeadally

Major: Computer Science

Degree: Master of Science

Although the current Internet Protocol known as IPv4 has served its purpose for over 20 years, its days are numbered. With IPv6 reaching a mature enough level, there is a need to evaluate the performance benefits or drawbacks that the new IPv6 protocol will have in comparison to the well established IPv4 protocol. Theoretically, the overhead between the two different protocols should be directly proportional to the difference in the packet's header size, however according to our findings, the empirical performance difference between IPv4 and IPv6, especially when the transition mechanisms are taken into consideration, is much larger than anticipated. We first examine the performance of each protocol independently. We then examined two transition mechanisms which perform the encapsulation at various points in the network: host-to-host and router-to-router (tunneling). This is a necessary and crucial step for IPv6's success since clear performance limitations and advantages should be well defined and agreed upon in advance before any major transitions take place. Our experiments were conducted using two dual stack (IPv4/IPv6) routers using end nodes running both Windows 2000 and Solaris 8.0 in order to compare two different IPv6 implementations side by side. Our tests were written in C++ and utilized metrics such as latency, throughput, CPU utilization, socket creation time, socket connection time, web server simulation, and a video client/server application for TCP/UDP in IPv4/IPv6 under both Windows 2000 and Solaris 8.0. Our goal was to perform an unbiased empirical performance evaluation between the two protocol stacks (IPv4 and IPv6), and how it related to the performance of the encapsulation methods utilized on identical hardware and under identical settings. Our empirical evaluation proved that IPv6 is not yet a mature enough technology and that it is still years away from having consistent and good enough implementations. The performance of IPv6 in many cases proved to be worse than IPv4, incurring an overhead much higher than its anticipated theoretical counterpart.

## AUTOBIOGRAPHICAL STATEMENT

IOAN RAICU

As a brief summary, I received my Bachelor of Science and Master of Science in Computer Science from Wayne State University in 2000 and 2002 respectively. My Master thesis title is “AN EMPIRICAL ANALYSIS OF INTERNET PROTOCOL VERSION 6 (IPV6)”, which dealt with quantifying the performance degradation of IPv6 compared to IPv4, its predecessor. I also did some work in wireless sensor networks, specifically on various routing algorithms.

During my undergraduate career in 1997, I founded a company called High Teck Computers. Owning a business has many lessons about responsibilities, punctuality, perseverance, communication skills, and last but not least, the development of a creative mind. The business was an excellent experience that could not have been obtained any other way, however it was an activity that required much dedicated time. In order to better concentrate on my research, in September 2000, I suspended all business activity.

In the fall of 2000, I started my Master of Science degree in Computer Science at Wayne State University under the supervision of my research advisor, Dr Sherali Zeadally. My graduate studies covered many topics over the broad category of networking: QoS, Resource Reservation Protocol (RSVP), IPv6, Wireless Sensor Networks. My experiences included everything from theoretical to practical; the High Speed Networking Lab offered me all the necessary equipment to conduct experiments to better understand how various network protocols (IPv4, IPv6, RSVP, etc) work and behave.

In regards to wireless sensor networks, I developed an indoor proximity detector in order to aggregate to an already existing GPS outdoor tracking system. By proving that distance can be incurred from the radio signal strength, I pursued analyzing various routing algorithms for wireless sensor networks based on diffusion in which decisions would be made based on distance of transmission and power levels in the sending and receiving nodes. Furthermore, variable size packets, and fragmentation mechanisms were used in order to minimize the power dissipation of the sensor nodes.

My long term goals involves completing a Ph.D. degree, join a university or industry lab, and eventually return back to square one in owning my own business, but on a much grander scale. Education is a never-ending process that keeps our intellectuality in top shape so we may have the ability to turn dreams into realities.

For more information, please refer to my web page at http://www.cs.wayne.edu/~iraicu/ or contact me at iraicu@cs.wayne.edu.